\documentclass{jfm}
\usepackage{graphicx}
\usepackage{siunitx}
\usepackage{gensymb}
\usepackage{epstopdf}
\usepackage{epsfig}
\usepackage{subcaption}
\usepackage{soul}
\usepackage{array}
\usepackage{cancel}
\usepackage{bm}
\usepackage{xcolor}
\definecolor{red}{RGB}{255,0,0}
\newcolumntype{L}{>{\centering\arraybackslash}m{3cm}}
\AppendGraphicsExtensions{.tif}

\graphicspath{{./figures/}}

\shorttitle{Velocity and spatial distribution of inertial particles in a turbulent channel flow }
\shortauthor{K. O. Fong, O. Amili and F. Coletti}

\title{Velocity and spatial distribution of inertial particles in a turbulent channel flow}

\author{Kee Onn Fong \aff{1}\textsuperscript{,}\aff{2}
  \corresp{\email{fongx065@umn.edu}}, Omid Amili\aff{1}
 \and Filippo Coletti\aff{1}\textsuperscript{,}\aff{2}}

\affiliation{\aff{1}Department of Aerospace Engineering and Mechanics, University of Minnesota,
Minneapolis, MN 55455, USA
\aff{2}St. Anthony Falls Laboratory, University of
Minnesota, Minneapolis, MN 55414, USA}

\begin{document}

\maketitle

\begin{abstract}
We present experimental observations of the velocity and spatial distribution of inertial particles dispersed in the turbulent downward flow through a vertical channel at friction Reynolds numbers $\Rey_\tau = 235$ and 335. The working fluid is air laden with size-selected glass micro-spheres, having Stokes numbers St = $\mathcal{O}$(10) and $\mathcal{O}$(100) when based on the Kolmogorov and viscous time scales, respectively. Cases at solid volume fractions $\phi_v = 3\times10^{-6}$ and $5\times10^{-5}$ are considered. In the more dilute regime, the particle concentration profile shows near-wall and centerline maxima compatible with a turbophoretic drift down the gradient of turbulence intensity; the particles travel at similar speed as the unladen flow except in the near-wall region; and their velocity fluctuations generally follow the unladen flow level over the channel core, exceeding it in the near-wall region. The denser regime presents substantial differences in all measured statistics: the near-wall concentration peak is much more pronounced, while the centerline maximum is absent; the mean particle velocity decreases over the logarithmic and buffer layers; and particle velocity fluctuations and deposition velocities are enhanced. An analysis of the spatial distributions of particle positions and velocities reveals different behaviors in the core and near-wall regions. In the channel core, dense clusters form which are somewhat elongated, tend to be preferentially aligned with the vertical/streamwise direction, and travel faster than the less concentrated particles. In the near-wall region, the particles arrange in highly elongated streaks associated to negative streamwise velocity fluctuations, several channel height in length and spaced by $\mathcal{O}$(100) wall units, supporting the view that these are coupled to fluid low-speed streaks typical of wall turbulence. The particle velocity fields contain a significant component of random uncorrelated motion, more prominent for higher St and in the near-wall region.

\end{abstract}

\section{Introduction}
Wall-bounded turbulent flows laden with inertial particles are relevant to a broad spectrum of environmental, biomedical and industrial processes. Examples include sediment transport in rivers \citep{nino1996}, aerosol inhalation in human airways \citep{kleinstreuer2010}, and reactors in chemical engineering processes \citep{capecelatro2014}. The motion of heavy particles in homogeneous turbulence is already complex, featuring well-known (though not fully understood) phenomena such as preferential concentration and consequent clustering (\citealt{eaton1994}, \citealt{monchaux2012}, \citealt{bragg2014}, \citealt{gustavsson2016}). The latter is thought to be maximized when the particle response time, $\tau_p$, is comparable to the Kolmogorov time scale, $\tau_\eta$, such that the Stokes number $St_\eta = \tau_p/\tau_\eta$ is of order unity (\citealt{wang1993}, \citealt{fessler1994}). In the presence of a wall, a mean drift of the particles sets up following the negative gradient of turbulence intensity (so-called turbophoresis) and causing a segregation of particles towards the wall (\citealt{caporaloni1975}, \citealt{reeks1983}, \citealt{young1997}, \citealt{guha2008}, \citealt{fouxon2018}). The relevant parameter is usually considered the Stokes number $St^+ = \tau_p/\tau_{\nu}$, where $\tau_{\nu}$ is the viscous time scale, with strong turbophoresis for $St^+ \sim 10\,–\,100$ (e.g., \citealt{marchioli2002}, \citealt{sardina2012a}, \citealt{bernardini2014}). Moreover, inertial particles were experimentally observed to arrange in long near-wall streaks (\citealt{kaftori1995a}, \citealt{kaftori1995b}, \citealt{nino1996}), and numerical simulations demonstrated the role of coherent turbulent structures in determining such behavior (\citealt{mclaughlin1989}; \citealt{zhang2000}; \citealt{rouson2001}; \citealt{marchioli2002}; \citealt{soldati2009}; \citealt{sardina2012a}; \citealt{nilsen2013}; \citealt{richter2013}; \citealt{bernardini2014}). For typical Reynolds numbers used in laboratory and computational studies, the parameter ranges leading to both clustering and turbophoresis overlap. In fact, it has been argued that both phenomena represent different aspects of the same process \citep{sardina2012a}. Moreover, although rarely discussed in particle-laden turbulence studies, inelastic particle collisions may also contribute to near-wall particle accumulation \citep{hrenya1997}. \textcolor{black}{Other relevant experiments with similar scope includes the vertical pipe study of \citet{varaksin2000} and the horizontal channel study of \citet{wu2006}.}

Most of the studies mentioned above considered very dilute particles smaller than all scales of the flow – a regime in which the backreaction of the dispersed phase on the carrier fluid is usually deemed negligible. According to \citet{elghobashi1994}, this condition (referred to as one-way coupling) is satisfied only for volume fractions $\phi_v \leq $ $\mathcal{O}$$(10^{-6})$, while at higher loadings the particles do influence the turbulence (two-way coupling). Such classification, however, was merely proposed as a guideline for numerical approaches; the boundary between both regimes is problem-dependent and is affected by other physical parameters, including the particle-to-fluid density ratio, $\rho_p/\rho_f$. In wall-bounded turbulence, the flow dynamics and the local particle concentration also evolve with wall distance, especially in \textcolor{black}{the} presence of turbophoresis. Several numerical studies investigated the two-way coupled regime by direct numerical simulation (DNS) of wall turbulence, representing the particles as material points that exchange momentum with the fluid (see, for example,  \citealt{pan1996}; \citealt{vreman2007}; \citealt{zhao2010}; \citealt{dritselis2011}; \citealt{richter2013}; \citealt{richter2014}; \citealt{li2016}; \citealt{wang2018}). The particles were found to affect the ejection-sweep cycle, the dynamics of streamwise vortices, the formation and strength of hairpin eddies, and in general to significantly modify the fluid Reynolds stresses. These modifications to the flow, in turn, altered the particle transport and thus their concentration and velocity statistics. With increasing of the particle mass loading, $\phi_m = \phi_v\times\rho_p/\rho_f$ , simulations also indicated the sizeable effect of inter-particle collision (four-way coupling), notably in reducing the near-wall concentration otherwise enhanced by turbophoresis (\citealt{li2001}; \citealt{vreman2007}, \citealt{nasr2009}, \citealt{kuerten2015}). Recently \citet{capecelatro2018} demonstrated a dramatic change from shear-production-dominated to drag-production-dominated regimes when the mass loading increased from $\mathcal{O}$(0.1) to $\mathcal{O}$(10).

Despite the remarkable insight offered by the point-particle approach, this method presents well-known limitations, partly related to the point-wise forcing on the fluid computational grid (\citealt{eaton2009}, \citealt{balachandar2010}). To overcome these shortcomings, advanced simulation strategies have been proposed (\citealt{capecelatro2013}, \citealt{gualtieri2015}, \citealt{horwitz2016}, \citealt{ireland2017}, \citealt{balachandar2019}). In general, our understanding of the physics of two-way coupled particle-laden turbulence is still incomplete, and as a result any simplified model may miss or misestimate significant aspects \citep{balachandar2010}. The availability of ever-increasing computational capabilities has allowed particle-resolved DNS to investigate relatively large numbers of particles in wall-bounded turbulent flows without the need of modeling the momentum exchange (\citealt{garcia2012}; \citealt{picano2015}; \citealt{lin2017}; \citealt{wang2017}). Those studies, however, can typically deal with $\mathcal{O}$$(10^{4})$ particles much larger than the viscous scales, as opposed to the millions of sub-Kolmogorov particles usually present in point-particle simulations. The latter situation is most relevant to gas-solid mixtures.

In this scenario, the importance of well-controlled laboratory experiments is paramount to reach a predictive understanding of these regimes, and to inform and validate numerical models. Unfortunately, similar studies are rare in the literature and cover limited portions of the parameter space. Several past experiments considered particle-laden water flows with $\rho_p/\rho_f =$ $\mathcal{O}$(1) and particle diameters of several wall units (\citealt{kaftori1995a}; \citealt{kaftori1995b}; \citealt{nino1996}; \citealt{kiger2002}; \citealt{righetti2004}; \citealt{rabencov2014}; \citealt{oliveira2017}; \citealt{shokri2017}). These conditions are relevant to sediment transport and pipelines, but not to other important applications such as dust and particulate transport in air. For gas-solid suspensions, a non-exhaustive list of previous experiments and their relevant physical parameters is provided in table 1. \textcolor{black}{Currently,} the main reference is \textcolor{black}{still} represented by the vertical channel flow measurements by Eaton and co-workers. In particular, \citet{fessler1994} and \citet{kulick1994} provided seminal insight into preferential concentration and turbulence modulation for a variety of regimes. However, \citet{benson2005} showed that their apparatus presented substantial wall roughness due to particle deposition on the walls, which according to \citet{vreman2015} partly explained the disagreement with simulations. \citet{benson2005} repeated the measurements with a smooth test section, only focusing on the most inertial particles that did not display turbophoresis. \citet{taniere1997} investigated a particle-laden boundary layer in a horizontal wind tunnel, focusing on particles with $St^+ > 270$. As in all horizontal flow configurations, the particle concentration profile was strongly impacted by gravity. \citet{kussin2002} measured particle motion and concentration as well as turbulence modulation in a horizontal channel flow with rough walls. Their particles were highly inertial and could not display turbophoresis or turbulence-induced clustering. \citet{caraman2003} considered a vertical particle-laden pipe flow in a regime where turbophoresis is expected, and carried out a detailed analysis of the moments of the particle velocity. Still, comparison with simulations was hampered by the measurement station being downstream of the pipe exit, the lack of concentration profiles, and possible wall roughness \citep{vreman2007}. \citet{khalitov2003} conducted measurements in a vertical channel flow laden with glass spheres of various sizes, covering a range of Stokes numbers for which turbophoresis is expected. They documented both particle-particle and gas-particle \textcolor{black}{velocity correlations in the streamwise and spanwise directions}, but not the concentration profiles. \citet{hadinoto2005} considered a vertical pipe flow laden with glass beads that were too inertial to segregate at the wall. \citet{li2012} imaged inertial particles at $St^+ \approx 100$ in a range of concentrations where two-way coupling effects are expected to be weak. Because their channel was horizontal and they only reported data near the bottom wall, possible turbophoresis was not distinguishable from gravitational settling.

Overall, there is a clear lack of laboratory observations of wall-bounded gas-solid flows in regimes where preferential concentration and turbophoresis are at play. In particular, little is known on the changes occurring when varying the loading across what is considered the boundary between one-way and two-way coupling. Liquid-solid flow studies cannot compensate for these deficiencies in the literature, as the momentum coupling is heavily affected by the density ratio. Importantly, concentration profiles are seldom reported, and therefore near-wall segregation (clearly evident in simulations) has not been fully documented. The seminal studies usually cited as experimental evidence of this phenomenon (e.g., \citealt{kaftori1995a}; \citealt{kaftori1995b}) were carried out in horizontal flumes where gravitational effects may be important. In order to bridge such knowledge gap, the present study experimentally investigates the transport of small solid particles in turbulent air flowing downward in a smooth-wall vertical channel. We focus on regimes (summarized in table 1) for which significant clustering and turbophoresis are expected, and use planar imaging to analyze the particle behavior for different levels of mass loading. The paper is organized as follows: in §2 we describe the laboratory facility and the methods used to conduct the experiments and analyze the data; in §3 we present the wall-normal profiles of particle concentration and velocities, and the spatial fields along wall-parallel planes at the channel core and near the wall; conclusions and an outlook for further research are provided in §4.
 
\begin{table}
  \begin{center}

  \begin{tabular}{@{}p{0.25\linewidth}p{0.06\linewidth}p{0.06\linewidth}p{0.07\linewidth}p{0.10\linewidth}p{0.08\linewidth}p{0.2\linewidth}p{0.1\linewidth}@{}}
        & $Re_\tau$  &  $St^+$ & $d^+$ & $\phi_v$ & \textcolor{black}{$\phi_m$} & Configuration & Wall quality \\[3pt]
       \citet{fessler1994} & 630 & 27-150 & 0.8-3 & $5\times10^{-5}$ -$2\times10^{-4}$ & \textcolor{black}{0.03-1} & Vertical channel & Rough \\[3pt]
       \citet{kulick1994}  & 630	& 1500-2000 & 2.5-3 & $5\times10^{-5}$ -$2\times10^{-4}$ & \textcolor{black}{0.02-0.8} & Vertical channel & Rough \\[3pt]
       \citet{taniere1997}  & 1700 & 270-540 & 1.6-2.3 & $5\times10^{-6}$ & \textcolor{black}{0.006-0.01} & Horizontal boundary layer & Smooth \\[3pt]
       \citet{kussin2002}  & 700-1300 & 570-7800 & 3-38 & $4\times10^{-4}$-$4\times10^{-3}$  & \textcolor{black}{0.1-1} & Horizontal channel & Rough \\[3pt]
       \citet{caraman2003}  & 133 & 70 & 0.8 & $5\times10^{-5}$ & \textcolor{black}{0.11} & Vertical pipe & Smooth \\[3pt]
       \citet{khalitov2003}  & 238 & 50-3190 & 0.6-5.1 & $5\times10^{-5}$ & \textcolor{black}{0.1} & Vertical channel & Smooth \\[3pt]
       \citet{hadinoto2005} & 253-544 & 730-2700 & 0.8-15 & $3\times10^{-4}$ &\textcolor{black}{0.7} & Vertical pipe & Smooth\\[3pt]
       \citet{benson2005}  & 630 & 1800 & 5 & $7\times10^{-5}$ & \textcolor{black}{0.15} & Vertical channel & Smooth, rough \\[3pt]
       \citet{li2012}  & 430 & 100 & 1.7 & $3\times10^{-7}$ -$5\times10^{-6}$ & \textcolor{black}{0.00025-0.005} & Horizontal channel & Smooth \\[3pt]
       Present study & 235-335 & 64-130 & 0.8-1.1 & $3\times10^{-6}$ -$5\times10^{-5}$ & \textcolor{black}{0.006-0.1} & Vertical channel & Smooth\\[3pt]
  \end{tabular}
  \caption{A list of experimental studies addressing gas-solid wall-bounded flows. When not explicitly reported in the referenced papers, the parameters are calculated using information therein.}
  \label{tab:past-exp}
  \end{center}
\end{table}

\section{Experimental method}\label{sec:method}
\subsection{Experimental facility and parameters}
\begin{figure}
  \centerline{\includegraphics[width=1.3\textwidth]{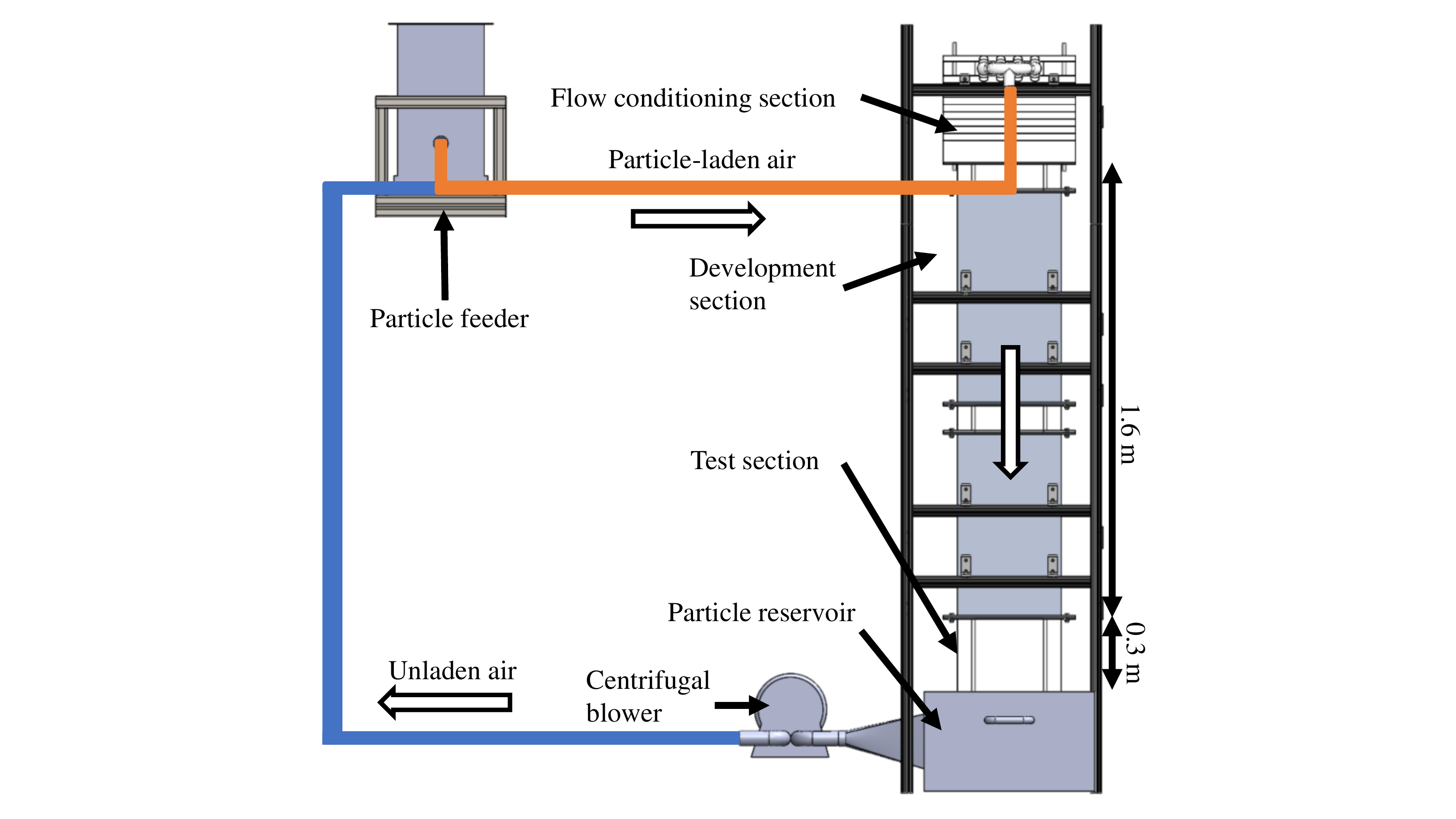}}
  \caption{A schematic diagram of the particle-laden channel flow facility and its main elements.}
\label{fig:solidworks_diagram}
\end{figure}

Experiments are conducted in a vertical recirculating wind tunnel depicted in figure \ref{fig:solidworks_diagram}, featuring a 1.9 m long rectangular channel with a 0.24 m by 0.03 m cross-section. A 1.5 kW centrifugal blower (Atlantic Blowers) controlled by a frequency converter drives air downwards, and the flow rate is continuously monitored via a Venturi flowmeter. Before the air enters the channel, size-selected glass beads (Mo-Sci Corp.) with a density of 2500 kg/m\textsuperscript{3} and diameter of 50 $\pm$ 6 \si{\um} (mean $\pm$ standard deviation measured by optical microscopy over $\mathcal{O}$$(10^4)$ samples) are injected into the flow through a precision screw-feeder (Vibra Screw Inc.).  \textcolor{black}{This corresponds to a ratio of channel width to particle diameter of $2h/d_p = 600$. }A flow conditioning section consisting of four screens and three honeycombs is placed at the channel inlet to disperse the particles uniformly. The measurement station consists of a 0.3 m long, fully transparent acrylic section that follows a 1.6 m long development section. The latter has a through-flow time 25 - 40 times larger than the particle response time, depending on the air flow regime. Integration of the particle equation of motion with the Schiller and Neumann’s correction \citep{clift2005} indicates that the particles reach their terminal velocity in about half the time it takes them to reach the test section. The particles exhausted from the channel are collected in a 109-liter settling chamber, allowing for the run times needed to achieve well-converged statistics without particles being ingested into the blower. Electrostatic dissipative acrylic (SciCron Technologies) is used to build the optical test section, and the channel is provided with static discharge wires grounded to structural supports. This prevents the particles from accumulating upon impaction and building up unwanted roughness, an effect that has impacted past experiments \citep{benson2005}. \textcolor{black}{This point is demonstrated and elaborated upon in Appendix A.} 

The fluid and particle parameters for the investigated cases are listed in table 2. Two flow rates are considered, associated \textcolor{black}{with} friction Reynolds numbers $Re_\tau$ = 235 and 335, respectively. The particle Reynolds number, defined with the still-air terminal velocity of the particles $V_t = \tau_pg = $ 0.17 m/s as a representative relative velocity with respect to the fluid, is $Re_p<1$ (here g is the gravitational acceleration). Using the particle root mean square (rms) of the particle velocity fluctuations as a velocity scale leads to higher values, but still within $Re_p = $ $\mathcal{O}$(1), suggesting that particle wakes negligibly affect the flow. The Froude number, defined as the ratio of the centerline fluid velocity over the still-air settling velocity, satisfies the condition $Fr\gg1$. It has been argued that this warrants a negligible influence of gravity (see the boundary layer study of \citet{sardina2012b}, where the freestream velocity is used to define $Fr$). However, this condition is not strictly applicable to the near-wall region where the fluid velocity vanishes. Moreover, for particle-turbulence interaction the relevant velocity scale is arguably the fluid rms fluctuation, which at the channel centerline is of the same order as the still-air settling velocity. Indeed, recent measurements in a vertical pipe from \citet{oliveira2017} at $Fr>10$ show large differences in particle behavior between downward and upward flow. Therefore, we will not generalize the present findings to other channel orientations, as gravity may play a significant role in this regime (as also discussed in §3).

\begin{table}
  \begin{center}
\def~{\hphantom{0}}
  \begin{tabular}{lcc}
         & LoSt & HiSt\\[3pt]
      \multicolumn{3}{l}{$Fluid\,\,phase\,\,parameters$} \\[3pt]
	Fluid density, $\rho_f$ (kg m\textsuperscript{-3}) & \multicolumn{2}{c}{1.2} \\[3pt]
	Kinematic viscosity, $\nu$ (m\textsuperscript{2} s\textsuperscript{-1}) & \multicolumn{2}{c}{$1.5\times10^{-5}$} \\[3pt]
	Channel half height, $h$ (mm) & \multicolumn{2}{c}{15} \\[3pt]
	Fluid centerline velocity, $U_c$ (m s\textsuperscript{-1}) & 4.41 & 6.51 \\ [3pt]
	Fluid bulk velocity, $U_{bulk}$ (m s\textsuperscript{-1}) & 3.01 & 4.66 \\ [3pt]
	Bulk Reynolds number, $Re_{bulk}=2hU_{bulk}/\nu$ & 6020 & 9320 \\ [3pt]
	Fluid friction velocity, $U_{\tau}$ (m s\textsuperscript{-1}) & 0.235 & 0.335 \\ [3pt]
	Friction Reynolds number, $Re_{\tau}=hU_{\tau}/\nu$ & 235 & 335 \\ [3pt]
	Viscous lengthscale, $\lambda_{\nu}$ (\si{\um}) & 64 & 45 \\ [3pt] 
	Viscous timescale, $\tau_{\nu}$ (ms) & 0.27 & 0.13 \\ [3pt]
	Kolmogorov scale close to centerline, $\eta_c$ (\si{\um}) & 200 & 150 \\[3pt]
	Kolmogorov timescale (at centerline), $\tau_{\eta}$ (ms) & 2.6 & 1.5 \\[3pt]
	\\[3pt]
	\multicolumn{3}{l}{$Particle\,\,characteristics$} \\[3pt]
	Density, $\rho_p$ (kg m\textsuperscript{-3}) & \multicolumn{2}{c}{2500} \\[3pt]
	Mean diameter, $d_p$ (\si{\um}) & \multicolumn{2}{c}{50} \\[3pt]
	Mean diameter in wall units, $d_p^+$ & 0.78 & 1.1 \\[3pt]
	Aerodynamic response time, $\tau_p$ (ms) & \multicolumn{2}{c}{17} \\[3pt]
	Reynolds number, $Re_p = d_p\tau_pg/\nu$ & \multicolumn{2}{c}{0.6} \\[3pt]
	Froude number, $Fr = U_c/(\tau_pg)$ & 26 & 38 \\ [3pt]
	Kolmogorov-based Stokes number, $St_{\eta}$ & 6.7 & 11.5 \\[3pt]
	Viscous Stokes number, $St^+$ & 64 & 130 \\[3pt]
	Restitution coefficient, $e$ & \multicolumn{2}{c}{0.73} \\[3pt]
  \end{tabular}
  \caption{Fluid and particle parameters for the investigated cases. \textcolor{black}{$\eta_c$ is calculated using the relation $\eta_c$ = $\lambda_\nu(\kappa Re_\tau)^{1/4}$ \citep{pope2000}.}}
  \label{tab:params}
  \end{center}
\end{table}

The regimes $Re_\tau = 235$ and 335 are associated \textcolor{black}{with} different fluid timescales, resulting in two different Stokes number cases referred to as LoSt and HiSt, respectively. The change in Reynolds number, while not inconsequential, is expected to have a lesser impact over the considered range compared to the variation in $St$. The range of $St^+$ and $St_\eta$ (the latter being defined with the Kolmogorov timescale at the channel centerline in the unladen flow) suggests significant turbophoresis and preferential concentration. For each Stokes number, two sets of measurements are carried out by changing the screw size in the particle feeder, resulting in volume fractions $\phi_v = 3\times10^{-6}$ (for a mass loading $\phi_m = 0.6\%$) and $5\times10^{-5}$ ($\phi_m = 10\%$), referred to as LoVF and HiVF, respectively. These correspond to global concentrations $C_0 = $ 46 and 880 particles/cm\textsuperscript{3}, respectively. The four-case matrix is summarized in table \ref{tab:expcases}. The fluid phase parameters in table \ref{tab:params} are based on the unladen flow, characterized by particle image velocimetry (PIV) as described in the following. While the mass loading for the LoVF cases is not expected to produce sizeable changes in the fluid flow, in the HiVF cases the turbulence is likely to be impacted \citep{kulick1994}, and therefore the listed values of the flow properties should be regarded as estimates.

\textcolor{black}{The coefficient of restitution, $e$ for particle-wall collisions is measured in a separate experiment. $e$ is defined as the ratio of the wall-normal particle velocity just after and just before the collision; details on the experimental measurements are elaborated upon in Appendix B. It is known that the value of $e$ is dependent on the particle Reynolds number and Stokes number based on the particle velocity relative to the fluid \citep{gondret2002}. However, the terminal velocity happens to be on the order of the rms wall-normal velocity of the particles in the vicinity of the wall (0.02$U_c$ – 0.05$U_c$, where $U_c$ is the centerline velocity), as will be shown later. Thus, the quoted restitution coefficient is relevant to the particle-laden flow condition. Still, the actual collision velocity in the channel flow experiment is expected to have significant scatter; possible consequences of such variance are discussed in §4.}

\begin{table}
  \begin{center}
\def~{\hphantom{0}}
  \begin{tabular}{ccc}
      Parameters & $St^+ = 64$ & $ St^+ = 130 $ \\[3pt]
       $ \phi_v = 3\times10^{-6} $ & LoSt-LoVF & HiSt-LoVF \\[3pt]
       $ \phi_v = 5\times10^{-5} $ & LoSt-HiVF & HiSt-HiVF \\[3pt]
  \end{tabular}
  \caption{Cases studied in this experiment and respective notation.}
  \label{tab:expcases}
  \end{center}
\end{table}

\subsection{Measurement methods}
\begin{figure}
\centering
\begin{subfigure}{.40\textwidth}
  \centerline{\includegraphics[scale=0.35]{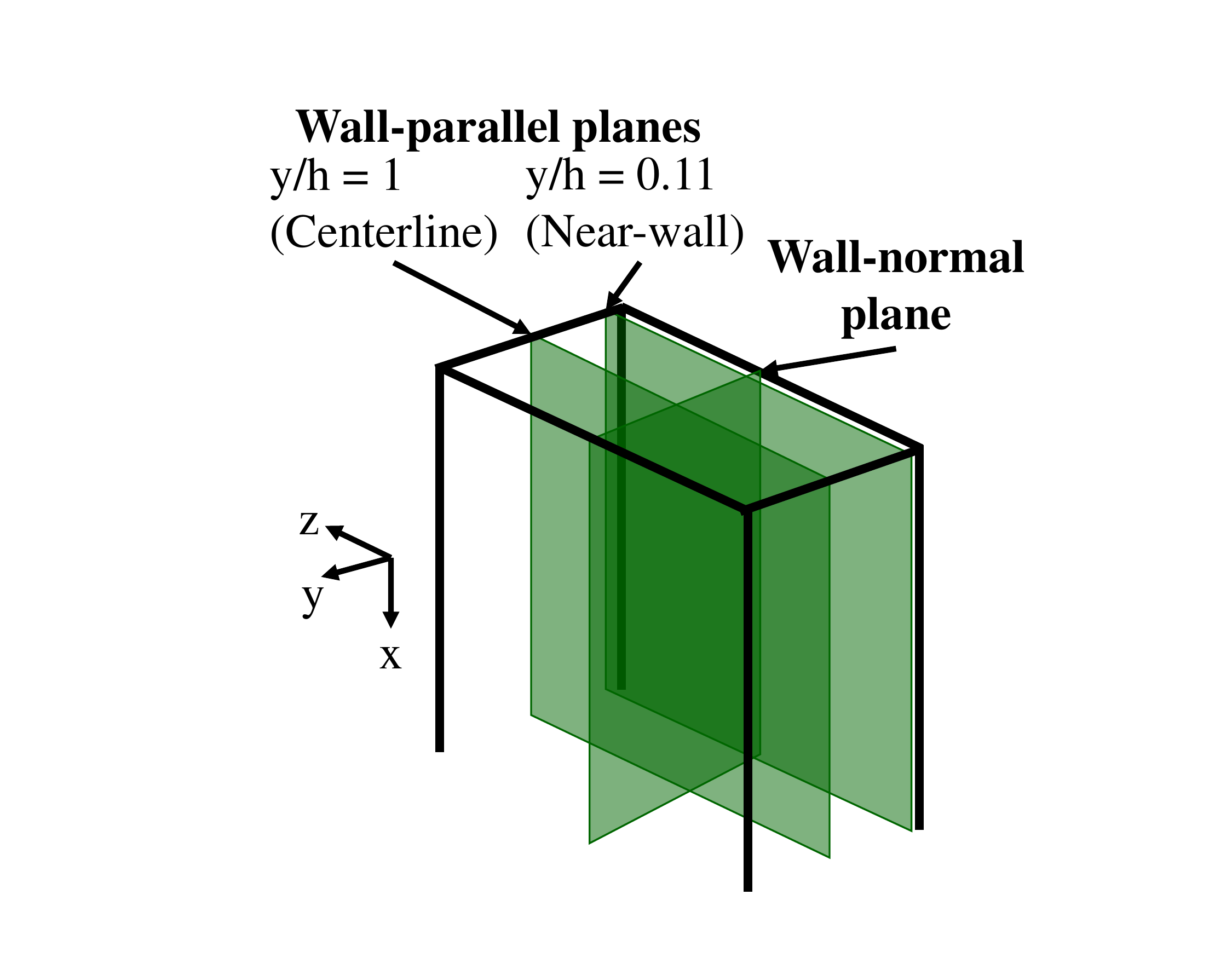}}
  \caption{}
\end{subfigure}
\begin{subfigure}{.59\textwidth}
  \centerline{\includegraphics[scale=0.425]{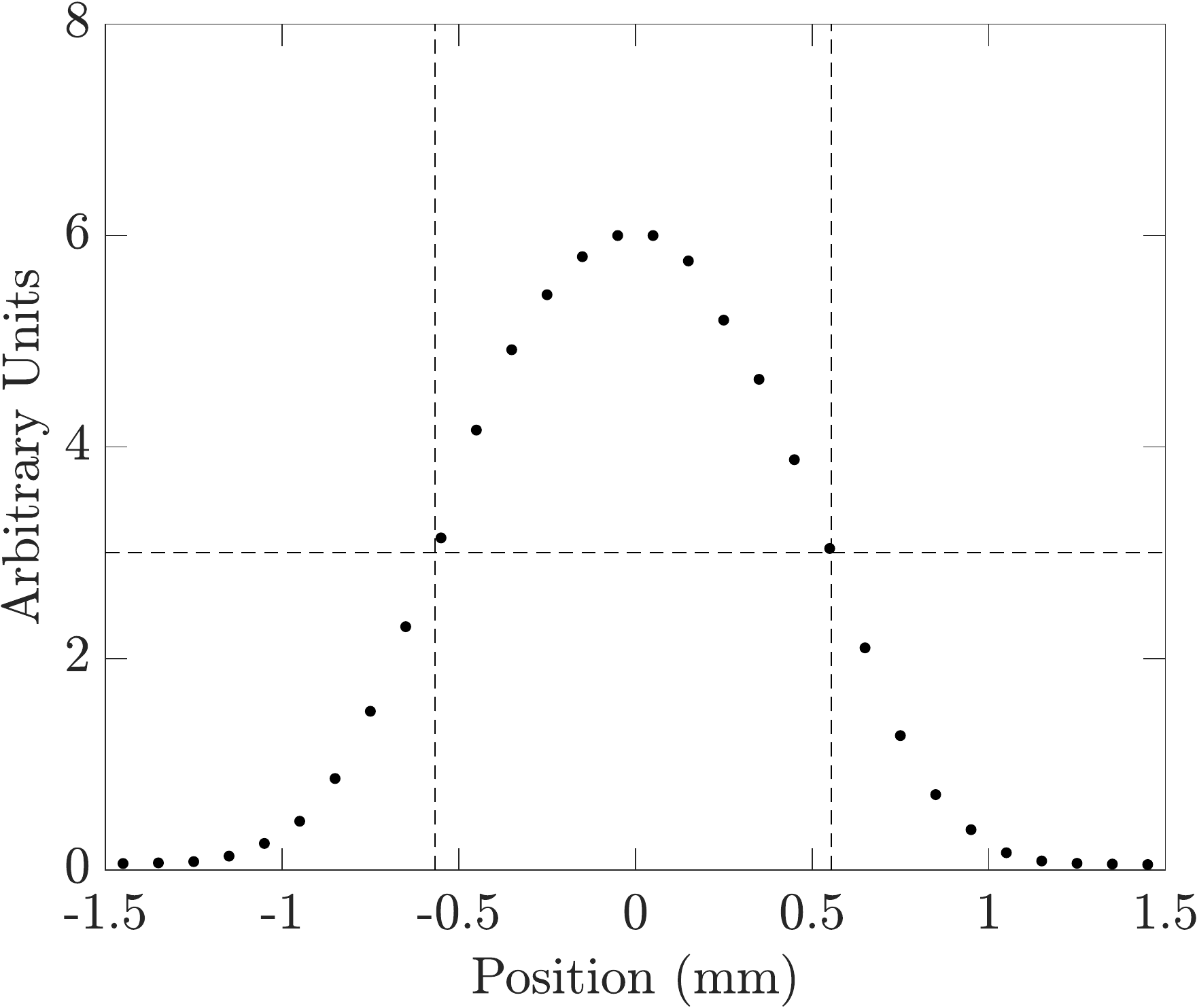}}
  \caption{}
\end{subfigure}
  \caption{(a) Main imaging planes investigated in the present study. (b) Light intensity profile as measured by a photodiode to estimate the laser sheet thickness. The dashed lines mark the full width at half maximum.}
\label{fig:planes}
\end{figure}

The imaging system consists of a double-pulsed Nd:YAG laser (30 mJ/pulse, New Wave Instruments) and a 1376 $\times$ 1040 pixel CCD camera (Sensicam, PCO Instruments) operated at 5 Hz and synchronized to the laser via a delay generator (BNC-5500, Berkeley Nucleonics). Planar measurements are obtained by shaping the laser beam into a sheet via an optical module combining cylindrical and spherical lenses. For all cases, the measurement locations include a wall-normal plane that contains the channel centerline. We denote with $x$, $y$, and $z$ the streamwise, wall-normal, and spanwise directions, the channel walls being located at $y = 0$ and $y = 2h$. For the HiVF cases, several wall-parallel planes are also imaged and compared, one at the centerline ($y = h$) and one in the near-wall region ($y = 0.11h$), see figure \ref{fig:planes}a. An additional wall-parallel plane (not shown in figure \ref{fig:planes}a) is imaged at $y = 0.2h$ for comparisons with wall-normal profiles. Spatial calibration and plane location are performed by imaging a target plate mounted on a micrometric traverse, inserted in the channel from the opening at the outflow end. The laser sheet thickness is evaluated using a photodiode (Thorlabs Inc.) coupled with a neutral density filter (CW Optics) and mounted on a traverse. The full width at half maximum is approximately 1.1 mm (Figure \ref{fig:planes}b).

The unladen (single-phase) case is characterized using 2C-2D PIV along the wall-normal symmetry plane. The flow is seeded with DEHS oil atomized by a Laskin nozzle into 1 - 2 \si{\um} droplets that faithfully follow the flow, their viscous Stokes number being $\mathcal{O}$($10^{-2})$. For those measurements, the CCD camera mounts a 200-mm Micro-Nikkor lens at f/5.6 to obtain a 13.3 $\times$ 10.1 mm\textsuperscript{2} field of view. The full channel height is covered by stitching together four slightly overlapping windows. The pulse delay is set to 10 \si{\us} and 15 \si{\us} for the higher and lower flow rate, respectively, keeping the typical tracer displacement to about 8 pixels. Image pairs are processed via a multi-pass cross-correlation algorithm, with a final interrogation window of 32 $\times$ 32 pixels and 50\% overlap. Flow statistics are obtained ensemble-averaging over 2000 uncorrelated realizations as well as in the homogeneous streamwise direction. The fully developed nature of the flow at the measurement station is confirmed by comparing profiles across the imaging window, with no appreciable changes for statistics up to second order.

In the particle-laden cases, the inertial particle position and velocity are characterized by Particle Tracking Velocimetry (PTV) using an in-house code developed based on the cross-correlation methods described in \citet{hassan1992} and \citet{ohmi2000}. Further details on the algorithm can be found in \citet{petersen2019}. For the measurements along the wall-normal plane, the CCD camera mounts a 105-mm Micro-Nikkor lens at f/16, \textcolor{black}{providing an imaging magnification of 36.1 px/mm} for a 34.3 $\times$ 25.9 mm\textsuperscript{2} field of view encompassing the full channel height. \textcolor{black}{The particle-per-pixel density is $4.0\times10^{-5}$ for the LoVF cases and $6.9\times10^{-4}$ for the HiVF cases.} Along the wall-parallel planes, an aperture of f/4 is used, \textcolor{black}{providing an imaging magnification of 19.0 px/mm for} a 55 $\times$ 70 mm\textsuperscript{2} window at the spanwise center of the channel. \textcolor{black}{The particle-per-pixel density is $1.8\times10^{-3}$ for the centerplane and $3.9\times10^{-3}$ for the near-wall plane.}The particle images are about 3 pixels, whose centroid is retrieved with an accuracy of approximately $\pm$0.1 pixels, as confirmed by tests on synthetic images. The pulse delay ranges between 70 \si{\us} and 100 \si{\us} depending on the cases and imaging locations, with typical particle displacements of 15 to 20 pixels. A pre-processing thresholding routine is applied to eliminate out-of-focus particles. The in-focus particles are tracked between image pairs using the relaxation method described by \citet{baek1996}. The local concentration is measured along the wall-normal plane by counting the number of particle centroids detected. This approach was used for inertial particles in air (\citealt{yang2005}; \citealt{sahu2014}; \citealt{sahu2016}) and in water \citep{kiger2002}. \citet{knowles2012} showed that, in water, laser-based measurements of particle concentration can be misestimated by as much as 30\%; however, they considered volume fractions one order of magnitude higher than the present case. Even at the higher loading investigated here, the average interparticle distance is $\sim$ 1 mm, which is much larger than the particle image. Due to clustering, the instantaneous local concentration can be higher, and some particles may go undetected. However, intense clustering usually pertains to a limited fraction of the particle set \citep{baker2017}; here the regions where such bias may be more significant are near the wall, due to turbophoresis. Still, the volume fraction evaluated from the particle count along the wall-normal plane (assuming a 1.1 mm thick imaging volume) agrees within 12 - 15\% with the value obtained by weighing the particles accumulated in the settling chamber during a given run time, lending confidence to the approach. Concentration and velocity statistics are based on ensemble-averaging over 2000 uncorrelated realizations (collected over four runs of 500 realizations each) as well as over the streamwise direction. The streamwise homogeneity of the particle statistics is discussed in §3.1.3. All statistics are verified to be steady-state during each run, with excellent repeatability in each run (within variations of the order of the statistical uncertainty). \textcolor{black}{It is noted that simultaneous fluid-phase measurements are not acquired in this experiment. This is due to the relatively high image concentration of the inertial particles, especially near the wall, reducing the signal from fluid tracers. The feasibility of obtaining simultaneous fluid measurements are discussed in §4.}

\section{Results}\label{sec:results}
Throughout this section, $\langle U \rangle$ and $\langle V \rangle$ denote streamwise and wall-normal components of the mean velocity. $u$ and $v$ are the corresponding fluctuating components, whose rms are denoted as $U_{rms}$ and $V_{rms}$, respectively. Error bars in the plots represent statistical uncertainties based on 95\% confidence levels \textcolor{black}{and using the random uncertainty of the dataset \citep{bendat2011}}. When the wall-normal profiles are shown over half of the channel height, the symmetry of the results along the center-line is within the statistical uncertainty. 
\subsection{Wall-normal measurements}
\subsubsection{Unladen air flow}\label{sec:unladen-flow}

\begin{figure}
  \centerline{\includegraphics[width=0.6\textwidth]{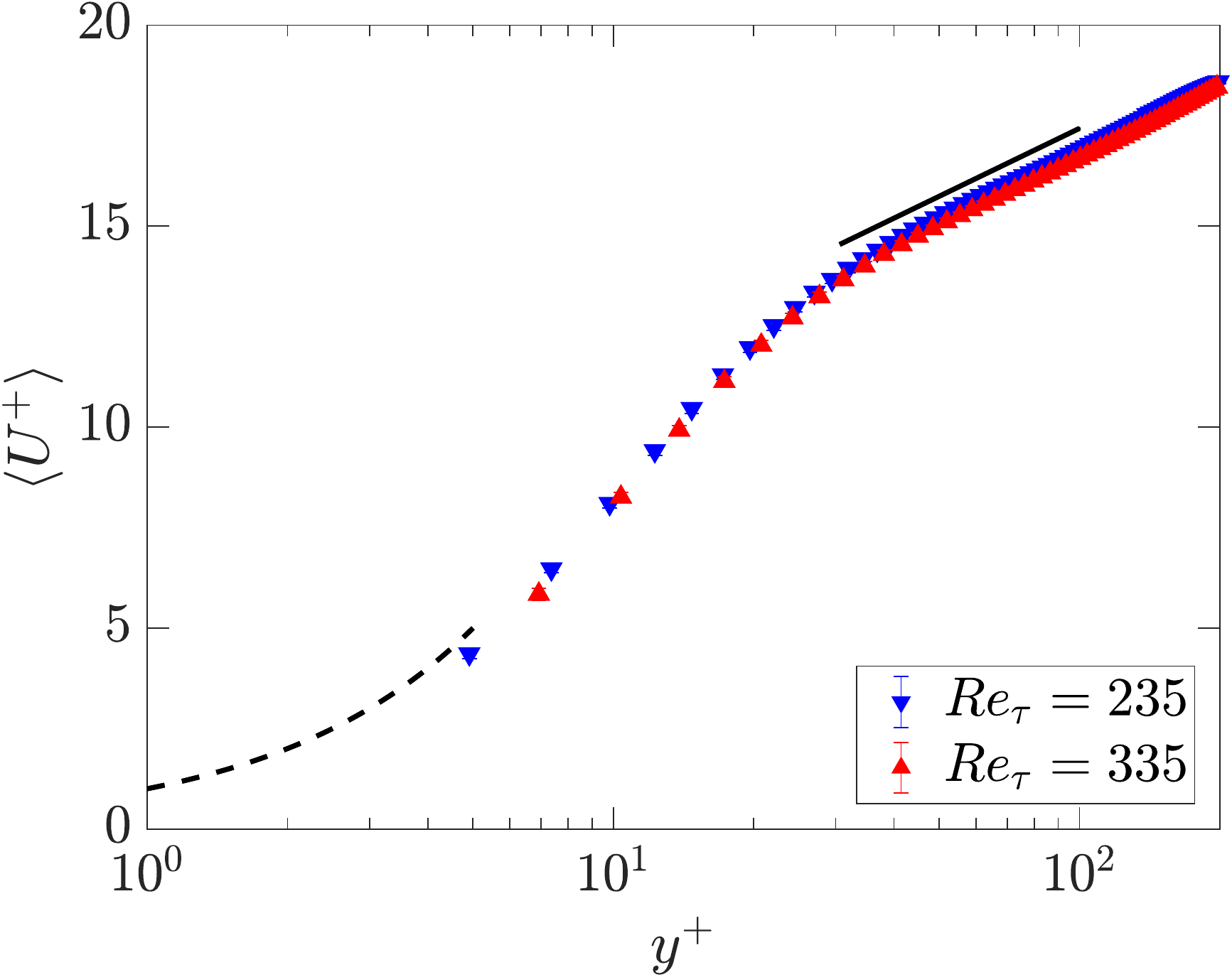}}
  \caption{The mean velocity profile shown in logarithmic scale for $Re_{\tau} = $ 235 and 335. The dashed black line represents the expected fluid velocity profile in the viscous sublayer ($y^+ < 5$). The solid black line indicates the log-law slope ($K=0.41, B = 5.2$, shifted up 1 unit in diagram for illustration) used to determine the friction Reynolds number from $y^+ = $ 30 to 100. }
\label{fig:unladen-log}
\end{figure}

\begin{figure}
\centering
\begin{subfigure}{.49\textwidth}
  \centerline{\includegraphics[scale=0.375]{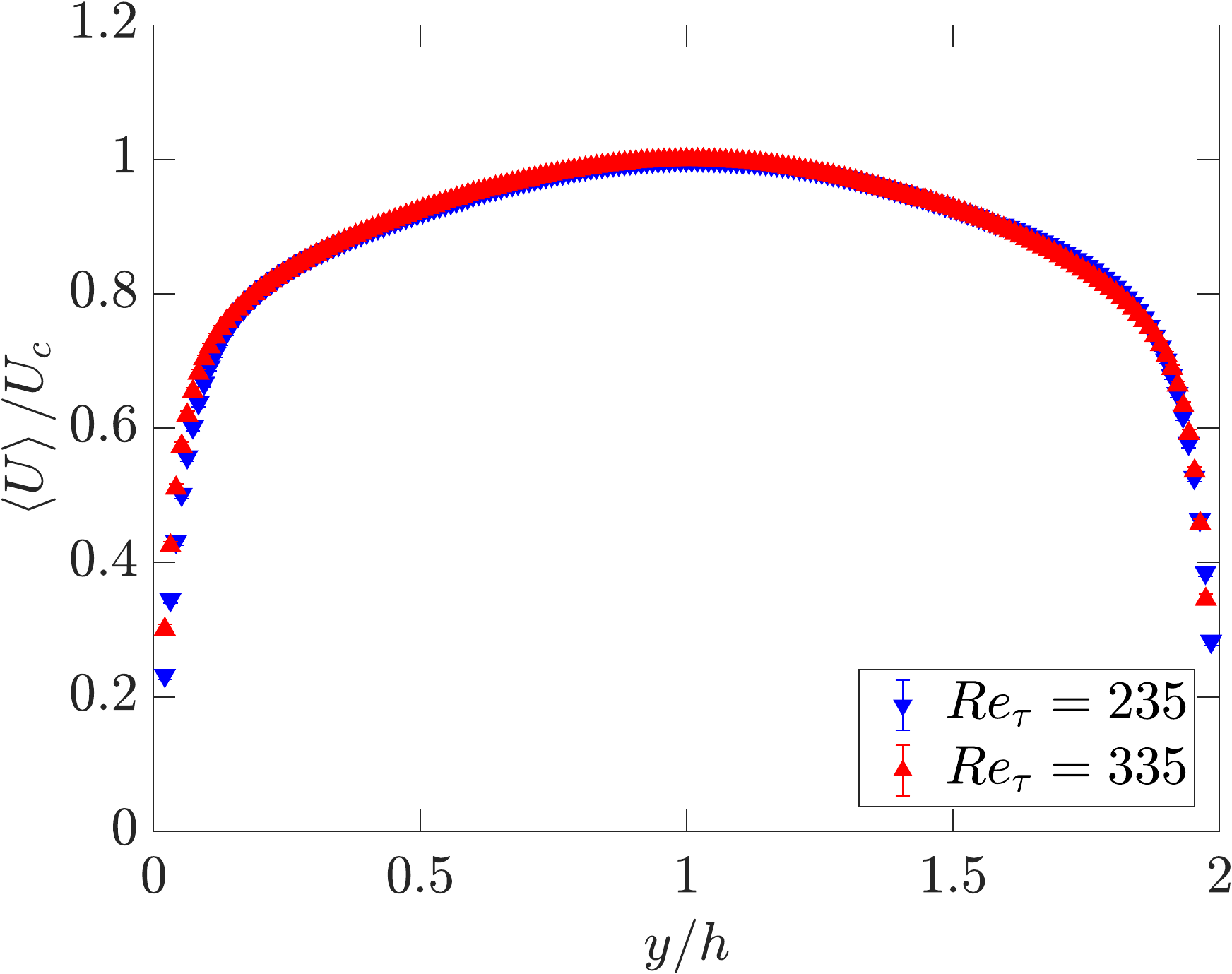}}
  \caption{}
 \end{subfigure}
 \begin{subfigure}{.49\textwidth}
   \centerline{\includegraphics[scale=0.375]{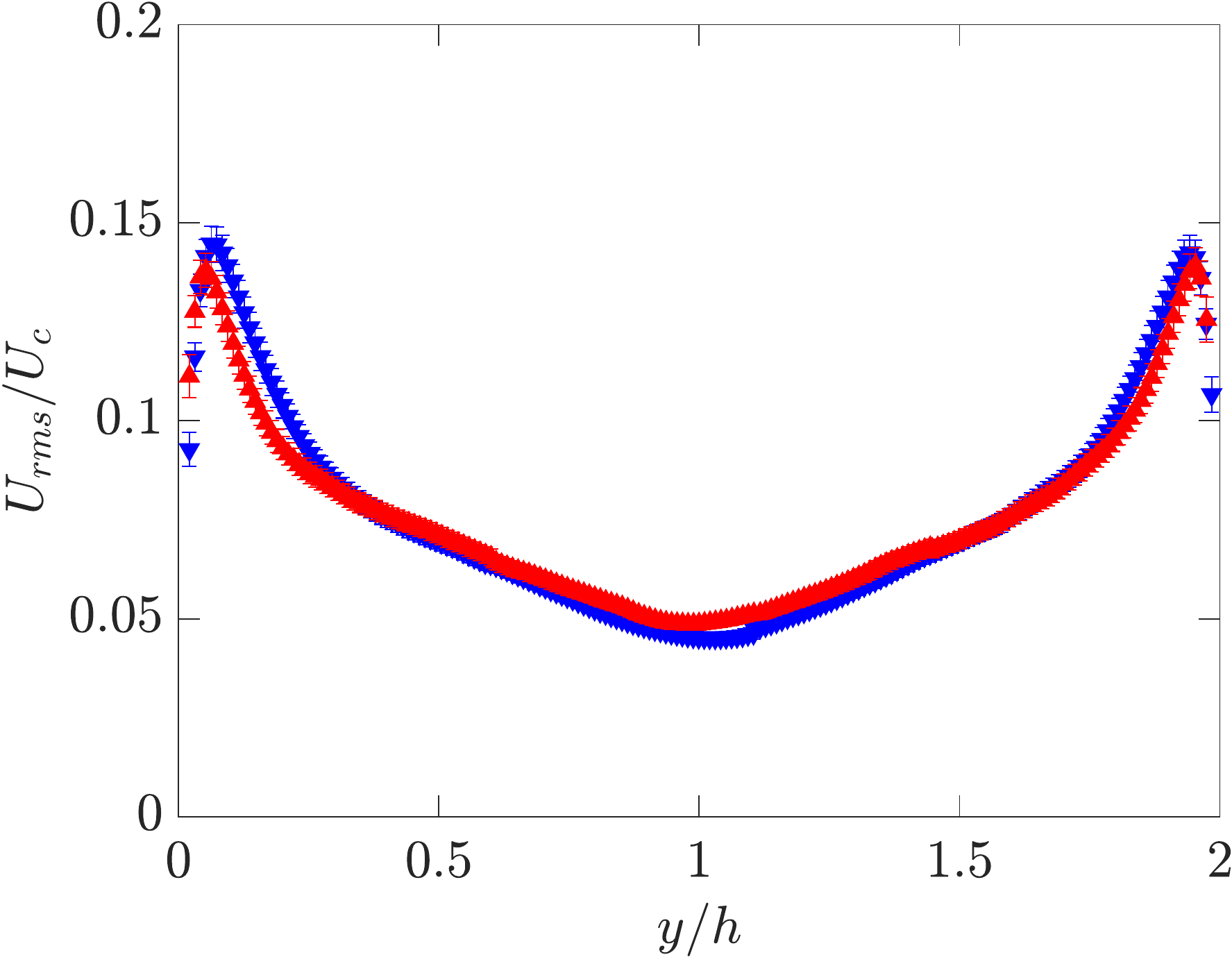}}
  \caption{}
 \end{subfigure}
 \begin{subfigure}{.49\textwidth}
   \centerline{\includegraphics[scale=0.375]{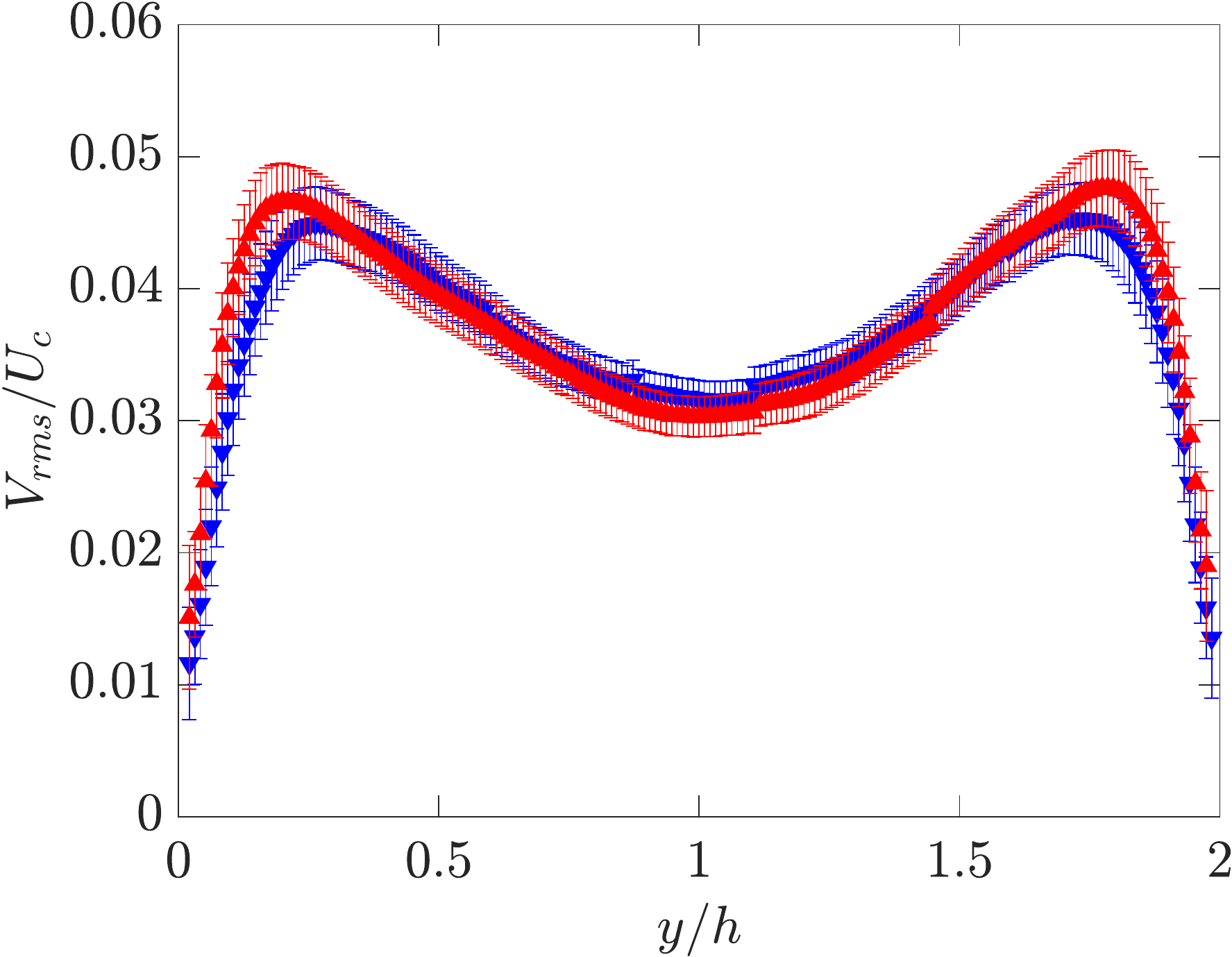}}
  \caption{}
 \end{subfigure}
 \begin{subfigure}{.5\textwidth}
   \centerline{\includegraphics[scale=0.375]{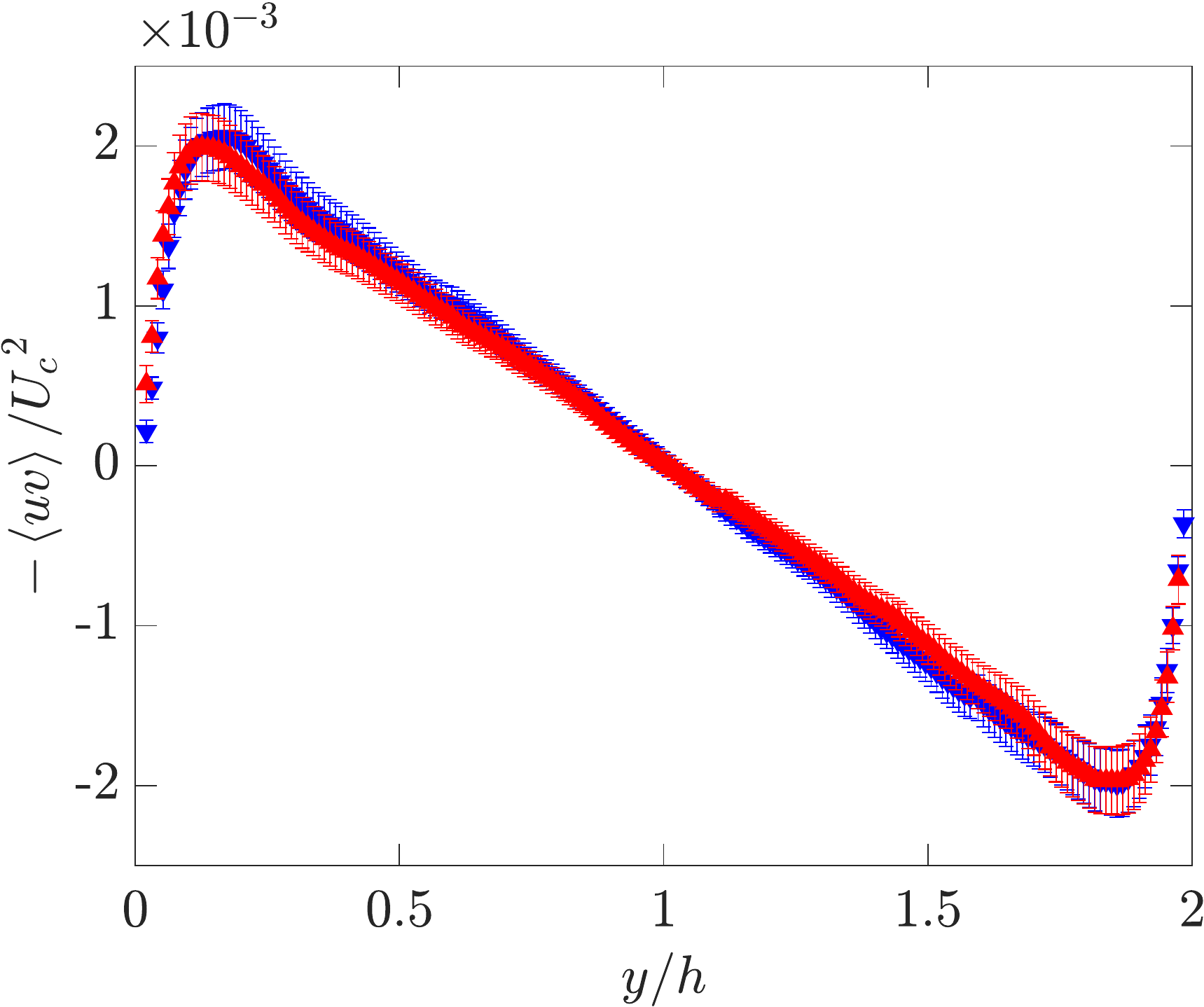}}
  \caption{}
 \end{subfigure}
  \caption{Wall-normal profiles for the unladen flow at both considered Reynolds numbers, normalized in outer units: (a) mean velocity, (b) streamwise rms velocity fluctuation; (c) wall-normal rms velocity fluctuation; (d) Reynolds shear stress.}
\label{fig:unladen}
\end{figure}

As a baseline, we first present the wall-normal profiles for the unladen fluid velocity. These are plotted in inner units in figure \ref{fig:unladen-log}, showing the expected logarithmic behavior above $y^+ \sim 30$. This is used to determine the friction velocity using a Clauser chart method (\citealt{clauser1956}; \citealt{wei2005}). The profiles for the mean velocity, streamwise and wall-normal rms fluctuation and Reynolds shear stress are plotted in figure \ref{fig:unladen} in outer units, i.e. normalized by the channel half-height and the centerline velocity. The agreement with DNS of spanwise-periodic channels at comparable Reynolds numbers (e.g. \citealt{moser1999}) suggests that the flow in the central part of the channel is not significantly impacted by the finite width of the cross-section. 

\subsubsection{Particle concentration}\label{sec:part-conc}

\begin{figure}
\centering
\begin{subfigure}{.49\textwidth}
  \centerline{\includegraphics[scale=0.39]{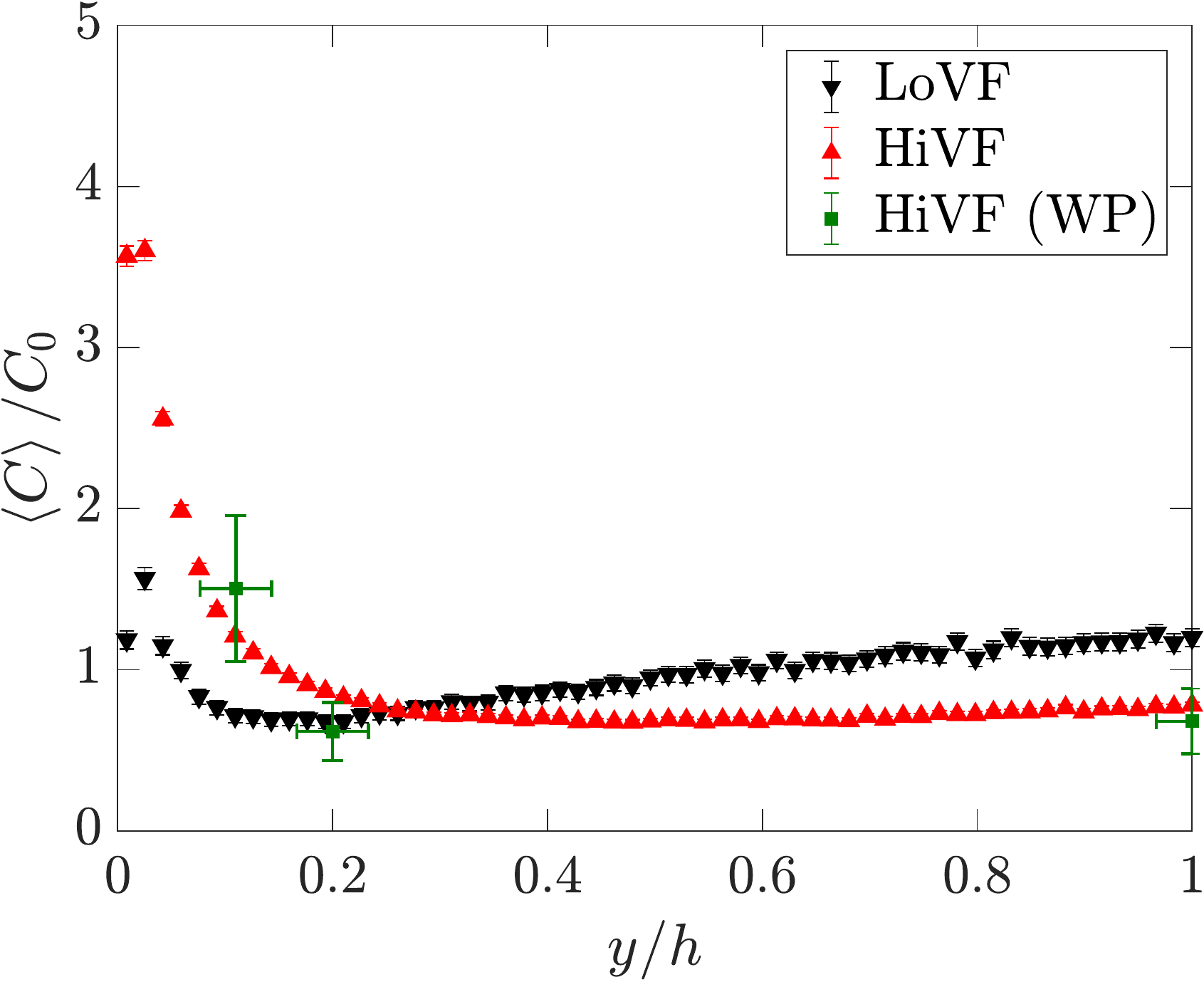}}
  \caption{}
 \end{subfigure}
 \begin{subfigure}{.49\textwidth}
   \centerline{\includegraphics[scale=0.39]{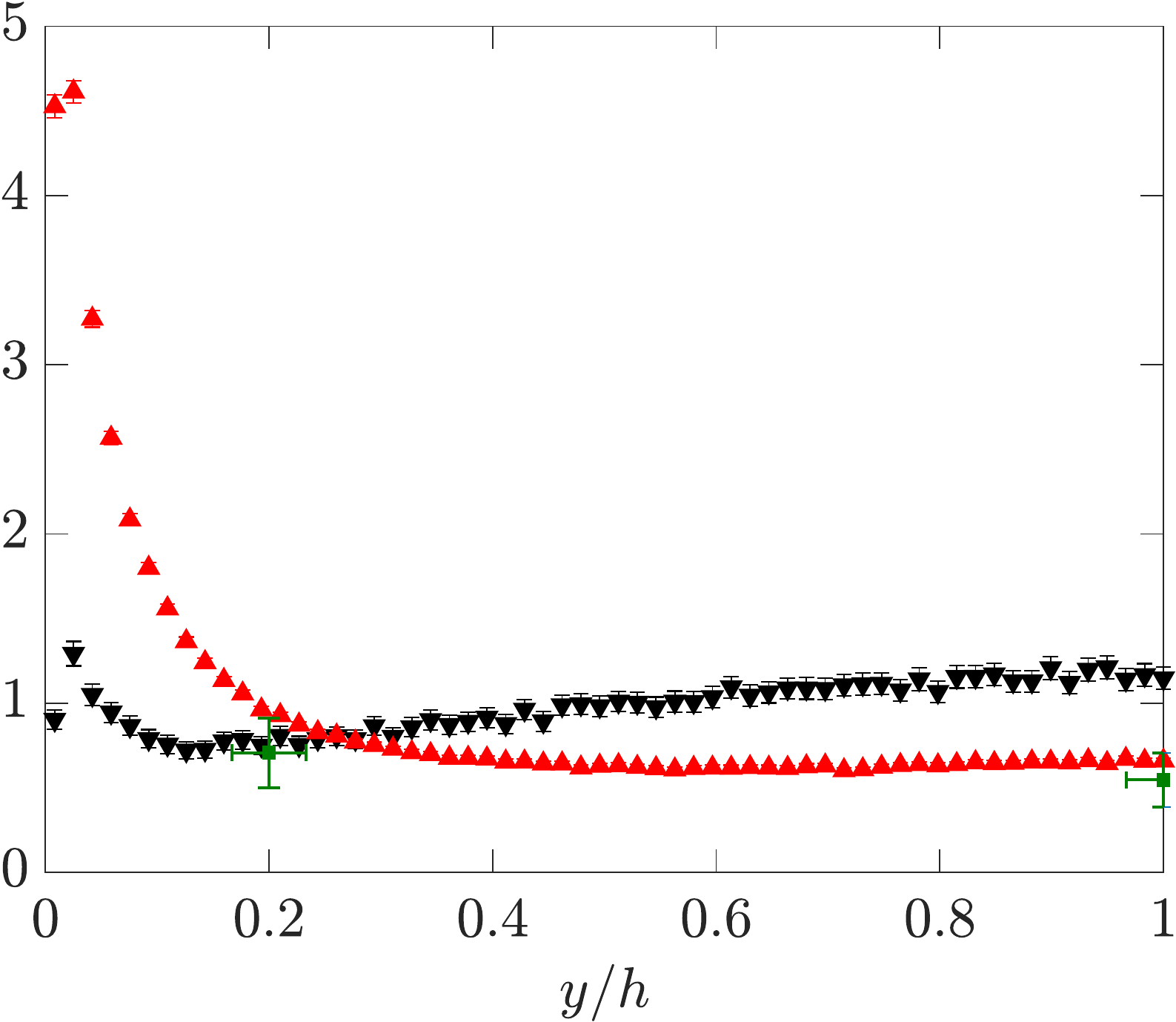}}
  \caption{}
 \end{subfigure}
  \caption{Mean particle concentration profiles normalized by the global concentration for (a) LoSt cases and (b) HiSt cases. Here and in the following plots, WP indicates data points from wall-parallel measurements, with the horizontal error bars indicating uncertainty on laser sheet position.}
\label{fig:1conc}
\end{figure}

Figure \ref{fig:1conc} displays the mean profiles of normalized particle concentration $C/C_0$ for the four particle-laden cases. Here and in the following wall-normal profile plots, data points in the profiles are plotted at the $y/h$ location at the center of the respective wall-normal bin, each bin having a width of 0.25mm (about 9 pixels). Data points measured for the HiVF cases along wall-parallel planes are also shown and found to agree closely to the wall-normal imaging results. Both LoVF and HiVF cases display a peak of concentration in the near-wall region, confirming that turbophoresis is active in the present regime. However, in the more dilute cases the peak is mild, and away from the wall the concentration gradually increases towards another local maximum at the centerline. On the other hand, the higher-loading cases display a much stronger peak of concentration near the wall, and the profile is essentially flat in the core region. The peaks appear to be at a finite standoff distance from the wall, which is however hard to quantify precisely.

As mentioned in the Introduction, previous measurements of near-wall segregation in similar regimes are lacking, and a comparison with past numerical simulations is in order. We refer to point-particle DNS studies, which are free from issues associated to turbulence modeling. Among those, several one-way-coupled simulations yielded near-wall concentration peaks two or more orders of magnitude above the channel mean, and mostly contained within the viscous sub-layer (\citealt{marchioli2002}; \citealt{marchioli2008}; \citealt{sardina2012a}; \citealt{bernardini2014}). Those results, while insightful, are influenced by the fact that point-particles can amass to arbitrary densities. In two-way-coupled simulations, the momentum back-reaction from the particles reduced the near-wall segregation, as did the inter-particle collisions, see \citet{li2001}, \citet{vreman2007}; \citet{nasr2009}. These authors did show concentrations reaching a minimum adjacent to the near-wall peak and increasing up to a centerline maximum, similarly to our LoVF profiles. However, they found that such reduction of the near-wall peak (and the simultaneous appearance of a centerline maximum) occurred for increasing mass loading, in contrast with the present results. On the other hand, the concentration profiles we observe for LoVF are consistent with the argument of \citet{young1997} that turbophoresis is driven by the gradient of fluid $V_{rms}$ (figure \ref{fig:unladen}c). \textcolor{black}{The concentration} is maximized in correspondence to the concentration minimum, decays steeply towards the wall and more mildly towards the centerline, \textcolor{black}{following approximately the same trend as the wall-normal gradient of $V_{rms}$}. We remark that the centerline concentration maximum was observed in several, but not all, one-way-coupled simulations, and was found to depend on the flow orientation: for example, \citet{nilsen2013} found it for downward and no-gravity flow, but not for upward flow. 

\subsubsection{Particle velocity}\label{part-vel}

\begin{figure}
\centering
\begin{subfigure}{.49\textwidth}
  \centerline{\includegraphics[scale=0.38]{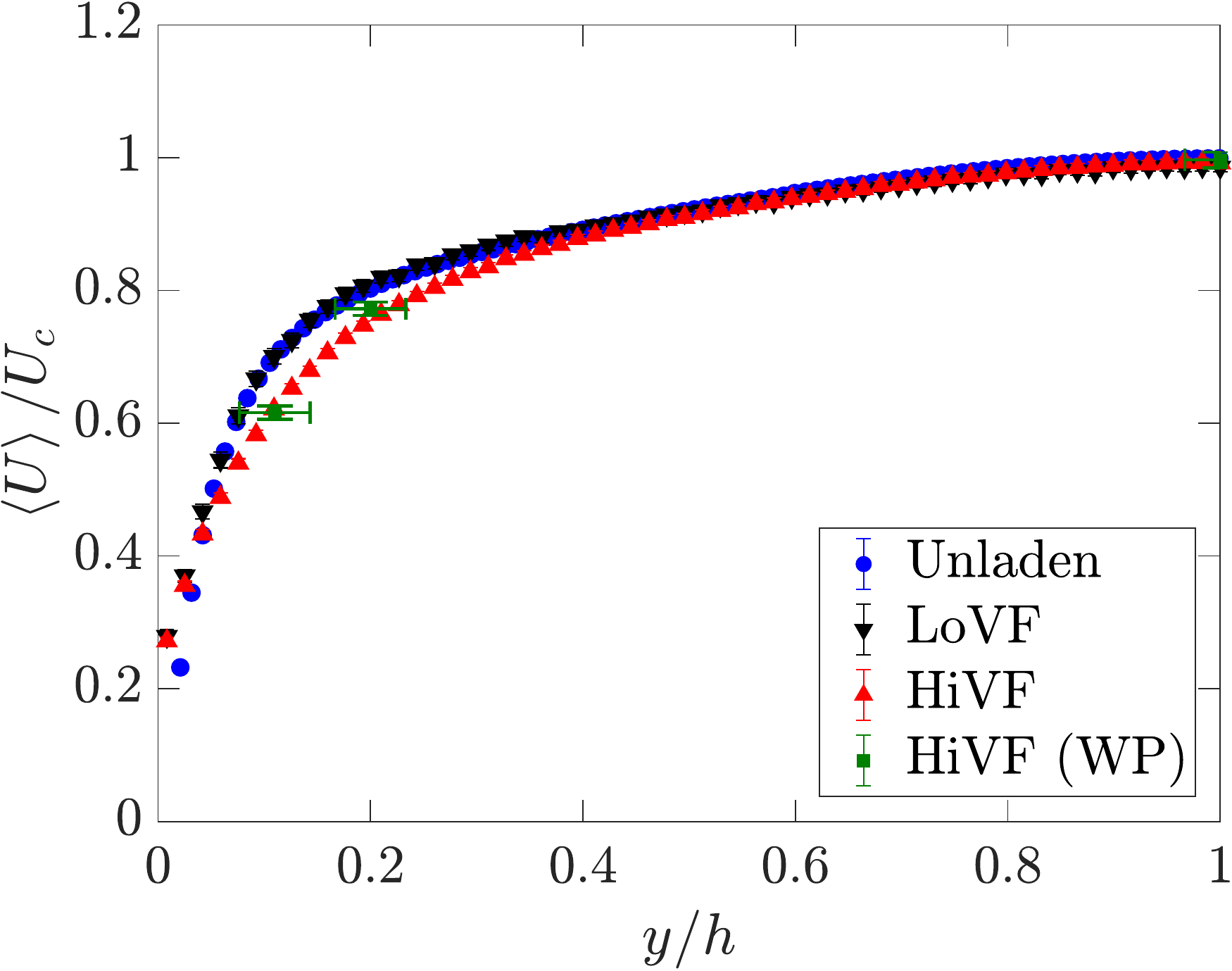}}
  \caption{}
 \end{subfigure}
 \begin{subfigure}{.49\textwidth}
   \centerline{\includegraphics[scale=0.38]{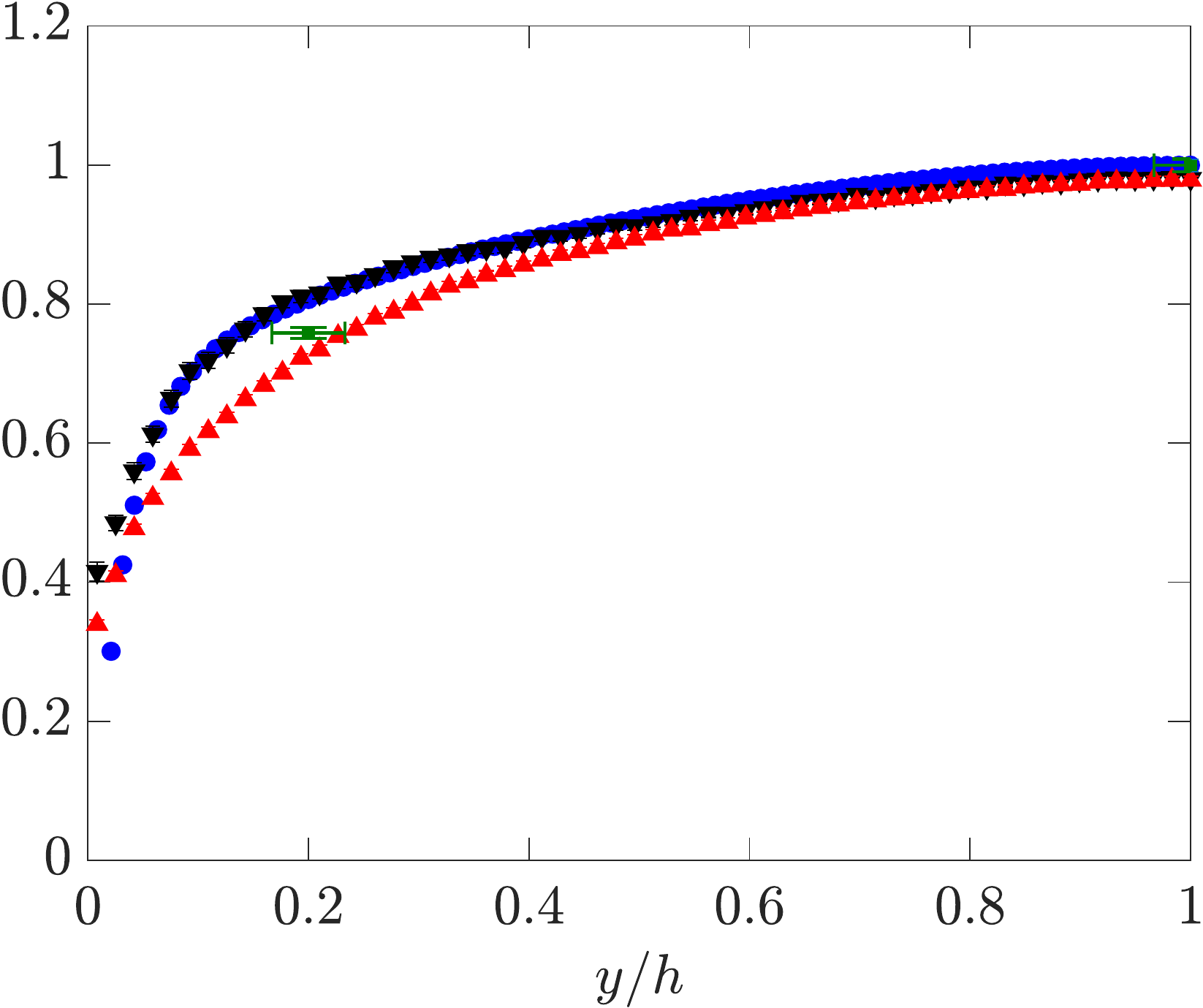}}
  \caption{}
 \end{subfigure}
 \begin{subfigure}{.49\textwidth}
   \centerline{\includegraphics[scale=0.38]{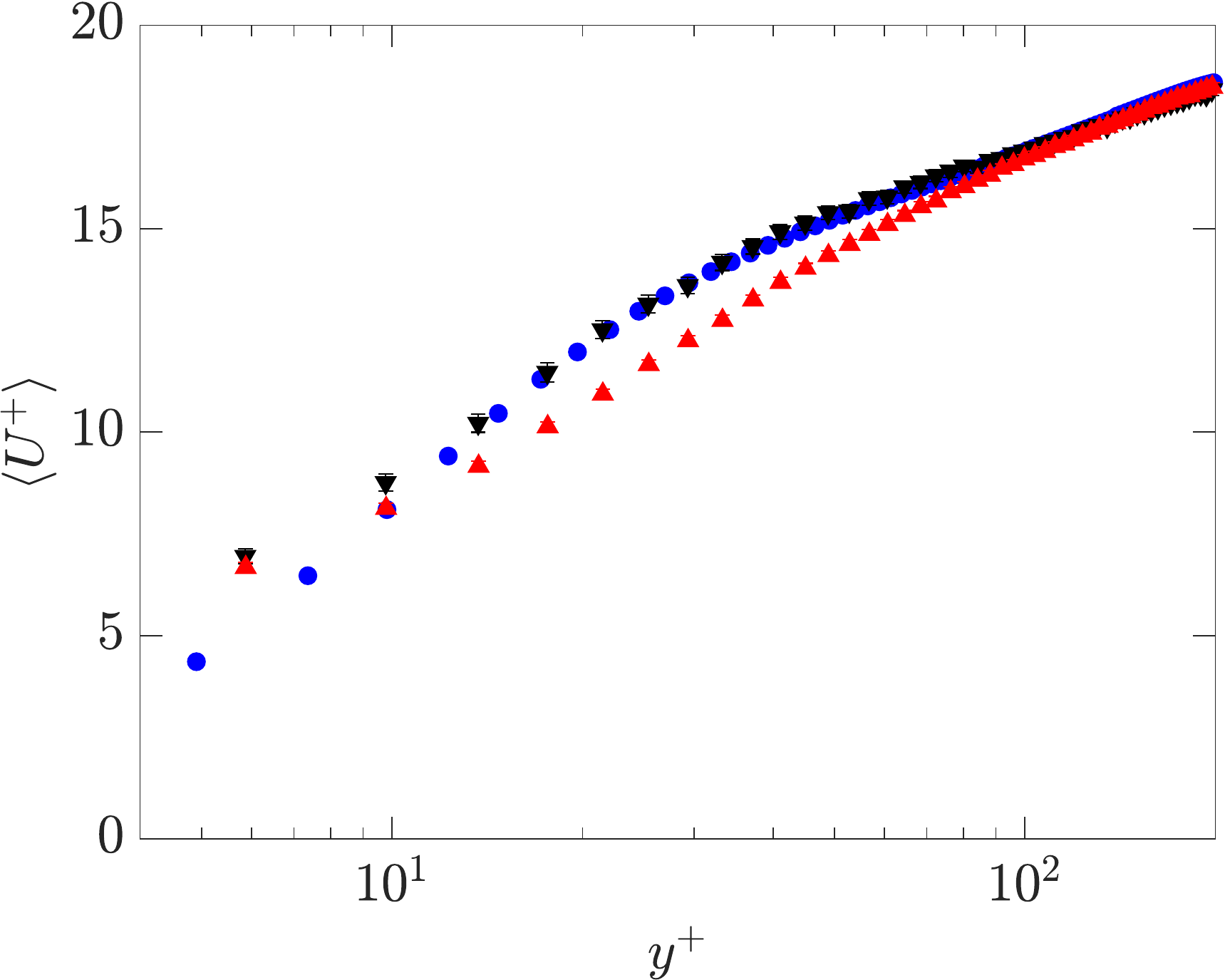}}
  \caption{}
 \end{subfigure}
 \begin{subfigure}{.49\textwidth}
   \centerline{\includegraphics[scale=0.38]{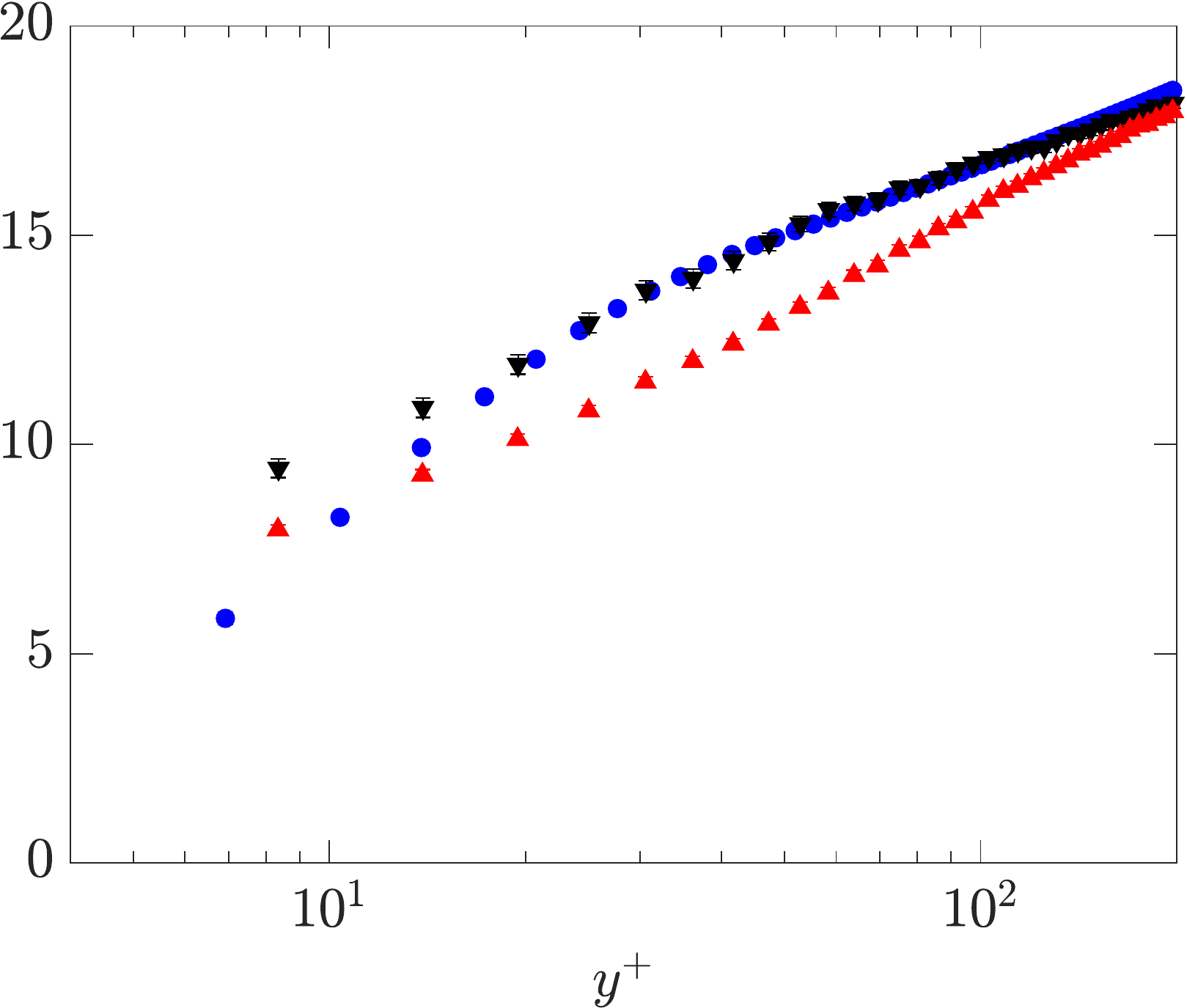}}
  \caption{}
 \end{subfigure}
  \caption{Profiles of mean streamwise particle velocity for (a, c) LoSt and (b, d) HiSt cases, normalized in outer units (top panels) and inner units (bottom panels). Unladen fluid profiles plotted for comparsion.}
\label{fig:2umean}
\end{figure}

Figure \ref{fig:2umean} displays mean velocity profiles compared to the unladen air velocity. The data is presented both in outer (a, b) and inner (c, d) units, normalizing by the unladen velocity scales. In the viscous and buffer layer, the particles travel faster than the unladen air. This is a consequence of fast-moving particles retaining part of their momentum when transported towards the wall by turbulent fluctuations, without being constrained by the no-slip boundary condition. Such behavior was already highlighted by \citet{kulick1994} and in several later experimental and numerical studies (e.g., \citealt{taniere1997}; \citealt{rouson2001}; \citealt{vreman2007}; \citealt{li2012}). \citet{righetti2004} explicitly commented on an effective slip boundary condition for the particle field. Further away from the wall, in the LoVF cases the particles travel at approximately the same speed as the unladen air, while in the HiVF cases they lag in the logarithmic and buffer layers, recovering \textcolor{black}{to} the unladen air velocity in the channel core. A decrease of mean velocity with increasing mass loading was also reported by \citet{kulick1994}, although for higher $St^+$. In that case the lag was visible up to the centerline, but this was likely due to the wall roughness \citep{benson2005}. In general, we observe less flat velocity profiles than in previous experiments, see e.g. \citet{kulick1994}; \citet{paris2001}; \citet{caraman2003}; \citet{benson2005}. \textcolor{black}{\citet{vreman2015}} argued that those were again influenced by some wall roughness that enhanced the wall-normal particle velocity fluctuations and in turn flattened the mean velocity profiles.

\begin{figure}
\centering
\begin{subfigure}{.49\textwidth}
  \centerline{\includegraphics[scale=0.375]{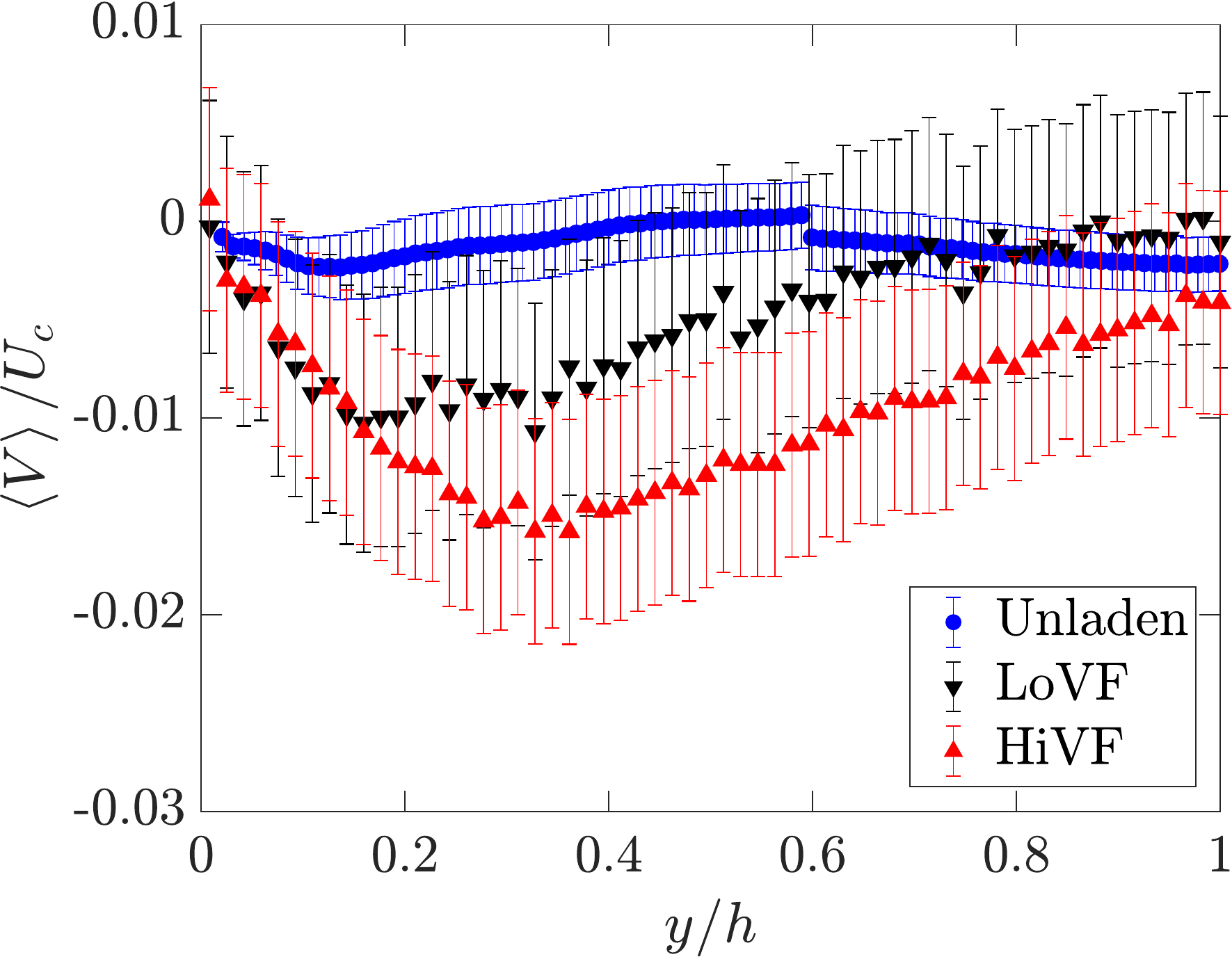}}
  \caption{}
 \end{subfigure}
 \begin{subfigure}{.49\textwidth}
   \centerline{\includegraphics[scale=0.375]{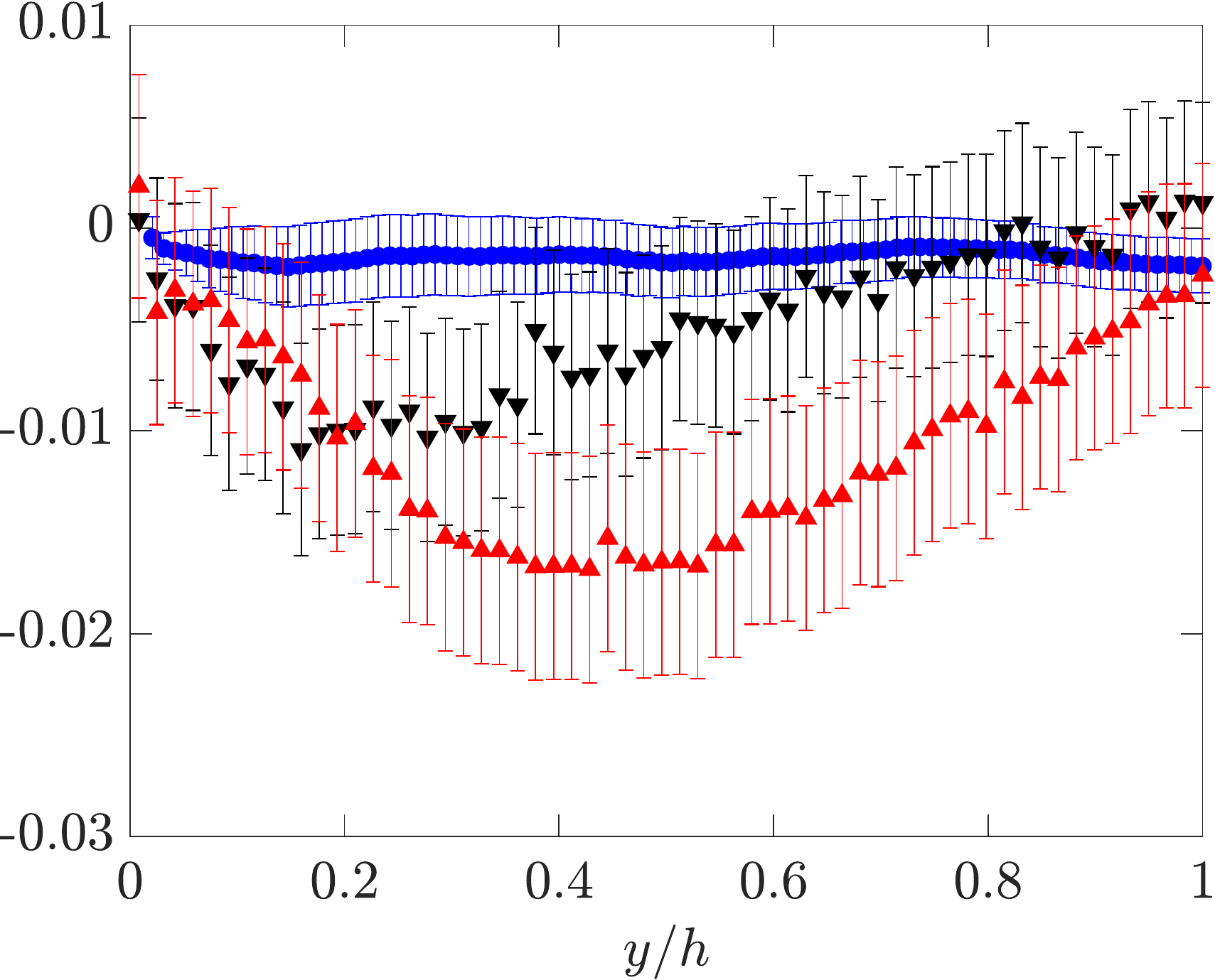}}
  \caption{}
 \end{subfigure}
  \caption{Profiles of mean wall-normal particle velocity for (a) LoSt and (b) HiSt cases. Unladen fluid profiles plotted for comparison.}
\label{fig:3vmean}
\end{figure}

Figure \ref{fig:3vmean} displays profiles of mean wall-normal particle velocity. In a fully developed state, this should be identically zero. This is the case (within error bounds) for the unladen fluid, while the particles do show some residual drift towards the wall. This is likely caused by turbophoresis; \textcolor{black}{the effect is exemplified in the LoVF case where the peak at $y/h \sim$ 0.2 approximately corresponds to the maximum of unladen fluid $V_{rms}$ and to the minimum of particle concentration, consistent with the theory of \citet{young1997} (see also \citealt{capecelatro2016})}. Numerical simulations at similar $St^+$ indicated that the turbophoretic drift continues to modify the particle field during $\mathcal{O}$($10^4$) viscous time scales, which over the considered range of $Re_{\tau}$ corresponds to $\mathcal{O}$($10^3$) channel heights (\citealt{marchioli2008}; \citealt{sardina2012a}; \citealt{bernardini2014}). While these estimates are influenced by the one-way coupled nature of the modeling, they clearly indicate that the particle field requires a much greater development length than the sole hydrodynamics. However, we also remark that the observed wall-normal mean velocities are about 1\% of the streamwise velocity, \textcolor{black}{and smaller than the rms fluctuation in the same direction} (reported below). Moreover, all statistics show no visible trend over the imaging windows (about 1.7$h$ and 3.7$h$ in streamwise direction for the wall-normal and wall-parallel measurements, respectively). Thus, also considering that the particles are expected to have reached terminal velocity much before entering the imaging section, the influence of the partial streamwise development is unlikely to qualitatively impact the reported trends.

\begin{figure}
\centering
\begin{subfigure}{.49\textwidth}
  \centerline{\includegraphics[scale=0.375]{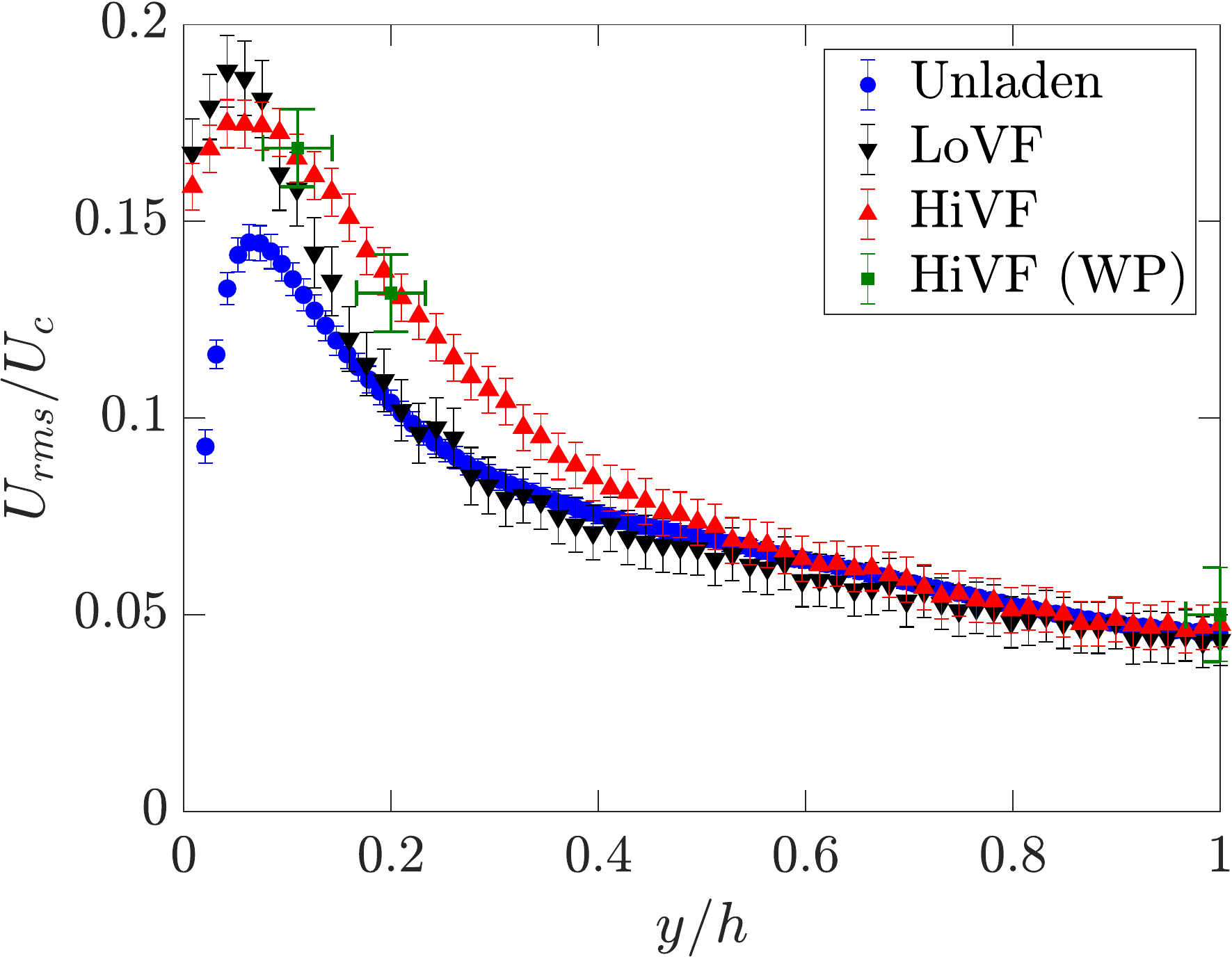}}
  \caption{}
 \end{subfigure}
 \begin{subfigure}{.49\textwidth}
   \centerline{\includegraphics[scale=0.375]{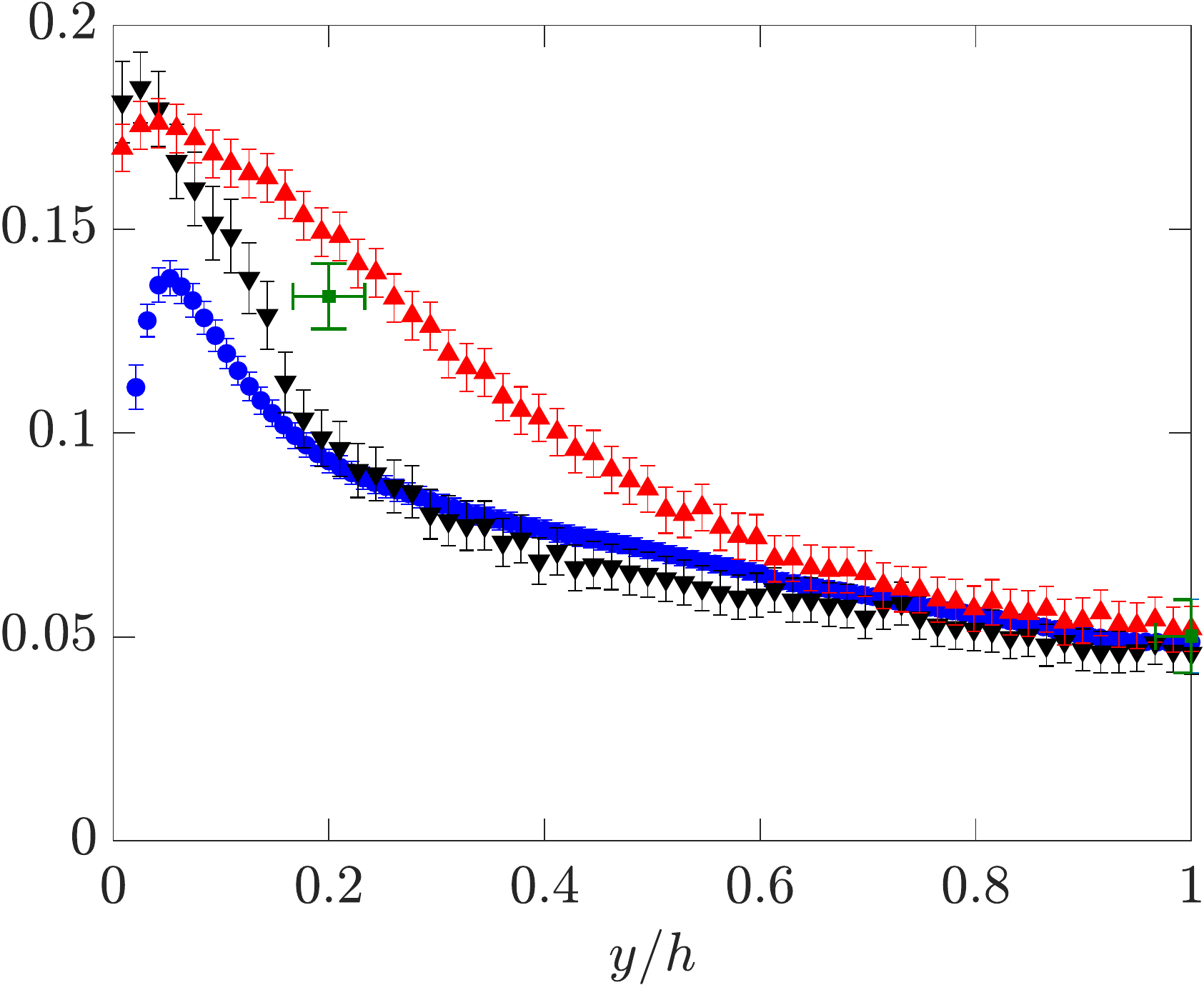}}
  \caption{}
 \end{subfigure}
  \caption{Profiles of rms streamwise particle velocity for (a) LoSt and (b) HiSt cases. Unladen fluid profiles plotted for comparison.}
\label{fig:4urms}
\end{figure}

The rms streamwise fluctuations of the particle velocity is plotted in figure \ref{fig:4urms}. The LoVF cases display profiles similar to the unladen flow in the channel core, and significantly more intense fluctuations (up to 20\% higher than the fluid) in the near-wall region, with little differences between LoSt and HiSt. Previous studies have found particles \textcolor{black}{with} velocity fluctuations stronger than the carrier fluid in several configurations, including particle-laden jets and homogeneous turbulence (e.g., \citealt{hardalupas1989}; \citealt{petersen2019}). Specifically in channel flows, the observed trend agrees with one-way coupled simulations as reported by \citet{marchioli2008} and \citet{nasr2009} for lower but still turbophoretic Stokes numbers. Following \citet{taniere1997}, the increase in rms velocities may be interpreted as a consequence of the spread in momentum of particles with different history: the ones arriving to a near-wall interrogation window from more distant locations retain some of their relatively high speed due to inertia; while those coming from a rebound on the wall have lost some of their kinetic energy in the collision. Remarkably, the HiVF cases show a significant increase of particle velocity fluctuations even at $y/h \sim$ 0.4, which is even more dramatic for HiSt. The near-wall peak is about the same as in LoVF, but the cross-section-average rms fluctuation is substantially augmented for the higher loading. This is in contrast with past two-way and four-way coupled point-particle DNS: \textcolor{black}{considering mass loadings higher than but comparable to the current HiVF cases,} \citet{li2001}, \citet{vreman2007}, and \citet{nasr2009} found a decrease in streamwise rms fluctuations (although the trend with increasing $\phi_v$ reported by Vreman was locally not monotonic at high loadings). 

\begin{figure}
\centering
\begin{subfigure}{.49\textwidth}
  \centerline{\includegraphics[scale=0.375]{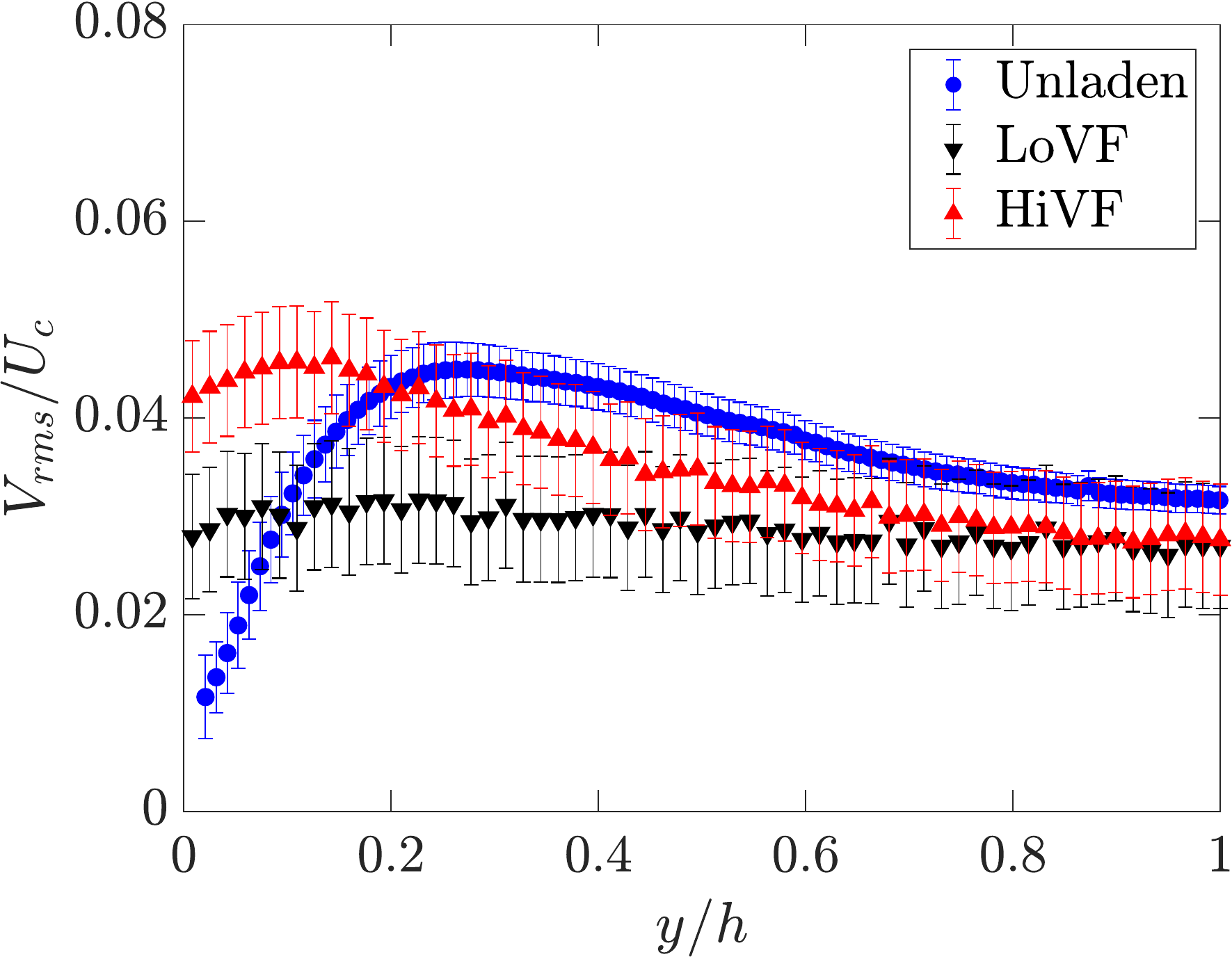}}
  \caption{}
 \end{subfigure}
 \begin{subfigure}{.49\textwidth}
   \centerline{\includegraphics[scale=0.375]{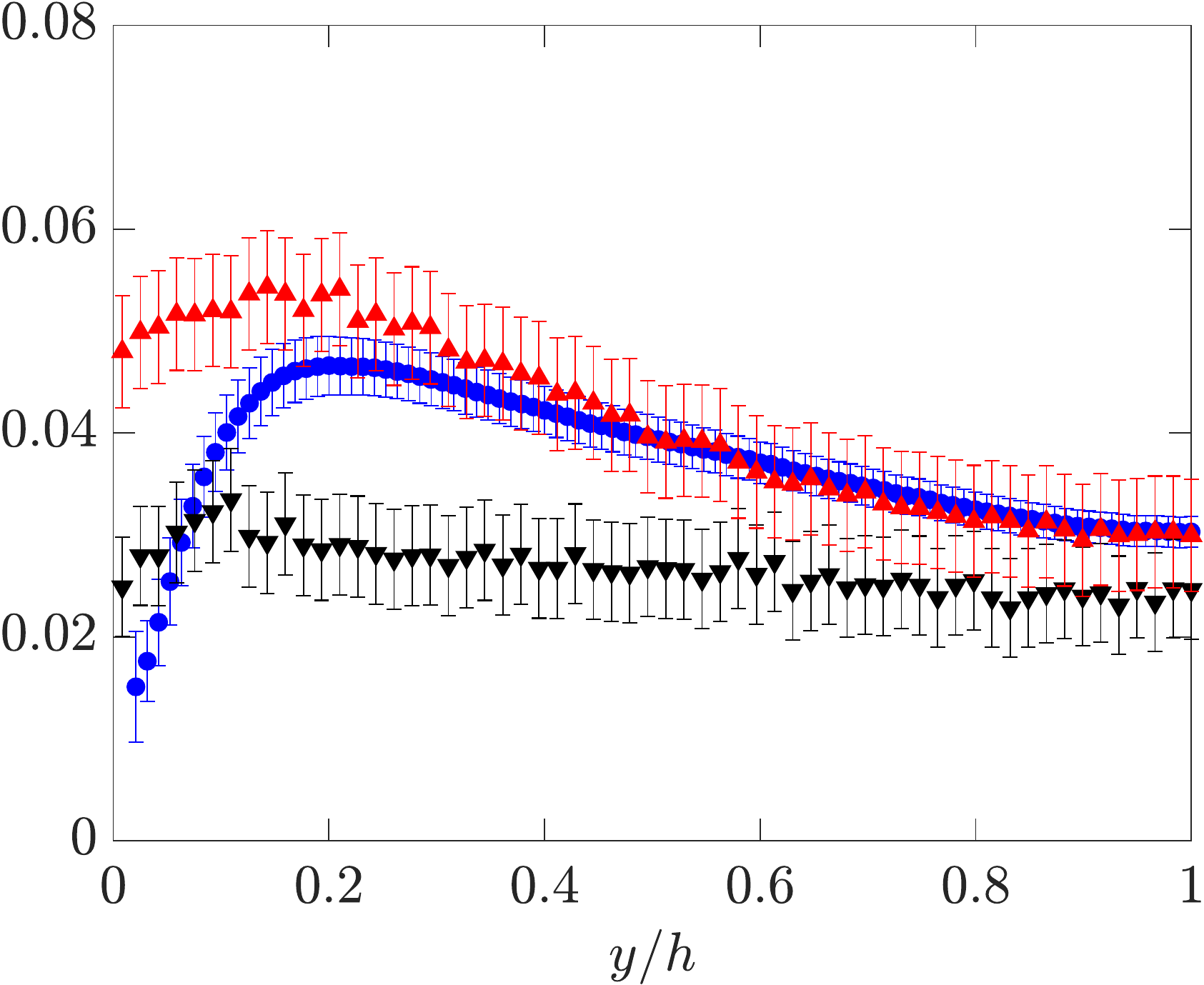}}
  \caption{}
 \end{subfigure}
  \caption{Profiles of rms wall-normal particle velocity for (a) LoSt and (b) HiSt cases. Unladen fluid profiles plotted for comparison.}
\label{fig:5vrms}
\end{figure}

Figure \ref{fig:5vrms} shows profiles of the wall-normal rms fluctuations of the particle velocities. For LoVF, the particle $V_{rms}$ is lower than the \textcolor{black}{unladen} fluid $V_{rms}$  in the channel core, but it remains fairly flat across the channel and largely exceeds the unladen fluid levels for $y/h < 0.1$. The effect of $St^+$ in the considered range is minor. Moreover, while the fluctuation level decreases approaching the wall, it does not appear to vanish. This is in contrast with one-way coupled simulations where the particle $V_{rms}$ is consistently lower than the fluid $V_{rms}$ (thus vanishing at the wall, \citealt{marchioli2008}), but is consistent with previous experiments with particles of similar $St^+$ \citep{li2012}. The approximately even redistribution of the lateral kinetic energy across the channel cross-section may partly be due to particle inertia, and partly to collisions with the wall (inter-particle collision being relatively unlikely at the lower volume fraction). At HiVF, $V_{rms}$ increases more significantly approaching the wall, and the tendency is stronger for HiSt. This behavior is similar as for $U_{rms}$, and indicates again that the particle-fluid dynamics has been altered: by the modification of the underlying turbulent flow and/or by the increase in particle-particle/wall-particle collisions.

\begin{figure}
\centering
\begin{subfigure}{.49\textwidth}
  \centerline{\includegraphics[scale=0.375]{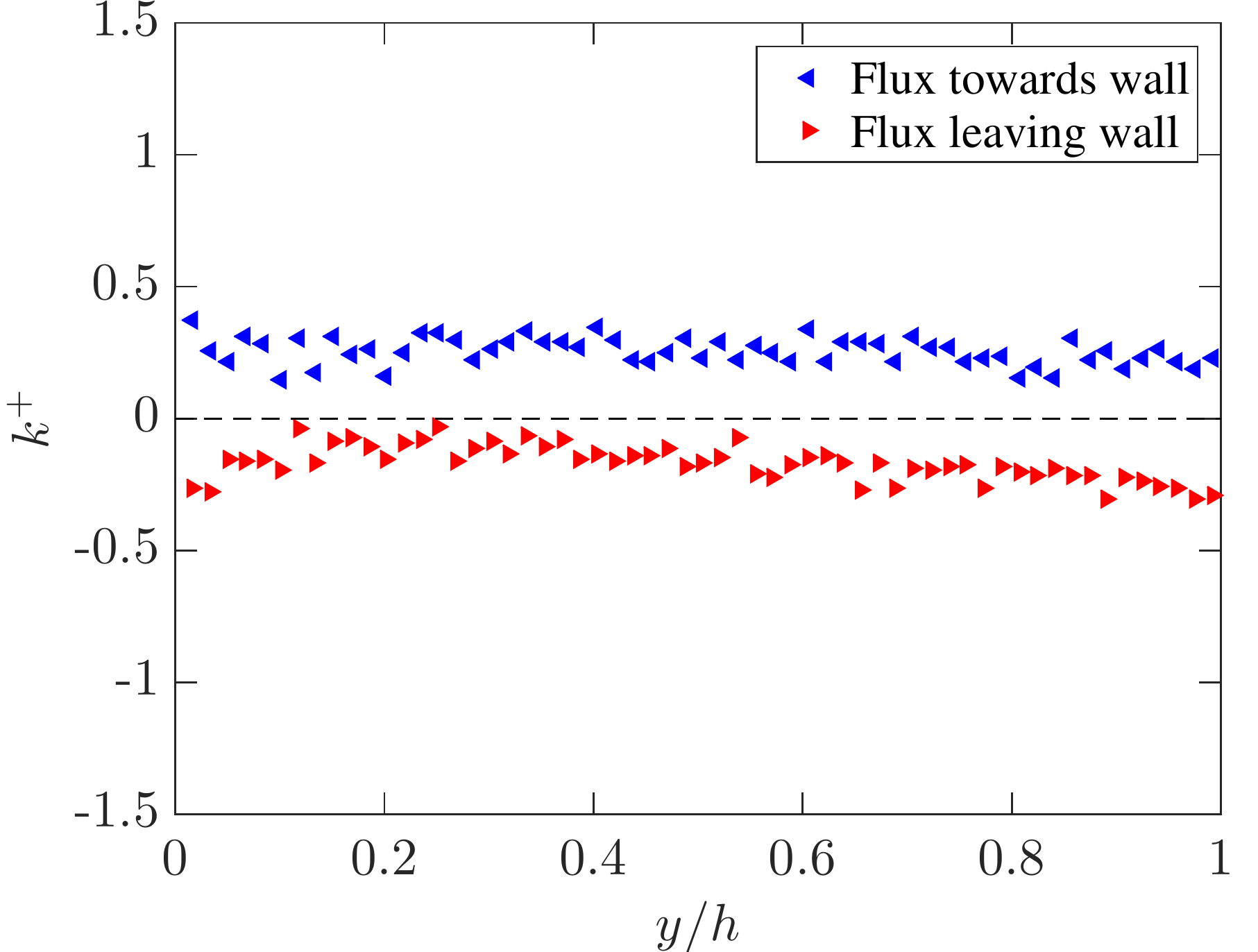}}
  \caption{}
 \end{subfigure}
 \begin{subfigure}{.49\textwidth}
   \centerline{\includegraphics[scale=0.375]{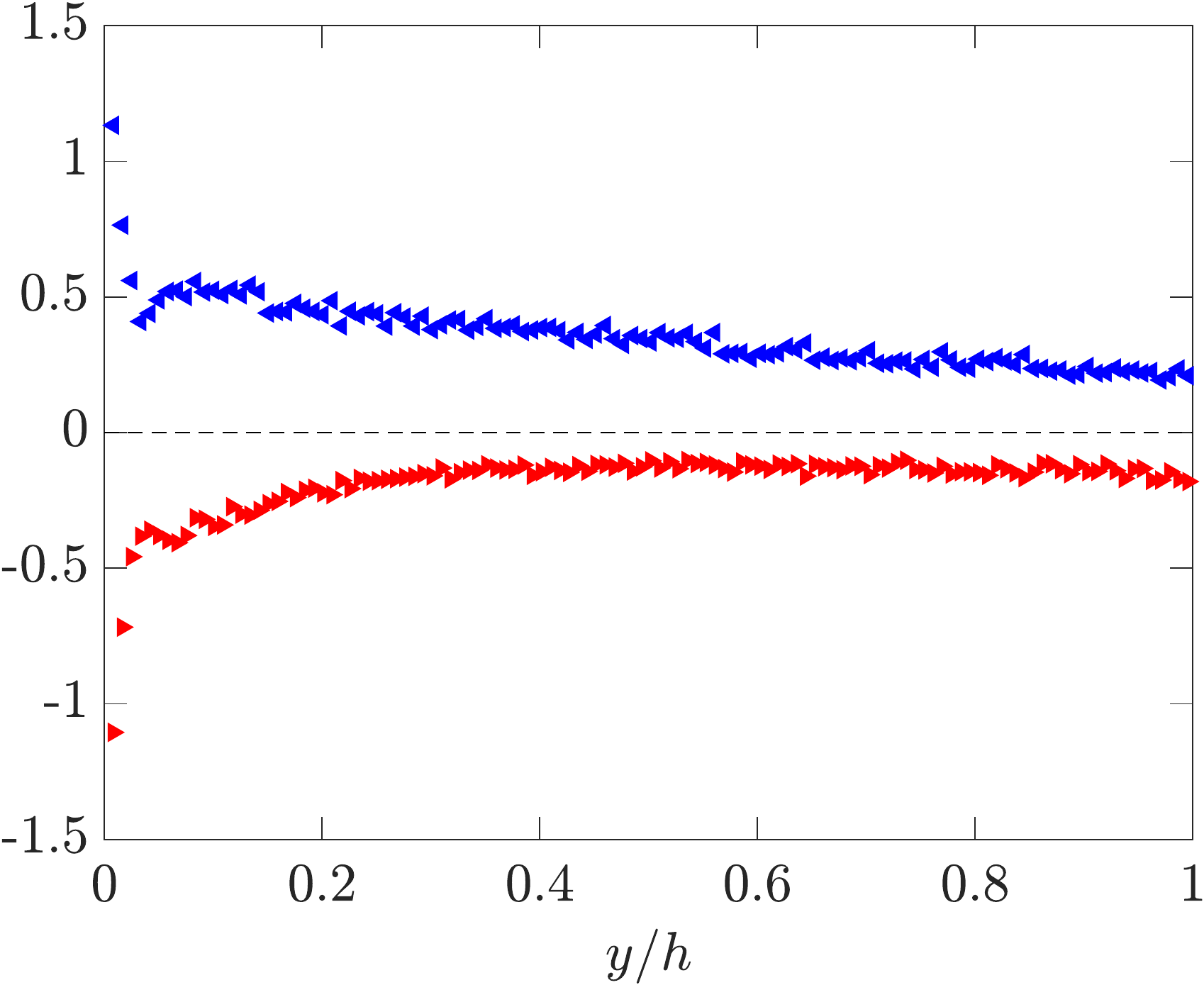}}
  \caption{}
 \end{subfigure}
  \caption{Profiles of characteristic flux velocity $k^+$ based on fluxes towards and away from the wall, for (a) LoSt-LoVF and (b) LoSt-HiVF cases.}
\label{fig:flux}
\end{figure}

The wall-normal velocity is tightly related to the particle flux towards the wall, which eventually may lead to particle deposition. The flux can be expressed as the rate of particles per unit area crossing a control plane, $J = (dN/dt)/A_s$, where $N$ is the number of particles on either side of the plane, $t$ indicates time and $A_s$ is the surface area of the control plane. Normalizing by the global concentration $C_0$ yields a characteristic velocity $k = J/C_0$; taking the control plane at the wall gives the commonly used deposition velocity $k_d$, which in turn can be made non-dimensional with a velocity scale usually taken as the friction velocity, $k_d^+ = k_d/u_{\tau}$ (\citealt{liu1974}; \citealt{young1997}; \citealt{bernardini2014}). Here the spanwise direction is assumed homogeneous and the above definitions are adapted to the two-dimensional measurements: the concentration is areal rather than volumic, and wall-parallel lines act as the control planes. Figure \ref{fig:flux} shows the non-dimensional characteristic velocity $k^+$ as a function of wall distance for the LoSt case (which shows similar trends to the HiSt case). We plot separately the fluxes towards and away from the wall. Because there is no net particle deposition, at the wall both fluxes are in balance (within experimental scatter). Wall collision cannot be directly detected with the present setup, but the absolute value of $k^+$ at the measurement location closest to the wall (i.e., control plane at $y^+ \sim 4$ for the LoSt case) is taken as a proxy of $k_d^+$. For the LoVF case, one retrieves $k_d^+ = $ $\mathcal{O}$(0.1), in agreement with previous observations (see, e.g., the collection of data in \citealt{young1997}). On the other hand, the HiVF case shows a sharp increase of flux in the inner layer and a much higher deposition velocity $k_d^+ =$ $\mathcal{O}$(1). This is consistent with the high $V_{rms}$ levels reported above, and indicates that the change in particle transport properties at high loading greatly impact wall collision and (for a non-reflective wall) deposition.

\begin{figure}
\centering
\begin{subfigure}{.49\textwidth}
  \centerline{\includegraphics[scale=0.375]{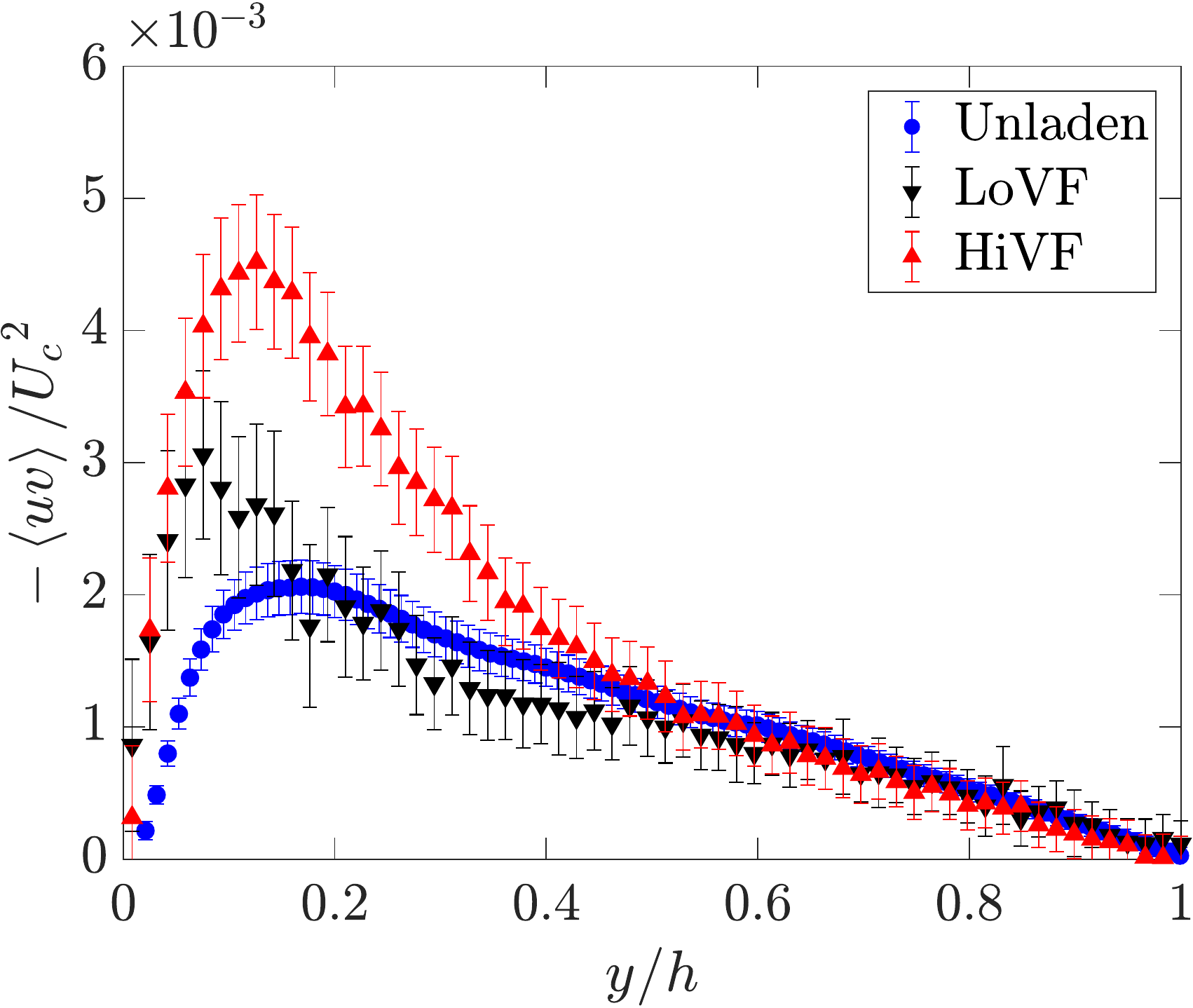}}
  \caption{}
 \end{subfigure}
 \begin{subfigure}{.49\textwidth}
   \centerline{\includegraphics[scale=0.375]{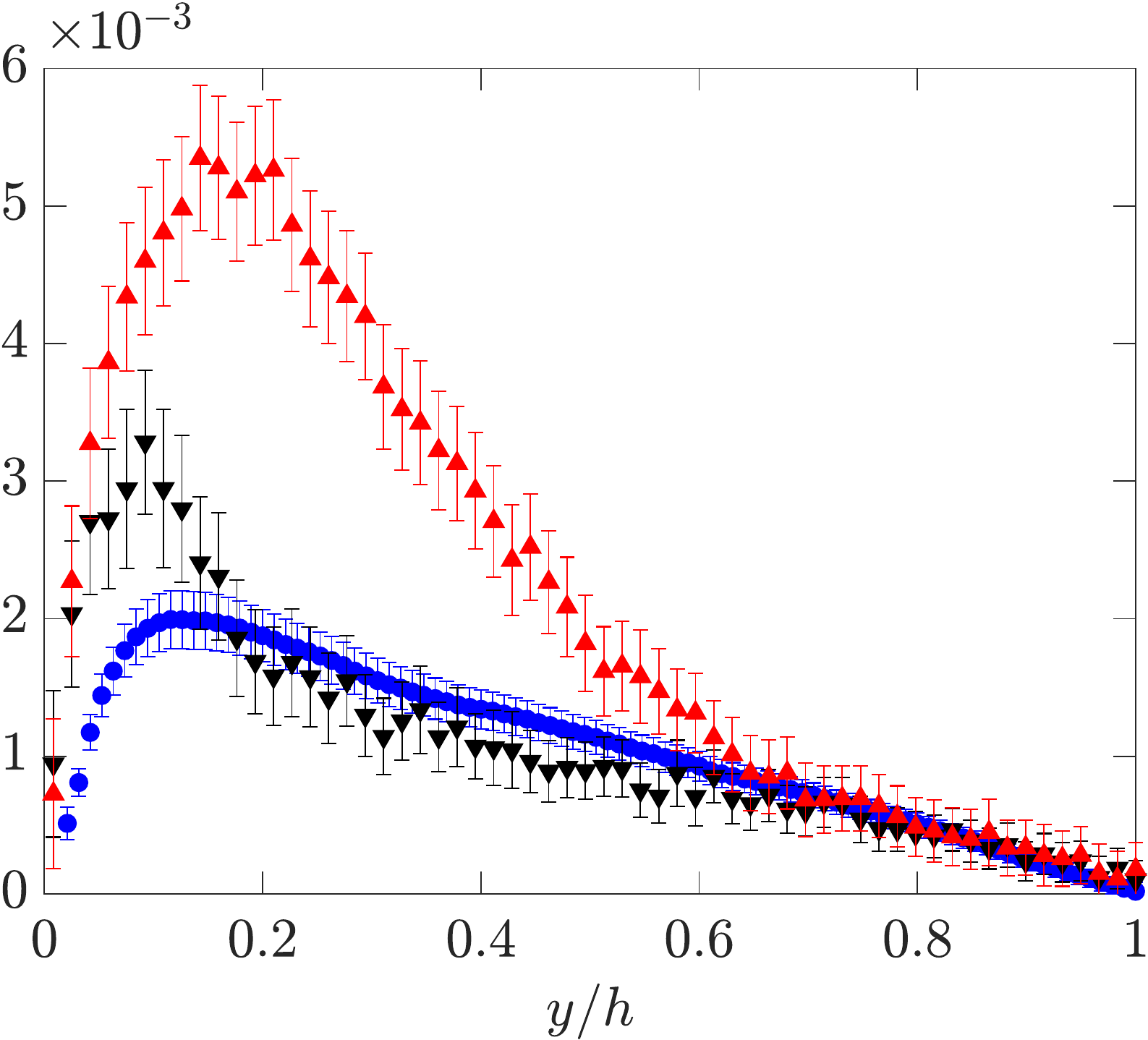}}
  \caption{}
 \end{subfigure}
  \caption{Profiles of particle Reynolds shear stress for (a) LoSt and (b) HiSt cases. Unladen fluid profiles plotted for comparison.}
\label{fig:6uvprime}
\end{figure}

In figure \ref{fig:6uvprime} we present profiles of the cross-correlation between the particle streamwise and wall-normal fluctuations, referred to as particle Reynolds shear stress, along with the \textcolor{black}{unladen} fluid counterpart. For LoVF, these are found to follow the trend of the unladen fluid in the channel core up to about $y/h =$ 0.2, but visibly exceed those values in the near wall region. A similar behavior was reported in the vertical pipe flow of \citet{caraman2003}, whereas in horizontal flow studies such as \citet{li2012} particle Reynolds stresses were above/below the fluid levels in the core/near-wall region. These discrepancies stress once more the consequential differences between configurations, in particular as it pertains the gravity force direction. The HiVF cases show again an earlier departure from the \textcolor{black}{unladen} fluid statistics and a more dramatic increase in correlation magnitude.

\begin{figure}
\centering
\begin{subfigure}{.55\textwidth}
  \centerline{\includegraphics[scale=0.38]{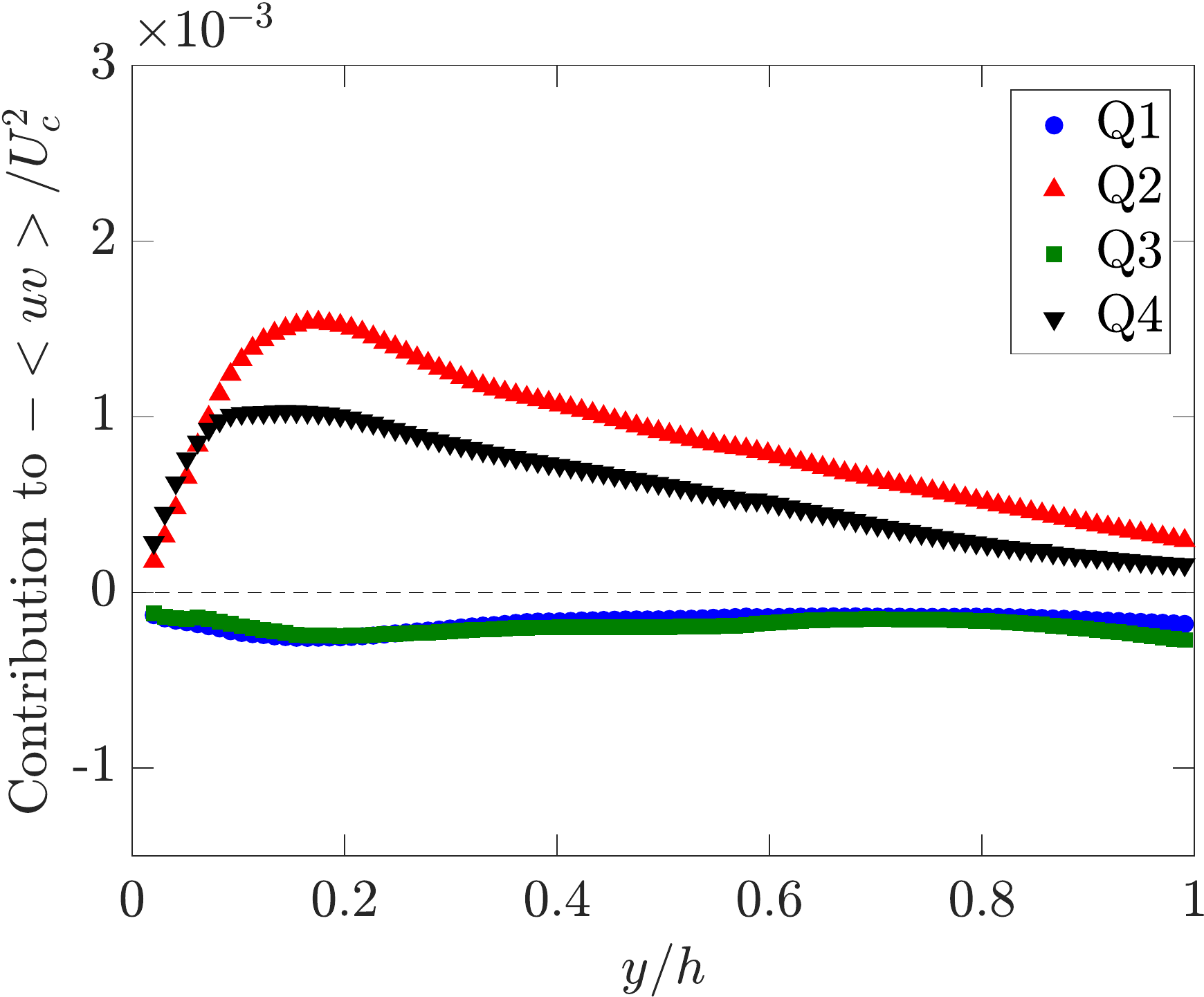}}
  \caption{}
 \end{subfigure}
\begin{subfigure}{.49\textwidth}
  \centerline{\includegraphics[scale=0.38]{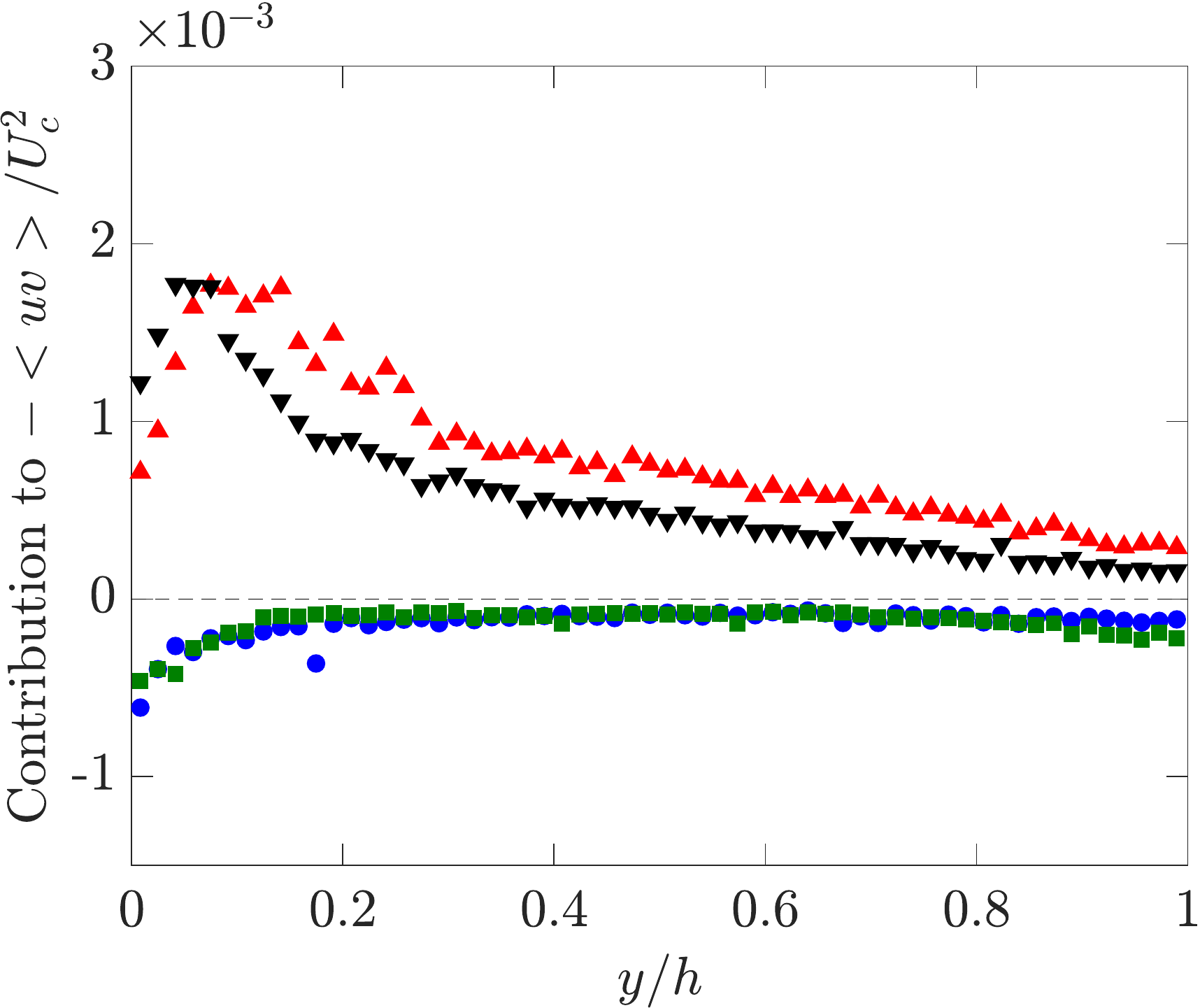}}
  \caption{}
 \end{subfigure}
 \begin{subfigure}{.49\textwidth}
   \centerline{\includegraphics[scale=0.38]{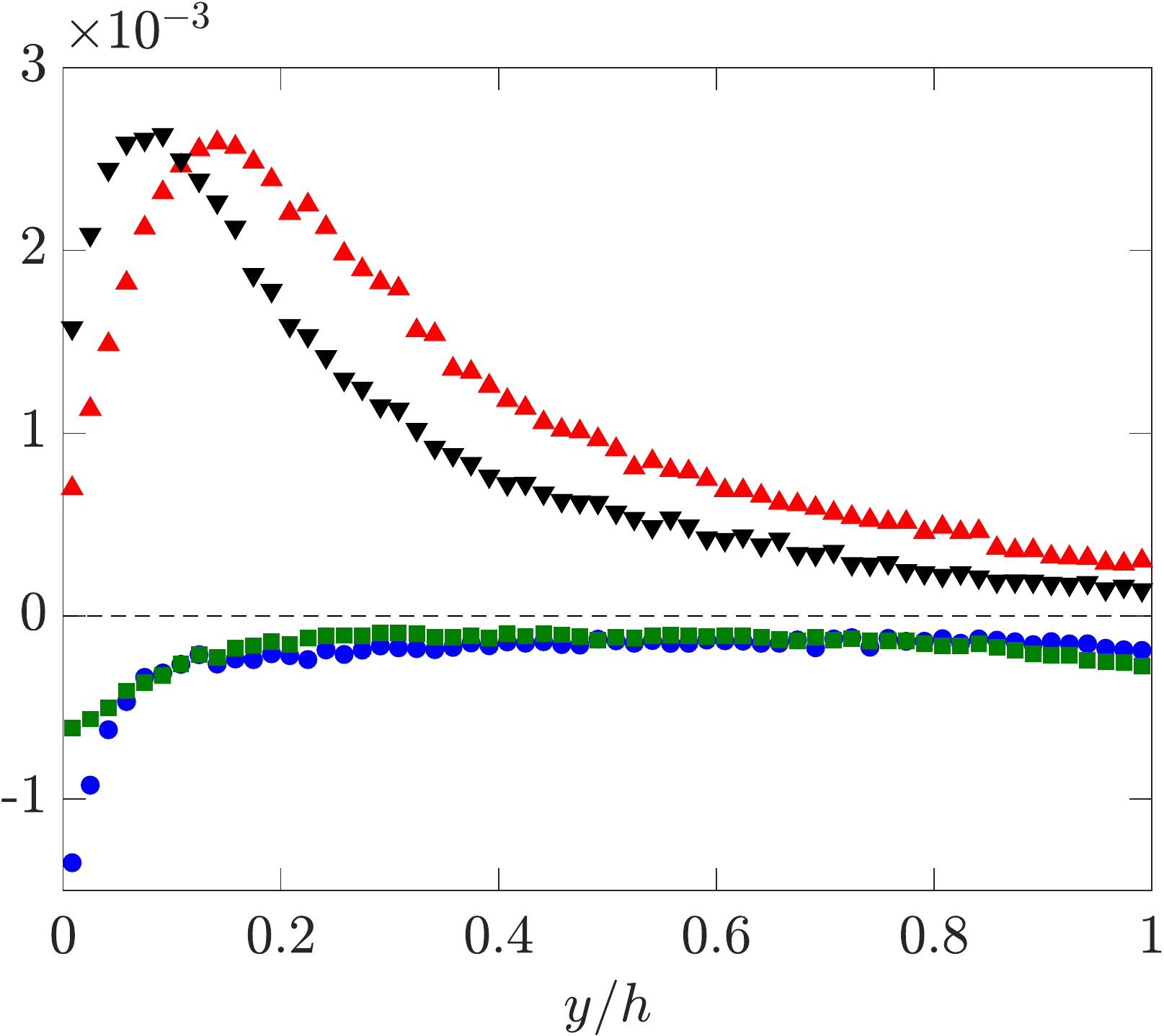}}
  \caption{}
 \end{subfigure}
  \caption{Contribution of each quadrant of the (u, v) plane to the Reynolds shear stresses for (a) unladen fluid and (b) inertial particles for the LoSt-LoVF case and (c) LoSt-HiVF case.}
\label{fig:quad}
\end{figure}

To explore this dynamic further, we perform a quadrant analysis in the ($u, v$) plane. Following classic notation utilized in wall turbulence studies, we label events belonging to the four quadrants as Q1 ($u >, v > 0$), Q2 ($u < 0, v > 0$), Q3 ($u < 0, v < 0$), and Q4 ($u >, v < 0$). We report on the LoSt cases, which behave similarly to the HiSt. Figure \ref{fig:quad}a shows, for reference, the contributions to the Reynolds stresses for the unladen fluid velocity at the same Reynolds number. This highlights the predominance of the Q2 and Q4 events which contribute to positive turbulence production, with Q4 prevailing on Q2 for $y/h < 0.06$ (or $y^+ < 15$), and vice versa further from the wall \citep{kim1987}. The particles (figure \ref{fig:quad}b,c) follow a similar trend, but with noteworthy differences. The prevalence of Q4 events in the near-wall region is much more pronounced, which is consistent with sweeps being crucial in the process of trapping the particles near the wall \textcolor{black}{\citep{marchioli2002}}. This is in stark contrast with the result of \citet{li2012}: they found overwhelmingly higher probability of Q2 events near the floor of their horizontal channel, where gravity caused much more frequent wall rebounds. At HiVF, both Q2 and Q4 contributions are similarly enhanced, but the cross-over point is farther from the wall compared to LoVF: the region where particles are \textcolor{black}{swept towards the wall} is wider, which corresponds to a more intense turbophoretic drift, a stronger near-wall peak of concentration, and a depletion of the centerline peak (see figure \ref{fig:1conc}). For this case also Q3 is remarkably large near the wall, probably a consequence of particles colliding with each other and with the wall \citep{righetti2004}.

\begin{figure}
  \centerline{\includegraphics[width=0.6\textwidth]{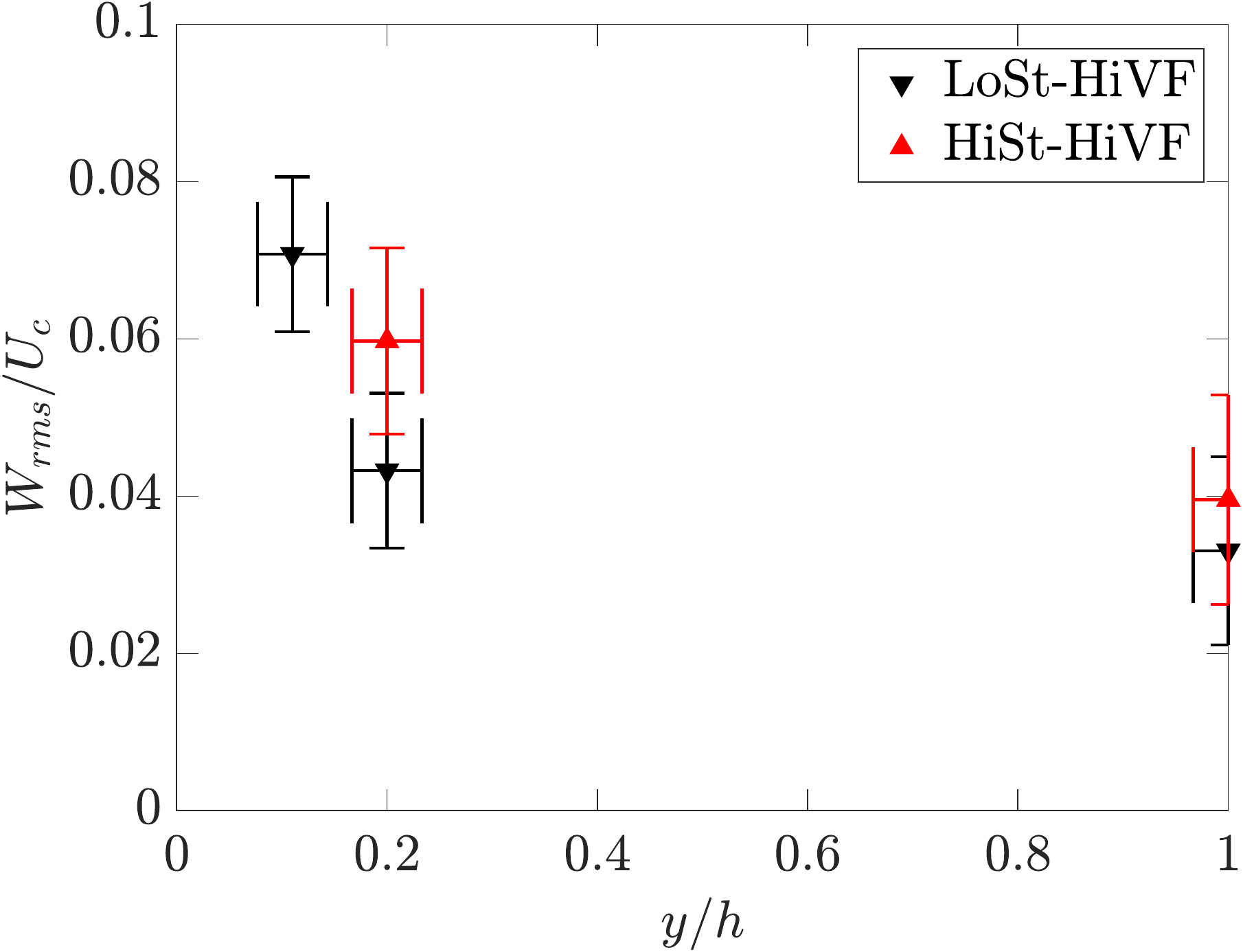}}
  \caption{Profile of rms spanwise particle velocity for LoSt-HiVF and HiSt-HiVF cases, obtained from the wall-parallel measurements expanded in §3.2.}
\label{fig:wrms}
\end{figure}

For completeness, we present in figure \ref{fig:wrms} the rms of the particle spanwise velocity fluctuations from the wall-parallel measurements. At $y/h$ = 0.2 and 1, the values are consistent with the fluid $W_{rms}$ in the DNS of \citet{moser1999} for $Re_{\tau} = $ 180 and 395 at the same wall-normal locations, which are expected to be close to the unladen fluid values in the present case. The sharp increase at $y/h = 0.11$ for the LoSt-HiVF case indicates again an \textcolor{black}{augmented fluctuation of the particle velocity} near the wall with higher loading, at odds with previous two-way coupled simulations (\citealt{li2001}; \citealt{nasr2009}).

\begin{figure}
  \centerline{\includegraphics[width=0.6\textwidth]{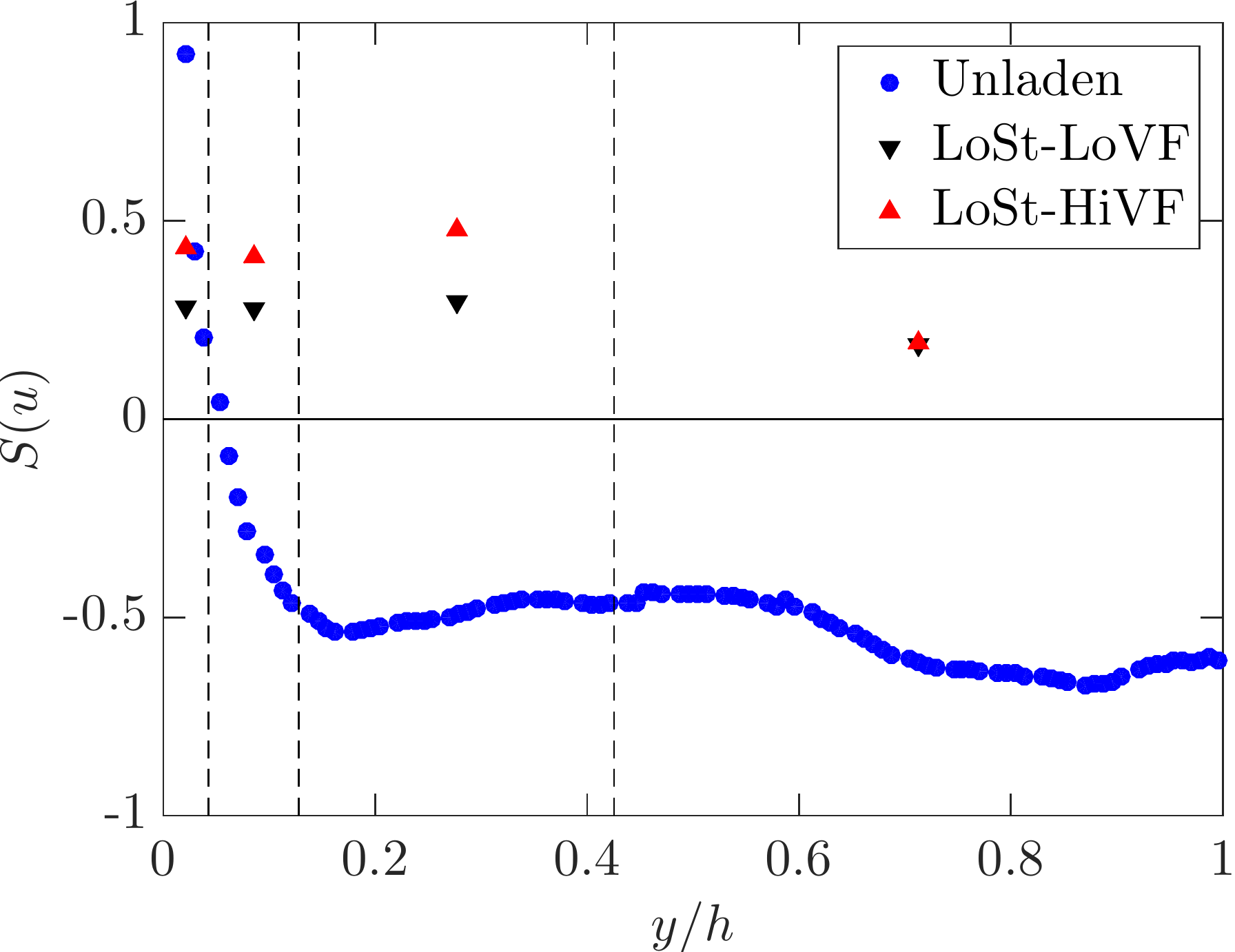}}
  \caption{Profile of the skewness of the particle streamwise velocity fluctuations in the LoSt-LoVF and LoSt-HiVF cases. Unladen fluid profiles plotted for comparison. The vertical dashed lines demarcate the regions over which the particle-laden data are averaged.}
\label{fig:LoSt-skew}
\end{figure}

We conclude this section considering the skewness of the streamwise velocity fluctuations for the inertial particles, in comparison with the unladen fluid. Figure \ref{fig:LoSt-skew} shows data for the LoSt cases (HiSt cases displaying the same trend). To improve convergence, the particle profiles are binned in four regions, each displaying fairly homogeneous behavior and roughly corresponding to the inner layer ($y^+ < 10$), buffer layer ($10 < y^+ < 30$), log layer ($30 < y^+ < 100$), and outer layer ($100 < y^+ < 235$). The \textcolor{black}{unladen} fluid streamwise fluctuations have positive skewness in the inner layer and part of the buffer layer, and negative elsewhere as expected (see e.g. \citealt{kim1987}). On the other hand, the inertial particles show positive skewness across the channel height, irrespective of volume fraction. Considering the flow is in the direction of gravity, this may be due to a tendency of the particles in the channel core to favor the downward side of turbulent eddies, as it is known to happen in homogeneous turbulence \citep{wang1993}. We will return to this point in §3.2.2.

\subsection{Wall-parallel measurements}

We leverage the wall-parallel plane imaging to investigate the instantaneous spatial organization of the particles and their velocity distribution. Streamwise/spanwise planes are especially suitable for this analysis, as they extend along homogeneous directions and thus allow for the efficient calculation of statistics that are unbiased by spatial gradients. We employ two-point quantities such as radial distribution functions and two-point Eulerian velocity correlations of streamwise velocity fluctuations, as well as tessellation techniques such as Voronoi diagrams and box-counting. We do not report here on the spanwise velocity fluctuation \textcolor{black}{ correlations in the wall-parallel planes,} as \textcolor{black}{the spanwise displacements} are not sufficiently larger than the uncertainty to yield accurate second-order statistics. These tools are used to investigate the wall-parallel plane at the centerline (center-plane) and the near-wall plane at $y/h = $ 0.11. The analysis is carried out only for the HiVF cases, for which the number of particles is sufficient to provide sufficient spatial resolution and statistical accuracy. While this does not allow direct assessment of the volume fraction effect on such quantities, it does bring useful insight on the particle spatial distribution in the regime for which the inter-phase coupling is expected to be more complex.

\subsubsection{Two-point statistics}

We use radial distribution functions (RDFs) to describe the scale-by-scale concentration in the area surrounding a generic particle, compared to a uniform distribution \citep{sundaram1997}. For 2D distributions such as those obtained by planar imaging, this can be written as (see, e.g., \citealt{wood2005}):

\begin{equation}
 g(r) = \frac{\langle N_r/A_r\rangle}{N/A} ,
 \label{rdf}
\end{equation}

where $N_r$ represents the number of particles within an annulus of radius $r$ and area $A_r$, $N$ is the total number of particles within the planar domain of area $A$. In presence of clustering, the RDF is expected to increase above 1 for decreasing $r$, and the range over which it remains significantly greater than unity approximately indicates the length scale over which clustering occurs. We compute RDFs by binning particle pairs in equally spaced annuli of radial width 0.5 mm ($0.03h$). An edge-correction strategy is needed for particles near the image boundaries. Omitting annuli that cross the image boundary limits the maximum separation to the radius of the domain-inscribed circle, reducing the number of usable particle pairs with increasing separations and thus affecting the large-scale characterization. We instead mirror the particle field across the image boundaries, so that the same number of annuli is used for each particle location. The maximum separation then equals the full image size, introducing only small biases near the boundaries (\citealt{salazar2008}; \citealt{dejong2010}; \citealt{petersen2019}). To avoid projection biases at separations below the illuminated volume thickness \citep{holtzer2002}, we only present $g(r)$ for $r >\,$1.1 mm.

\begin{figure}
\centering
\begin{subfigure}{.49\textwidth}
  \centerline{\includegraphics[scale=0.38]{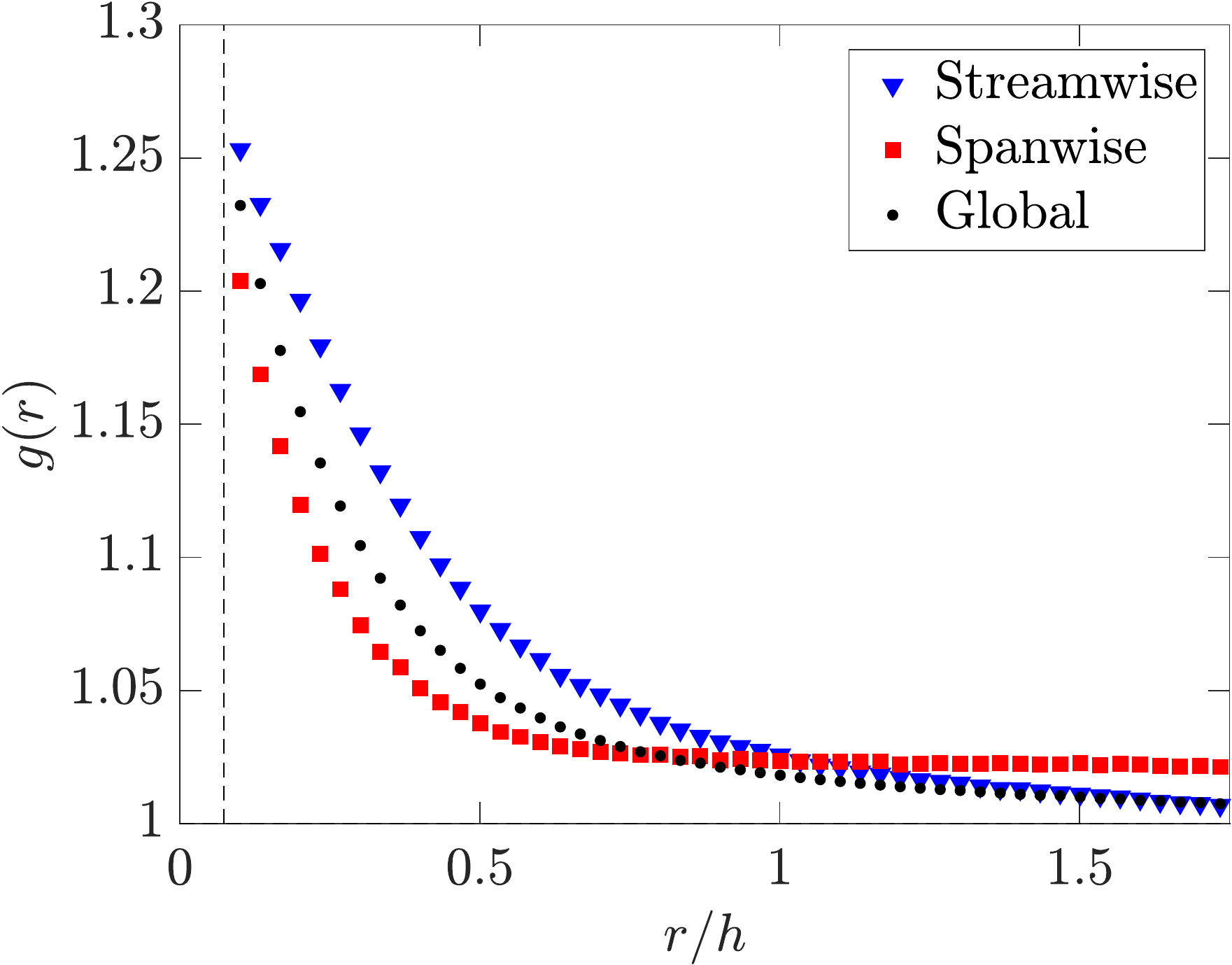}}
  \caption{}
 \end{subfigure}
 \begin{subfigure}{.49\textwidth}
   \centerline{\includegraphics[scale=0.38]{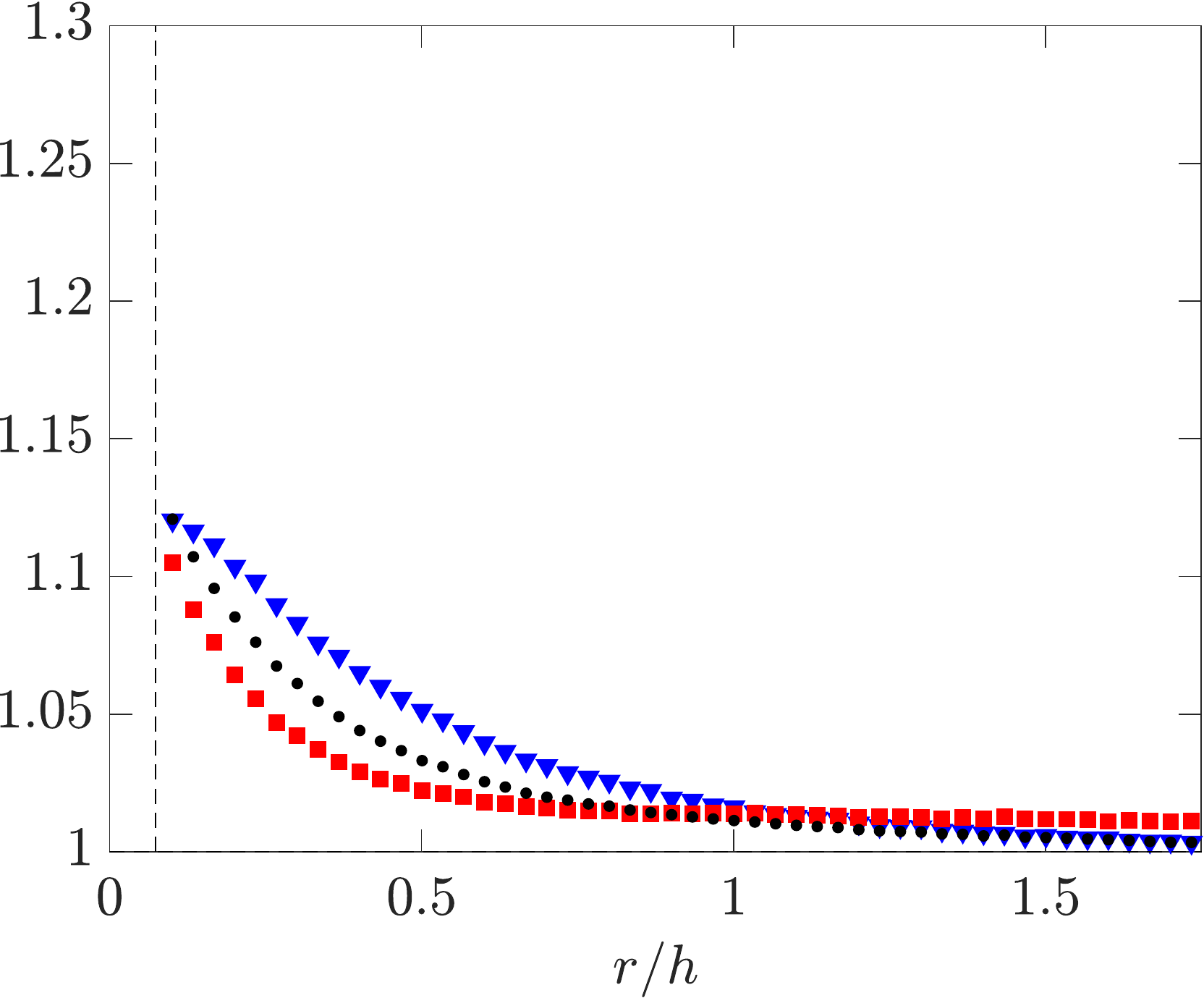}}
  \caption{}
 \end{subfigure}
  \caption{Global and directional RDFs along the center-plane for (a) LoSt-HiVF and (b) HiSt-HiVF. The vertical dashed line indicates the laser sheet thickness.}
\label{fig:centerplane-rdf}
\end{figure}

\begin{figure}
\centering
\begin{subfigure}{.49\textwidth}
  \centerline{\includegraphics[scale=0.36]{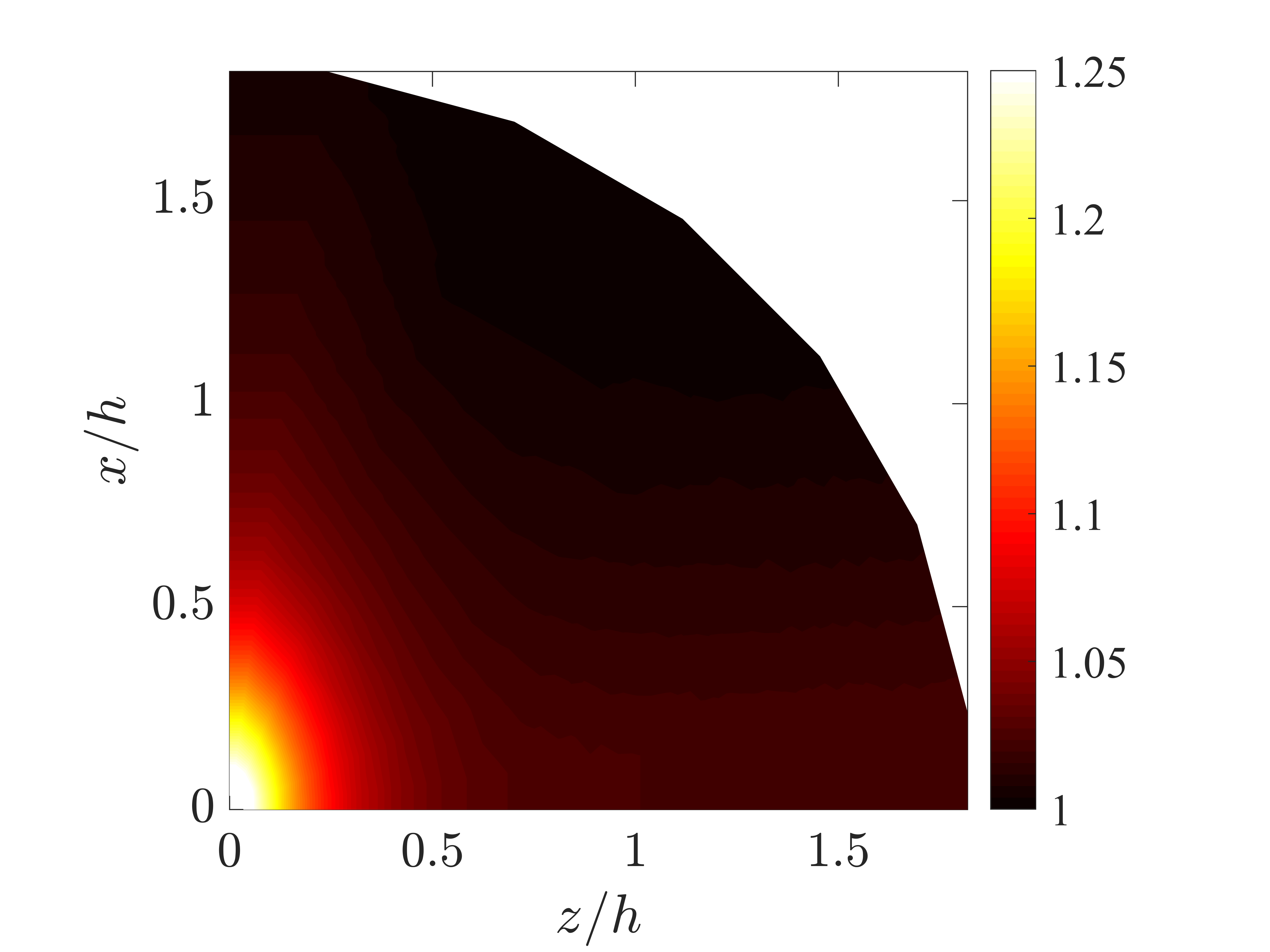}}
  \caption{}
 \end{subfigure}
 \begin{subfigure}{.49\textwidth}
   \centerline{\includegraphics[scale=0.36]{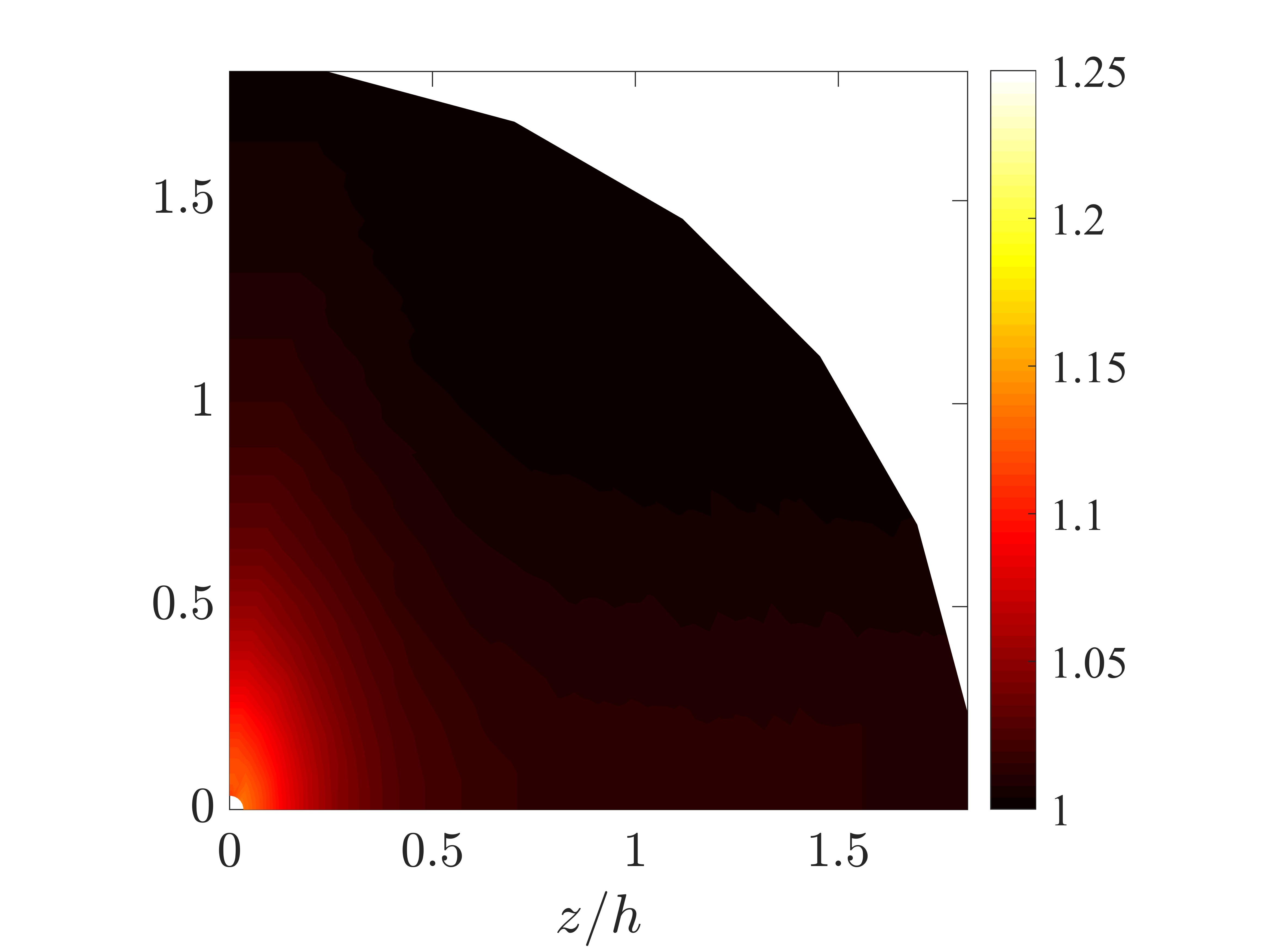}}
  \caption{}
 \end{subfigure}
  \caption{Angular distribution functions (ADFs) along the center-plane for (a) LoSt-HiVF and (b) HiSt-HiVF.}
\label{fig:centerplane-adf}
\end{figure}

This “global” (i.e., omnidirectional) definition of RDF does not discriminate between different directions of the separation $r$. We also calculate “directional” RDFs, in which the separations are oriented either streamwise or spanwise. This allows us to characterize the streamwise and spanwise extent of the highly concentrated particle structures. Additionally, we calculate the angular distribution function (ADF, see \citealt{gualtieri2009}, \textcolor{black}{\citealt{nicolai2013}}) which is obtained by binning the planar domain in polar coordinates ($r, \theta$):

\begin{equation}
 g(r,\theta) = \langle\frac{N_{r,\theta}(r,\theta)}{A_{r,\theta}(r,\theta)}\rangle\,/\,(\frac{N}{A}) ,
 \label{adf}
\end{equation}

Here $\theta = 0$ and $\theta = 90\degree$ correspond to spanwise and streamwise directions, respectively. We use equally spaced annuli of radial width 0.03$h$ and divide each of them in 24 azimuthal sectors of area $A$ (in which we count $N$ particles). Streamwise and spanwise homogeneity are leveraged to limit the analysis to one quarter of the $(r, \theta)$ circle.

We first consider the wall-parallel plane at the center-plane. Figure \ref{fig:centerplane-rdf} shows RDFs (global and directional) for LoSt and HiSt cases. The global RDFs indicate that clustering extends over similar length scales for both cases, but it is significantly more pronounced for LoSt. This is not unexpected since the latter is closer to the condition $St_\eta \sim 1$, which was shown to produce more intense clustering in homogeneous turbulence (\citealt{wang1993}; \citealt{wood2005}) and at the center-plane of channel flows \citep{fessler1994}. The directional RDFs also indicate that the clusters are more elongated in streamwise than in spanwise direction. The spanwise RDF remains somewhat above unity throughout the field, indicating that some structure in the particle distribution persists over large scales in that direction. This general picture is confirmed by the ADFs in figure \ref{fig:centerplane-adf}, which also show how the particle field becomes more quickly decorrelated for separations in direction $\theta \sim 45\degree$.

We next consider the two-point Eulerian velocity correlations, which provide information on the level of spatial coherence of the particle motion. We follow \citet{fevrier2005}, who in turn borrowed the formalism proposed by \citet{sundaram1999}, and write the general expression for the correlation between the streamwise velocity fluctuations of particles $m$ and $n$, normalized by their velocity variance:

\begin{equation}
 R_{uu}(r) = \frac{\langle u^{(m)}u^{(n)}\,|\,\bm{x}=\bm{x_p^{(m)}};\bm{x}+\bm{r}=\bm{x_p^{(n)}}\rangle}{\langle u^2\rangle} ,
 \label{ruu}
\end{equation}

Here $\bm{x}$ is the location within the measurement plane, $\bm{x_p^{(i)}}$ is the position of the generic $\bm{i}$th particle, $\bm{r}$ is the separation vector connecting the particle pair ($m,n$), and angle brackets represent ensemble-averaging over all particle pairs. Boldface denotes vectorial quantities. The calculation is implemented with the same processing routine used for the RDFs and ADFs (which contains the information on the particle pair mutual positions). Again, we calculate both “directional” correlations, in which the separation vector is either streamwise or spanwise, and polar correlations, which span the ($r,\theta$) space.

\begin{figure}
\centering
\begin{subfigure}{.49\textwidth}
  \centerline{\includegraphics[scale=0.38]{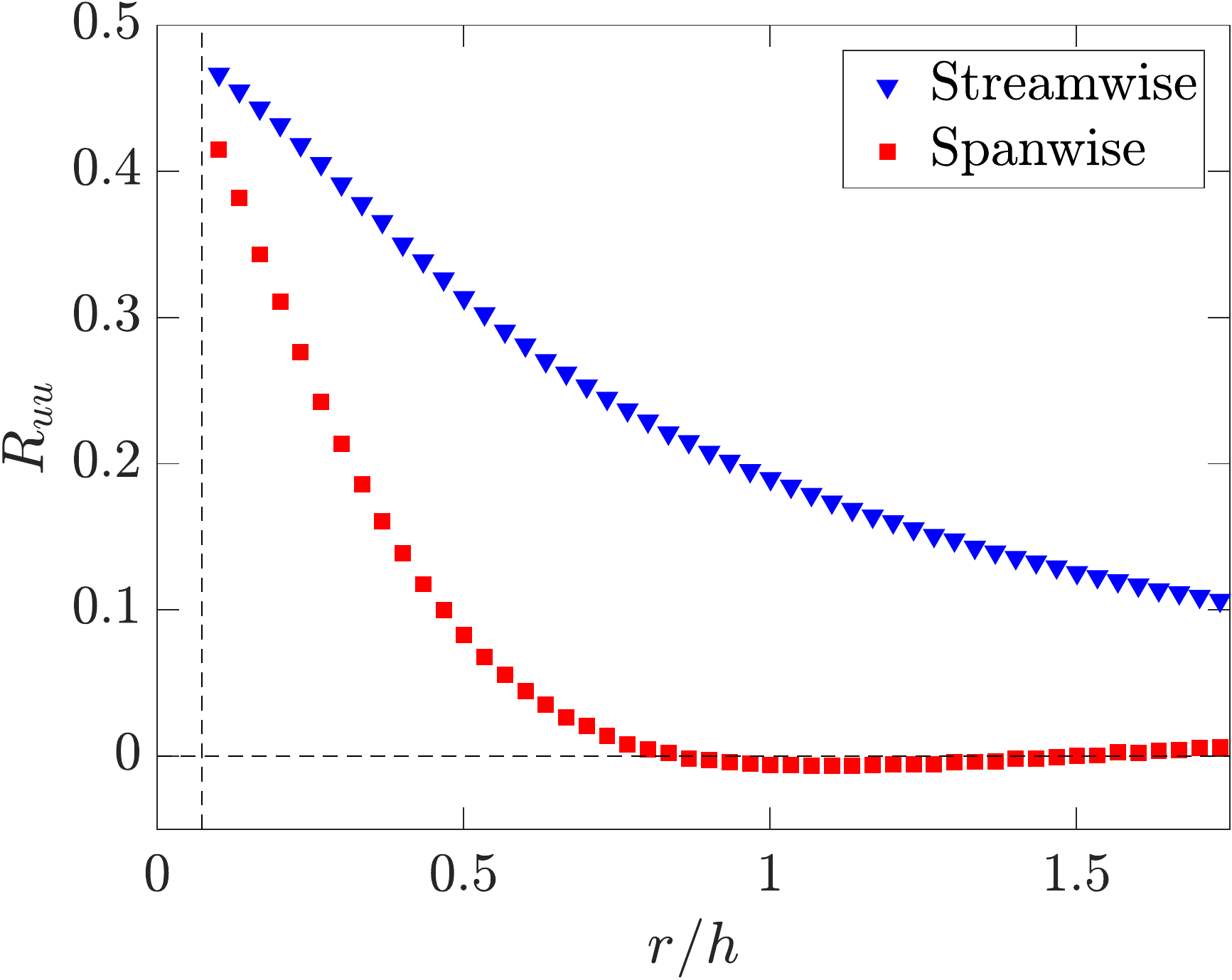}}
  \caption{}
 \end{subfigure}
 \begin{subfigure}{.49\textwidth}
   \centerline{\includegraphics[scale=0.38]{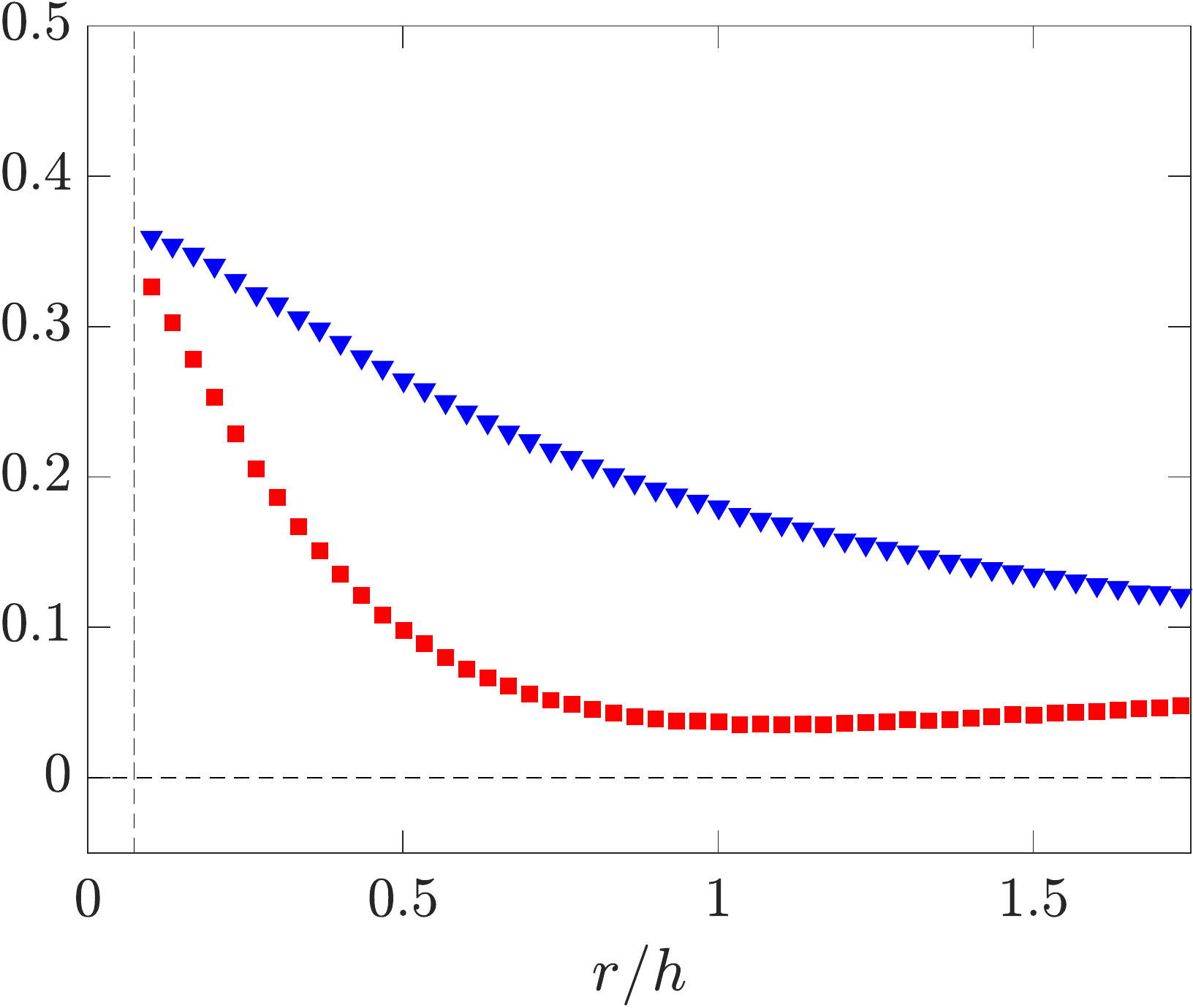}}
  \caption{}
 \end{subfigure}
  \caption{Two-point correlation of streamwise velocity fluctuations with separations in streamwise and spanwise directions along the center-plane, for (a) LoSt-HiVF and (b) HiSt-HiVF. The vertical dashed line indicates the laser sheet thickness.}
\label{fig:centerplane-ruu}
\end{figure}

Figure \ref{fig:centerplane-ruu} displays the directional velocity correlations evaluated at the center-plane. For both LoSt and HiSt cases, the normalized values do not approach unity for vanishingly small separations. (This is also confirmed by data points for separations smaller than the laser sheet thickness, not shown because inherently less accurate.) This indicates that a significant portion of the particle velocity is not spatially correlated. This is in line with the mesoscopic Eulerian formalism introduced by \citet{fevrier2005}, according to which inertial particle motion consists of two components: a contribution from the underlying turbulent velocity field, spatially correlated; and a quasi-Brownian velocity distribution, random and as such spatially uncorrelated. The latter is rooted in the particle inertia, in particular the memory of interactions with distant eddies. This results in different velocities of arbitrarily close particles, possibly enhancing collision rates, and is consistent with the concepts of caustics and sling effect (\citealt{wilkinson2005}; \citealt{bewley2013}; \citealt{reeks2014}). The gap between unity and $R_{uu}$ for vanishing separations is a measure of the fraction of random uncorrelated motion (\citealt{fevrier2005}; \citealt{vance2006}). This framework has been employed in numerous theoretical and numerical studies to analyze and model different particle-laden flows, from turbulent channels \citep{vance2006} to homogenous turbulence \citep{meneguz2011} and planar jets \citep{masi2014}. However, experimental observations of Eulerian particle velocity correlations have been rarely reported, \citet{khalitov2003} and \citet{sahu2014} being notable exceptions. Figure \ref{fig:centerplane-ruu} indicates that HiSt particles display a larger uncorrelated component of the motion than LoSt, consistently with the mesoscopic Eulerian formalism. 

\begin{figure}
\centering
\begin{subfigure}{.49\textwidth}
  \centerline{\includegraphics[scale=0.36]{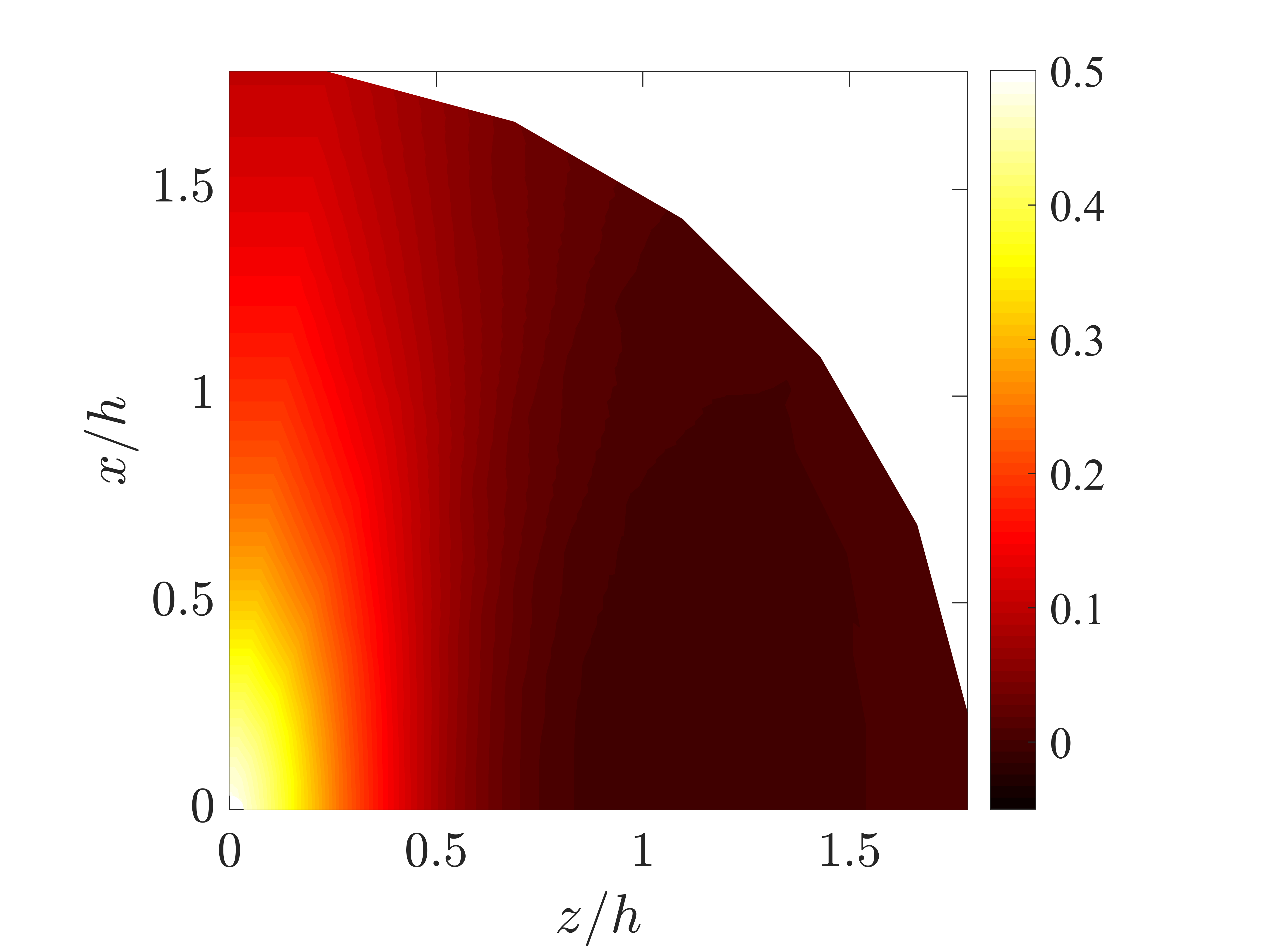}}
  \caption{}
 \end{subfigure}
 \begin{subfigure}{.49\textwidth}
   \centerline{\includegraphics[scale=0.36]{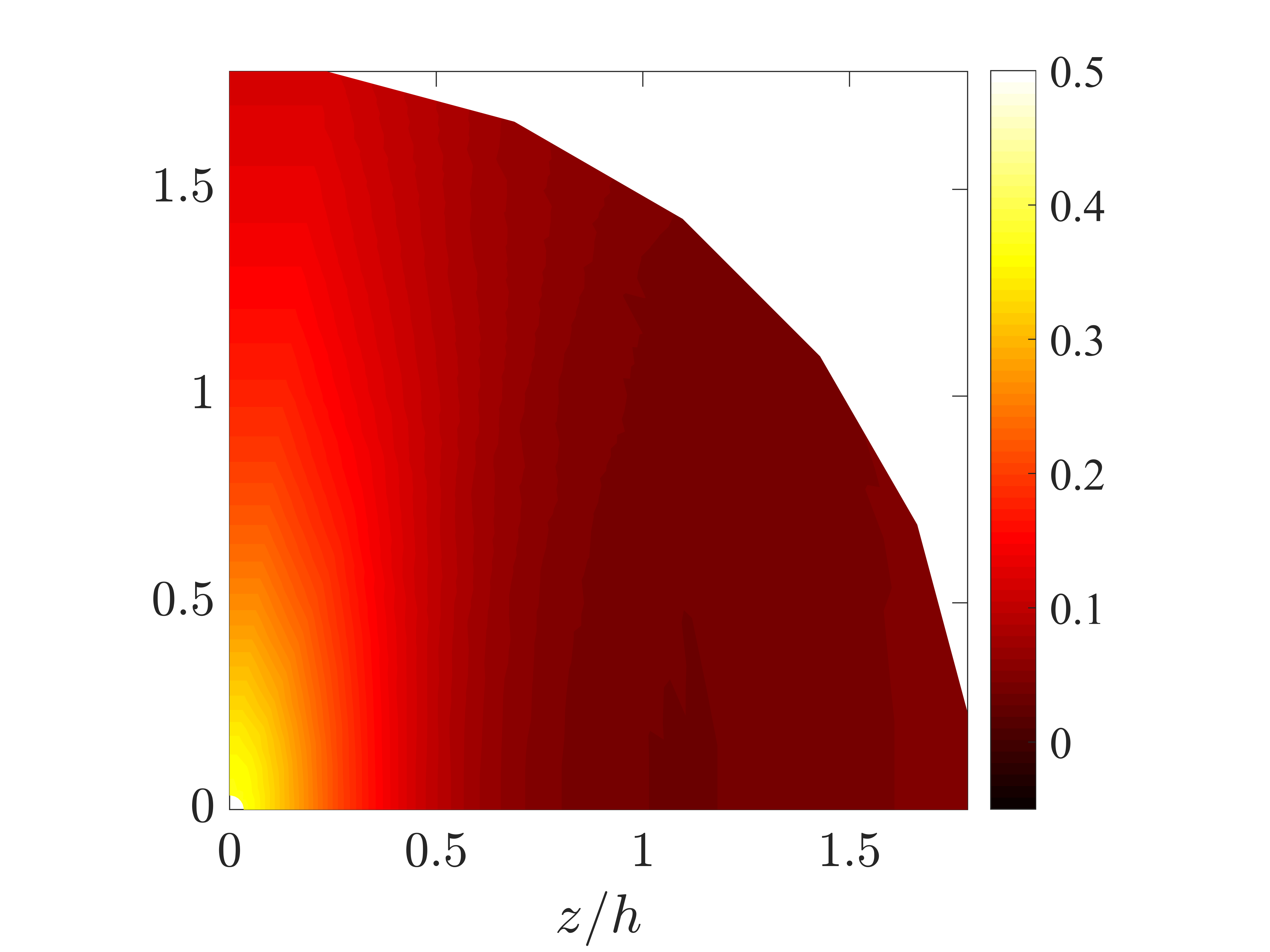}}
  \caption{}
 \end{subfigure}
  \caption{Polar map of streamwise velocity two-point correlation along the center-plane, for (a) LoSt-HiVF and (b) HiSt-HiVF.}
\label{fig:centerplane-quu}
\end{figure}

From figure \ref{fig:centerplane-ruu} one also observes that more inertial particles show a slower decay of velocity correlation with increasing separation, according to the picture of high-$St$ particles responding to larger turbulent scales. Moreover, the streamwise fluctuations are significantly more correlated in streamwise than spanwise directions. This is consistent with the idea that the correlated particle motion is dictated by the turbulent flow. Indeed, if one defines integral scales of the fluctuating particle velocity based on the separation at which the correlation drops by 50\%, the transverse scale appears to be roughly half the longitudinal one, similar to the expected behavior of the underlying turbulence. The polar diagrams of $R_{uu}$ in figure  \ref{fig:centerplane-quu} confirm this picture, and further suggest that the particle motion is organized in large streamwise-elongated structures, whose half-width is about 0.5$h$. This is consistent with the spatial particle distributions as deduced from the RDFs.

\begin{figure}
\centering
\begin{subfigure}{.49\textwidth}
  \centerline{\includegraphics[scale=0.38]{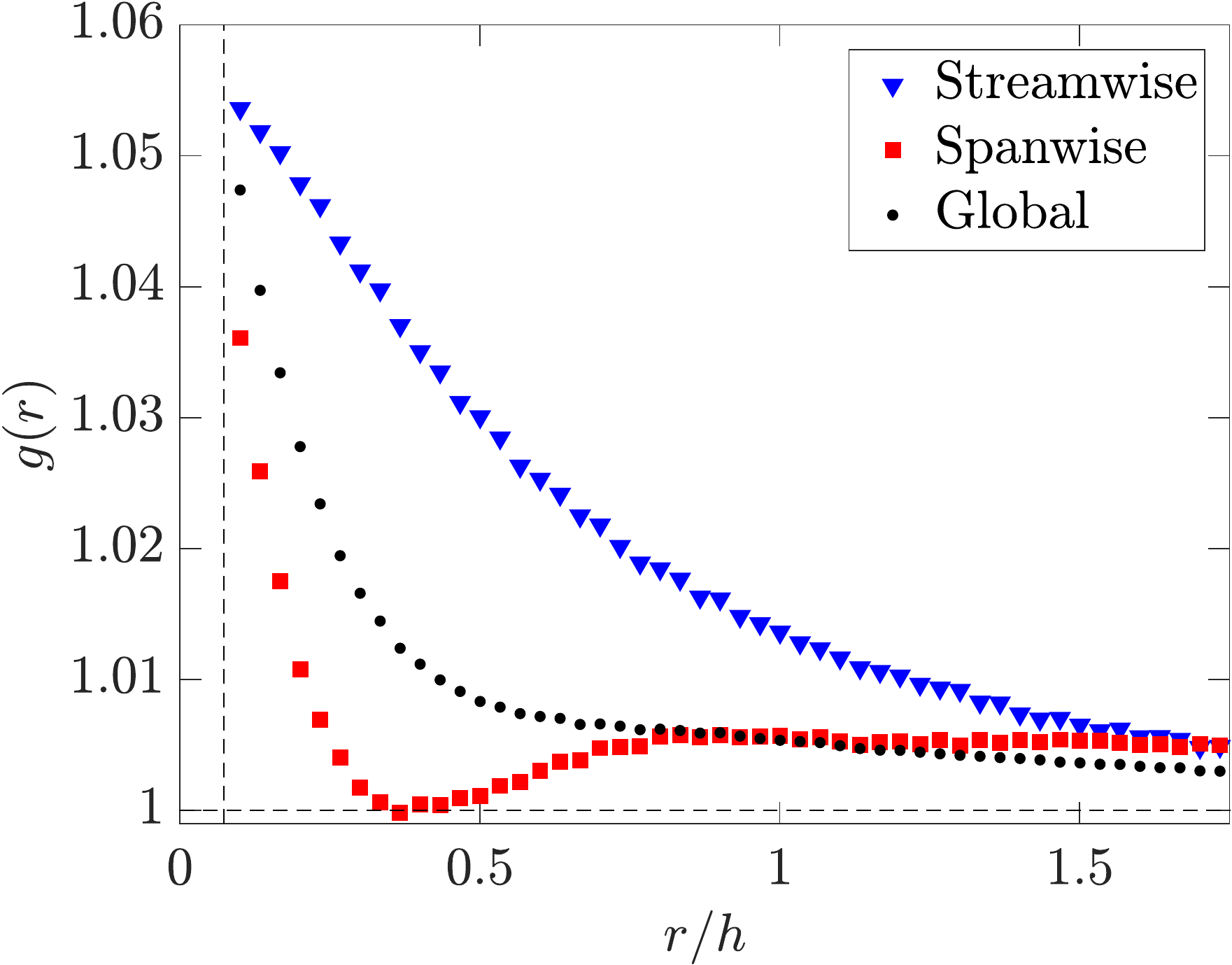}}
  \caption{}
 \end{subfigure}
 \begin{subfigure}{.50\textwidth}
   \centerline{\includegraphics[scale=0.34]{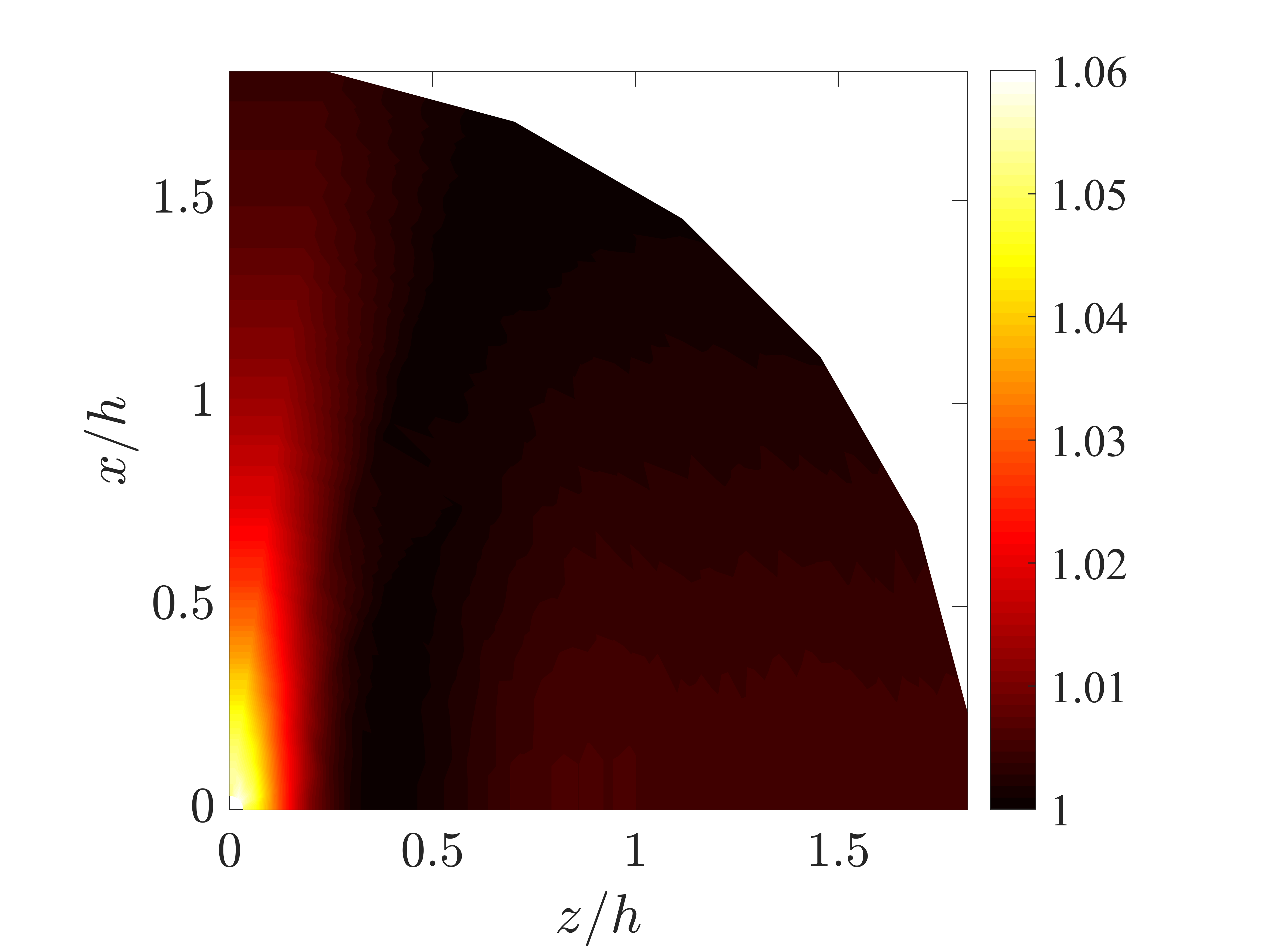}}
  \caption{}
 \end{subfigure}
  \caption{(a) Global and directional RDFs and (b) ADF for the LoSt-HiVF case along the near-wall plane.}
\label{fig:nearwall-rdf}
\end{figure}

After considering two-point statistics at the center-plane, we move our attention to the near-wall plane. We focus on the LoSt case, for which more significant preferential concentration is expected. Figure \ref{fig:nearwall-rdf} displays global and directional RDFs and ADFs, which indicate how the particles are arranged in elongated streaks, multiple channel heights in length. Indeed, due to the highly anisotropic spatial distribution of the particles in this region, the global RDFs provide limited insight compared to the directional representations. The amplitude of the peak is significantly smaller than at the center-plane, indicating generally weaker clustering. This is consistent with the fact that particles have much larger response times than the near-wall turbulent scales. As mentioned in the Introduction, several authors used point-particle simulations to investigate the near-wall structure of the particle distributions in regimes for which turbophoresis and preferential concentration are intense (\citealt{mclaughlin1989}; \citealt{zhang2000}; \citealt{rouson2001}; \citealt{marchioli2002}; \citealt{soldati2009}; \citealt{sardina2012a}; \citealt{bernardini2014}). They found thin streaks separated by $\mathcal{O}$(100) wall units, which roughly correspond to fluid-phase low-speed streaks in wall-bounded flows \citep{robinson1991}, and are even longer than the fluid streaks. Experimental observations of particle streaks have been sporadic, and mostly limited to snapshot realizations (\citealt{kaftori1995a}; \citealt{nino1996}). The present measurements provide quantitative information on such structures: the spanwise RDF shows a minimum at separations of $\sim 0.3h$ or 70 wall units (which can be interpreted as a measure of the streaks width) and recovery to a local maximum at $\sim0.75h$ or 175 wall units (a measure of the streak spacing). These values are somewhat larger than what reported by numerical studies at similar regimes. Moreover, the RDF amplitude we observe at small separations is much smaller than in computations, \textcolor{black}{as will be confirmed by instantaneous particle distributions shown later}. Beside the above-mentioned limitations of the point-particle modeling approach, the differences can be partly attributed to the location and thickness of the observation region. Most numerical studies report on streaks along thin slices within the viscous sublayer, which are challenging to isolate in laboratory experiments. Here the imaged particles are contained in a slab of thickness $\sim$ 1.1 mm centered at $y/h \sim0.11$, thus in the approximate range $y^+$ = 17 – 34. The projection through this thickness may significantly influence the apparent concentration in a region of large wall-normal gradients. Future quantitative comparisons with simulations should take into account such finite thickness of the illumination volume. 

\begin{figure}
\centering
\begin{subfigure}{.49\textwidth}
  \centerline{\includegraphics[scale=0.38]{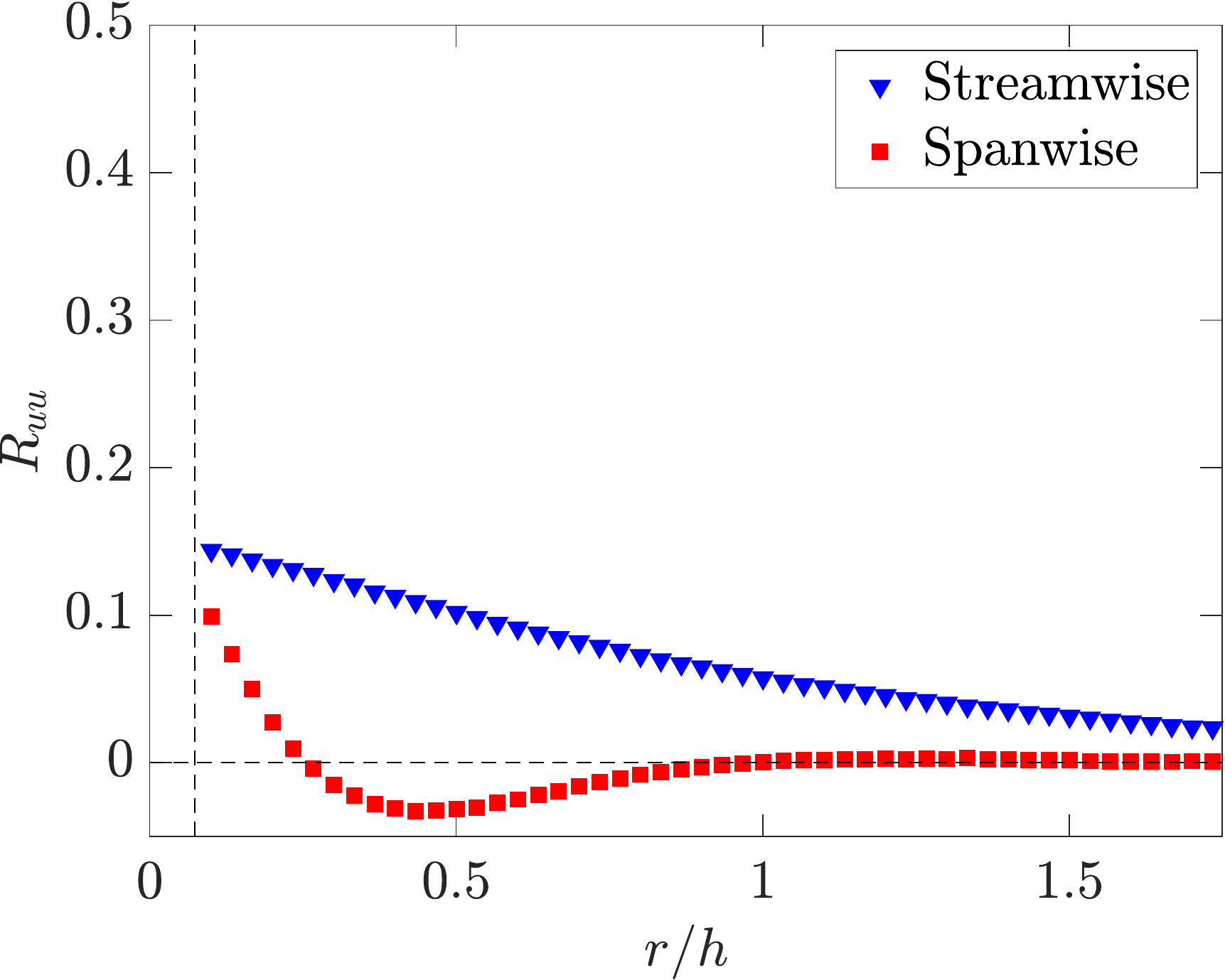}}
  \caption{}
 \end{subfigure}
 \begin{subfigure}{.50\textwidth}
   \centerline{\includegraphics[scale=0.34]{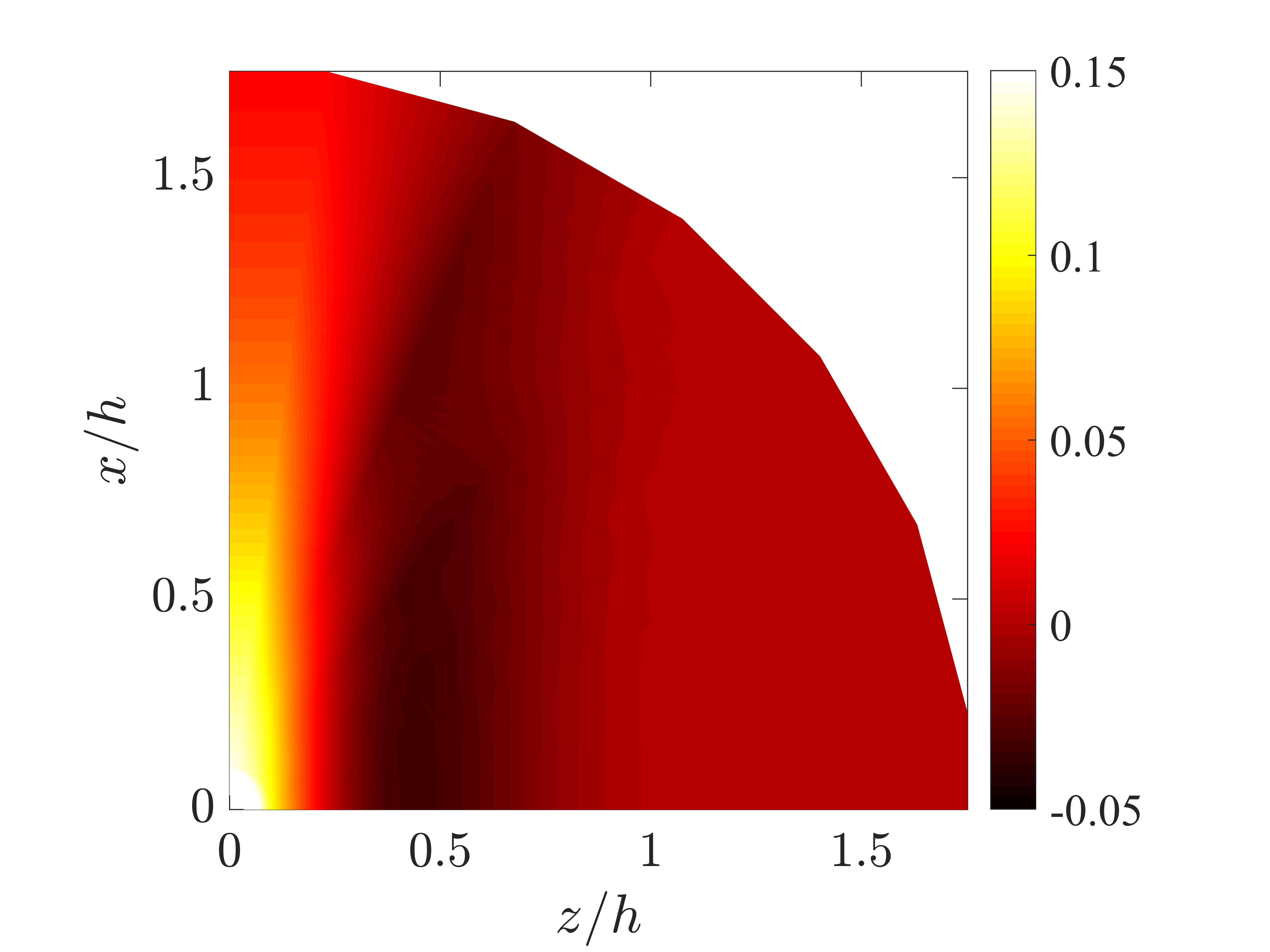}}
  \caption{}
 \end{subfigure}
  \caption{(a) Two-point correlation of streamwise velocity fluctuations with separations in near-wall plane.(b) Spatial velocity correlation map for streamwise velocity fluctuations of inertial particles in the near-wall plane.}
\label{fig:nearwall-ruu}
\end{figure}

When compared to the center-plane, the \textcolor{black}{two-point} velocity correlations at $y/h$ = 0.11 show an even stronger uncorrelated component of the motion as shown in figure \ref{fig:nearwall-ruu}. \textcolor{black}{The uncorrelated velocity component near the wall} is expected to increase with increasing particle inertia \citep{fevrier2005}, and indeed in the near-wall region the particle response time is much larger than the local time scale of the turbulence (i.e., $St^+ \gg 1$). Besides inertia, inter-particle and wall-particle collisions may also contribute to the random particle motion \citep{vance2006}. In the HiVF regime considered, both near-wall concentration and deposition velocity are relatively high (§3.1), thus collisions may play a significant role in the observed partitioning between correlated and uncorrelated velocity. The negative lobe of velocity correlation along the spanwise direction indicates that the particles are arranged in a streaky fashion, alternating positive and negative streamwise velocity fluctuations. The longitudinal extent of those features cannot be precisely assessed from the present measurements, but the long tail of the correlation function in streamwise direction suggests they can extend beyond the field of view. The trends in figure \ref{fig:nearwall-ruu} are quantitatively similar to the RDFs and ADFs in figure \ref{fig:nearwall-rdf}, implying that the fluctuations of particle velocity and concentration are simultaneous. We will elaborate on this point in the next subsection.

\subsubsection{Domain tessellation}

In order to further investigate the instantaneous distribution of the inertial particle positions and velocities, we apply domain tessellation methods along the wall-parallel planes. These have been widely used to study clustering of inertial particles in turbulence \citep{monchaux2012}. These approaches should be considered complementary to RDFs, since the latter are strictly two-point quantities, while tessellations are sensitive to the multi-particle arrangement. The simplest method is perhaps box-counting, which consists of dividing the domain into boxes of equal size, counting the particles in each box, and comparing the PDF of the number of particles per box against the Poisson distribution expected for randomly distributed particles. This technique provides a simple scalar measure of the amount of clustering and has been fruitfully exploited in experimental studies (\citealt{fessler1994}; \citealt{aliseda2002}). In recent years, the Voronoi tessellation method \citep{monchaux2010} has gained broader favor: the domain (in our case the two-dimensional image) is divided into cells associated to individual particles, each cell containing the set of points closer to that particle than to any other. The inverse of the area of each cell equals the local instantaneous concentration, $C = 1/A_{cell}$. The method has been used in several experimental and numerical studies of wall-bounded particle-laden flows (\citealt{garcia2012}; \citealt{nilsen2013}; \textcolor{black}{\citealt{nicolai2013}}; \citealt{rabencov2014}). Compared to the box-counting method, it has the advantage of not requiring an extrinsic/arbitrary length scale (the box size). 

\begin{figure}
\centering
\begin{subfigure}{.54\textwidth}
  \centerline{\includegraphics[scale=0.55]{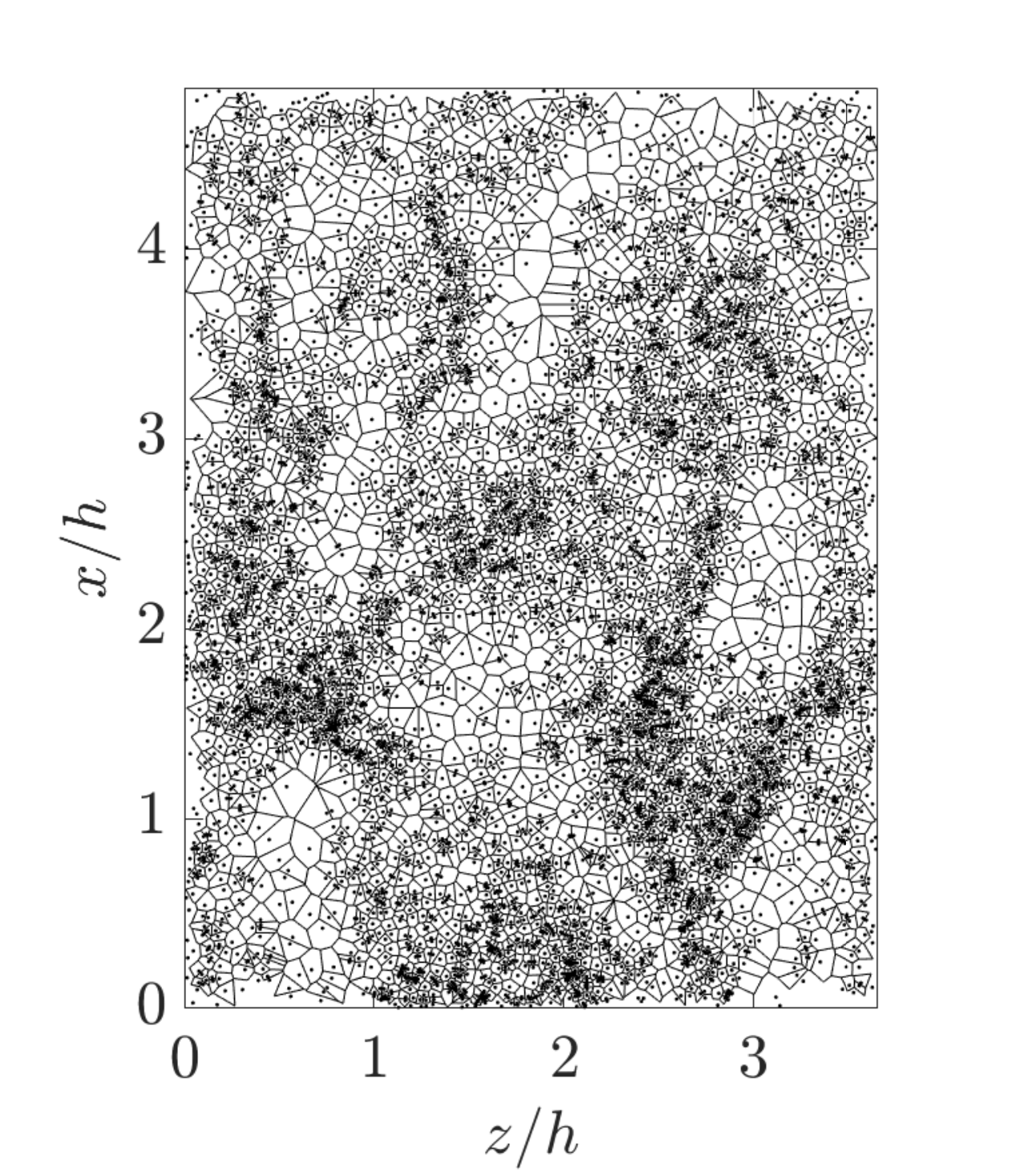}}
  \caption{}
 \end{subfigure}
 \begin{subfigure}{.44\textwidth}
   \centerline{\includegraphics[scale=0.55]{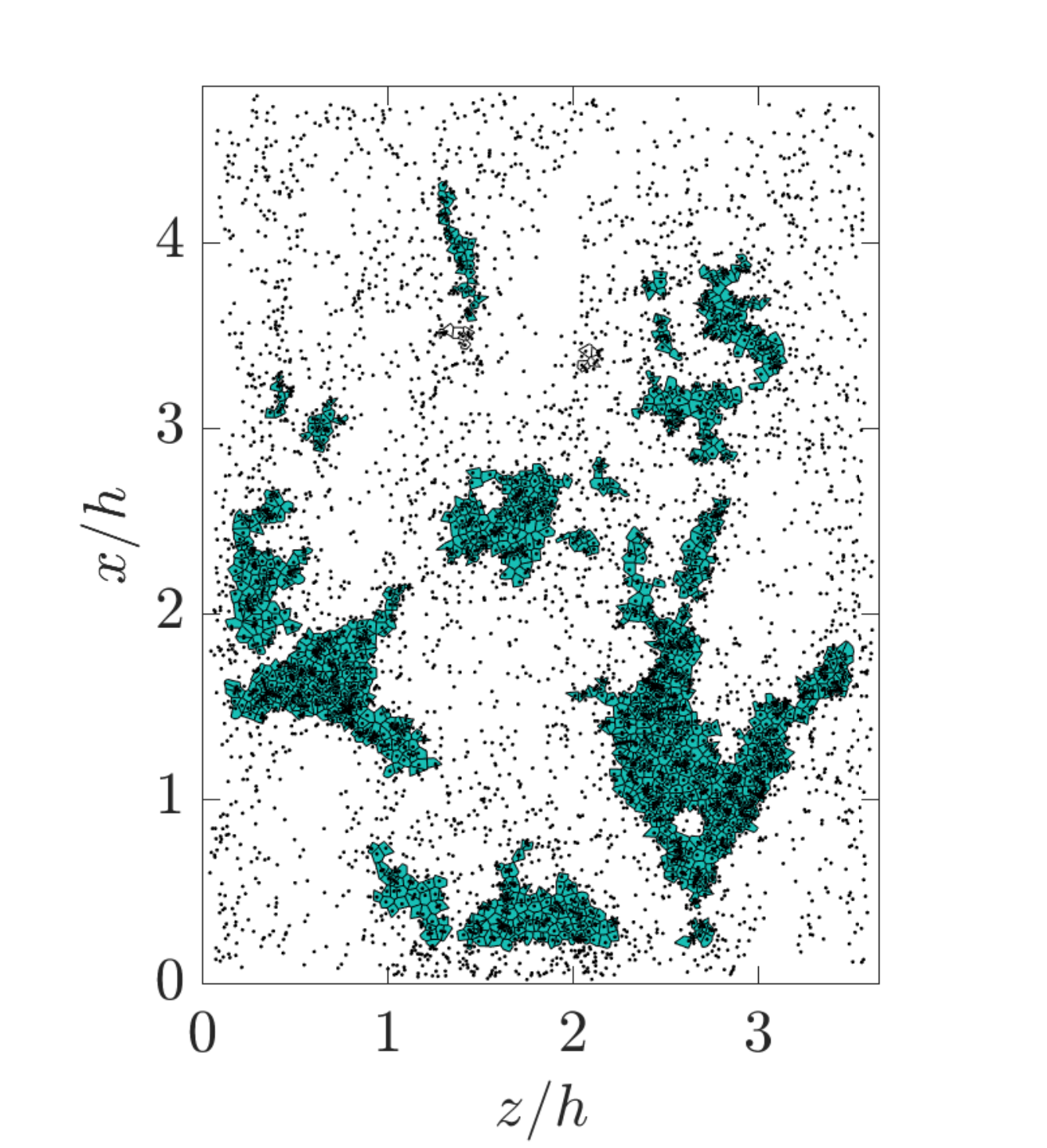}}
  \caption{}
 \end{subfigure}
 \begin{subfigure}{.49\textwidth}
   \centerline{\includegraphics[scale=0.4]{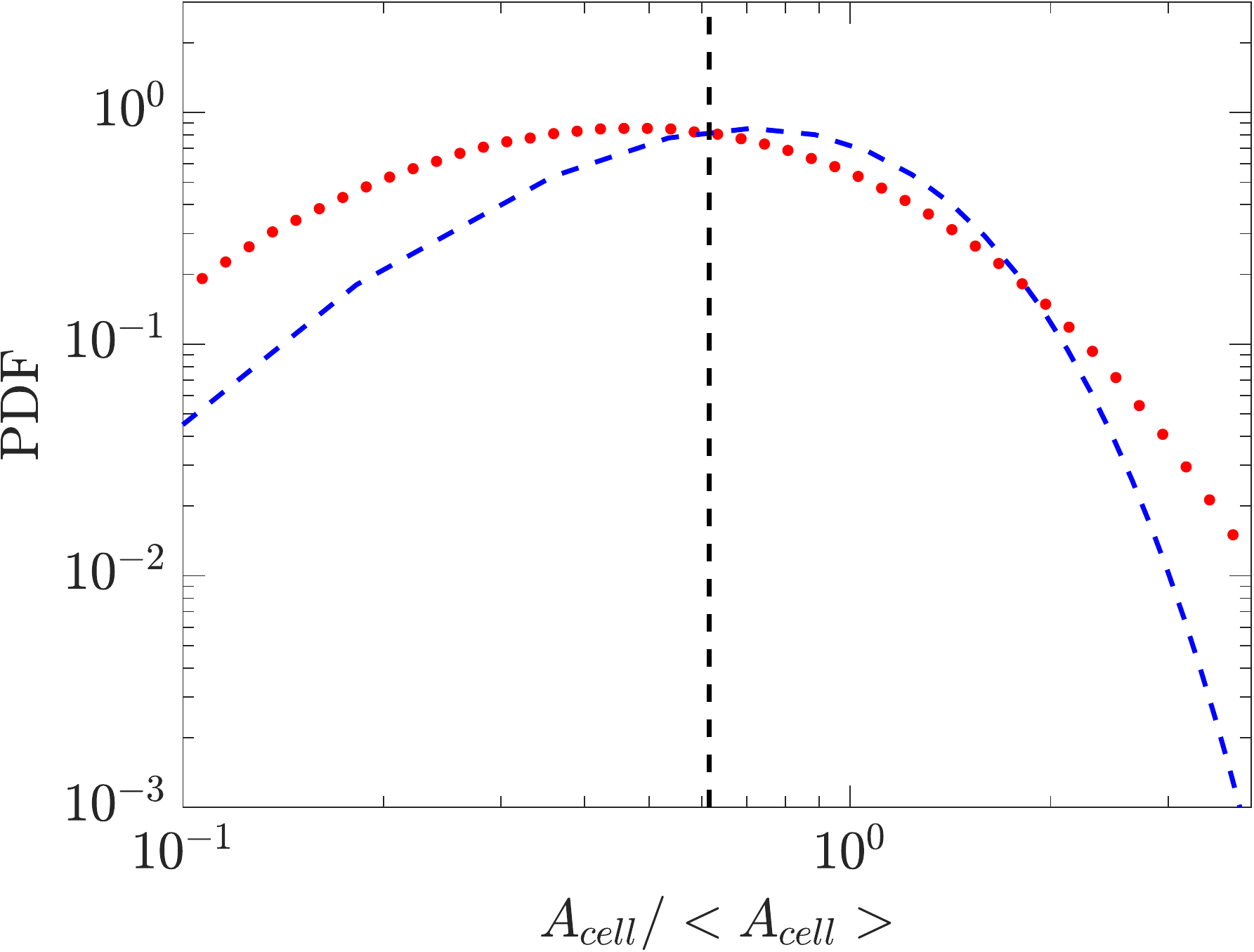}}
  \caption{}
 \end{subfigure}
  \caption{(a) Voronoi tessellation diagram in a sample center-plane realization, with (b) highlighted clusters (coherent clusters in cyan). (c) PDF of the Voronoi cell areas $A_{cell}$ along the center-plane (red circles), compared with a $\Gamma$ distribution (blue dashed line); the vertical dashed line indicates the threshold $A^*_{cell}$.}
\label{fig:voronoi-ctr}
\end{figure}

Here we adopt the Voronoi tessellation to investigate the particle distribution along the center-plane. Figure \ref{fig:voronoi-ctr}a shows a sample instantaneous realization for the LoSt-HiVF case, with Voronoi cells drawn around each particle. In figure \ref{fig:voronoi-ctr}c the PDF of the cell areas (normalized by its ensemble average) is plotted. As typical of inertial particles clustered by turbulence, the distribution is much wider compared to a random Poisson process, which is well approximated by a $\Gamma$ distribution \citep{ferenc2007}. The PDF of the Voronoi cells is found to closely follow a log-normal distribution (\citealt{monchaux2010}; \citealt{petersen2019}), which allows us to characterize the curve by its standard deviation $\sigma_A$. The latter is a metric of the amount of clustering: LoSt and HiSt cases are found to have $\sigma_A/\langle A_{cell}\rangle =$ \textcolor{black}{0.78 and 0.70}, respectively, confirming that the former has stronger tendency to produce clusters.

The topology and behavior of clusters of highly concentrated particles are relevant to the interphase coupling, especially in view of significant collective backreaction of the dispersed phase on the carrier fluid. We therefore analyze individual clusters \textcolor{black}{(colored in figure \ref{fig:voronoi-ctr}b)}, defined as connected groups of particles whose Voronoi cell areas are smaller than a threshold value $A_{cell}^*$ (figure \ref{fig:voronoi-ctr}c): the latter is taken as the value below which the probability of finding small cell areas is higher than for randomly distributed particles \citep{monchaux2010}. To avoid spurious edge effects, we apply the additional constraint that the area of the cells neighboring a cluster also be smaller than $A_{cell}^*$ \citep{zamansky2016}. The sum of the areas of all cells belonging to each cluster is taken as its “cluster area”, $A_C$. 


\begin{figure}
\centering
\begin{subfigure}{.45\textwidth}
  \centerline{\includegraphics[scale=0.41]{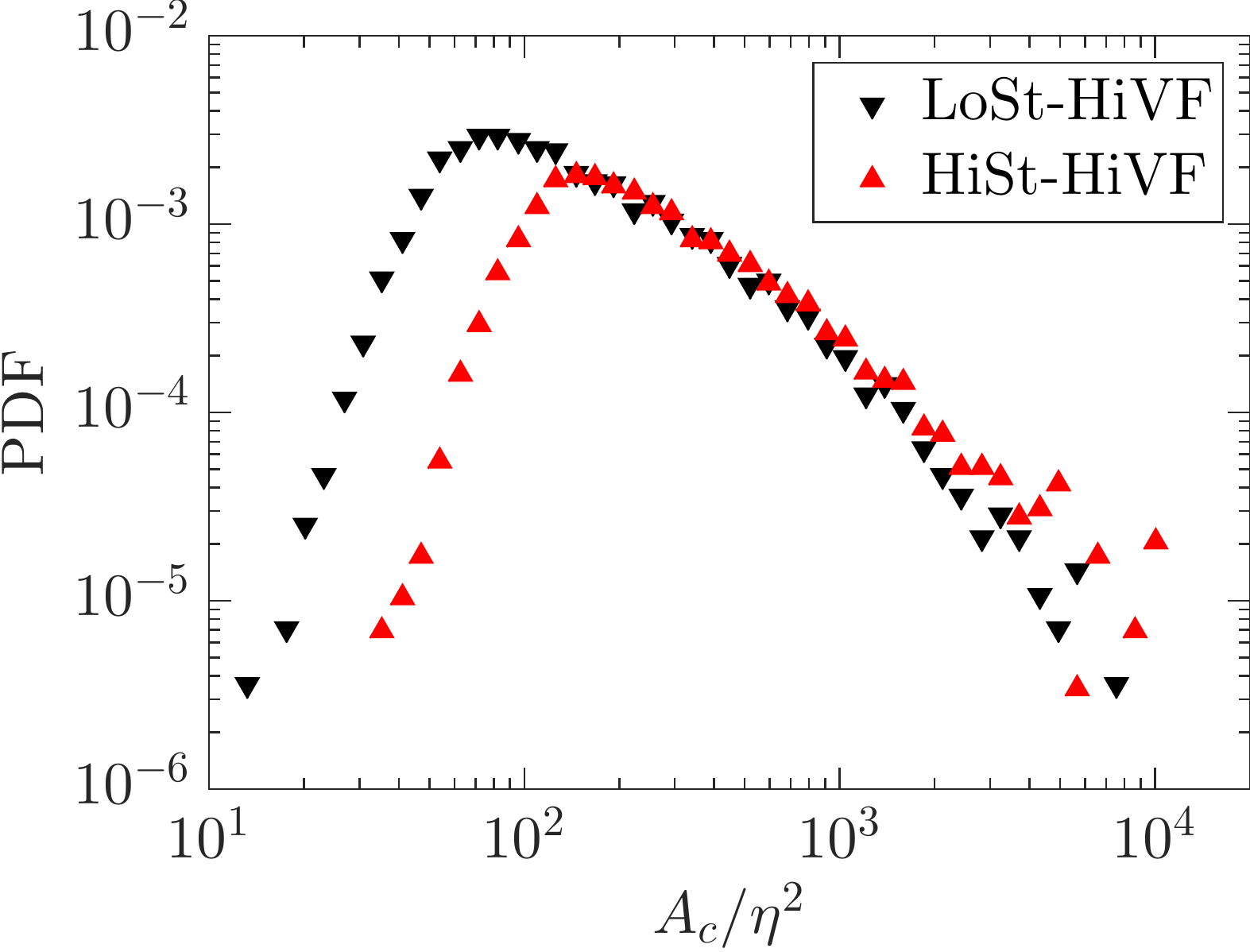}}
  \caption{}
 \end{subfigure}
 \begin{subfigure}{.54\textwidth}
   \centerline{\includegraphics[scale=0.41]{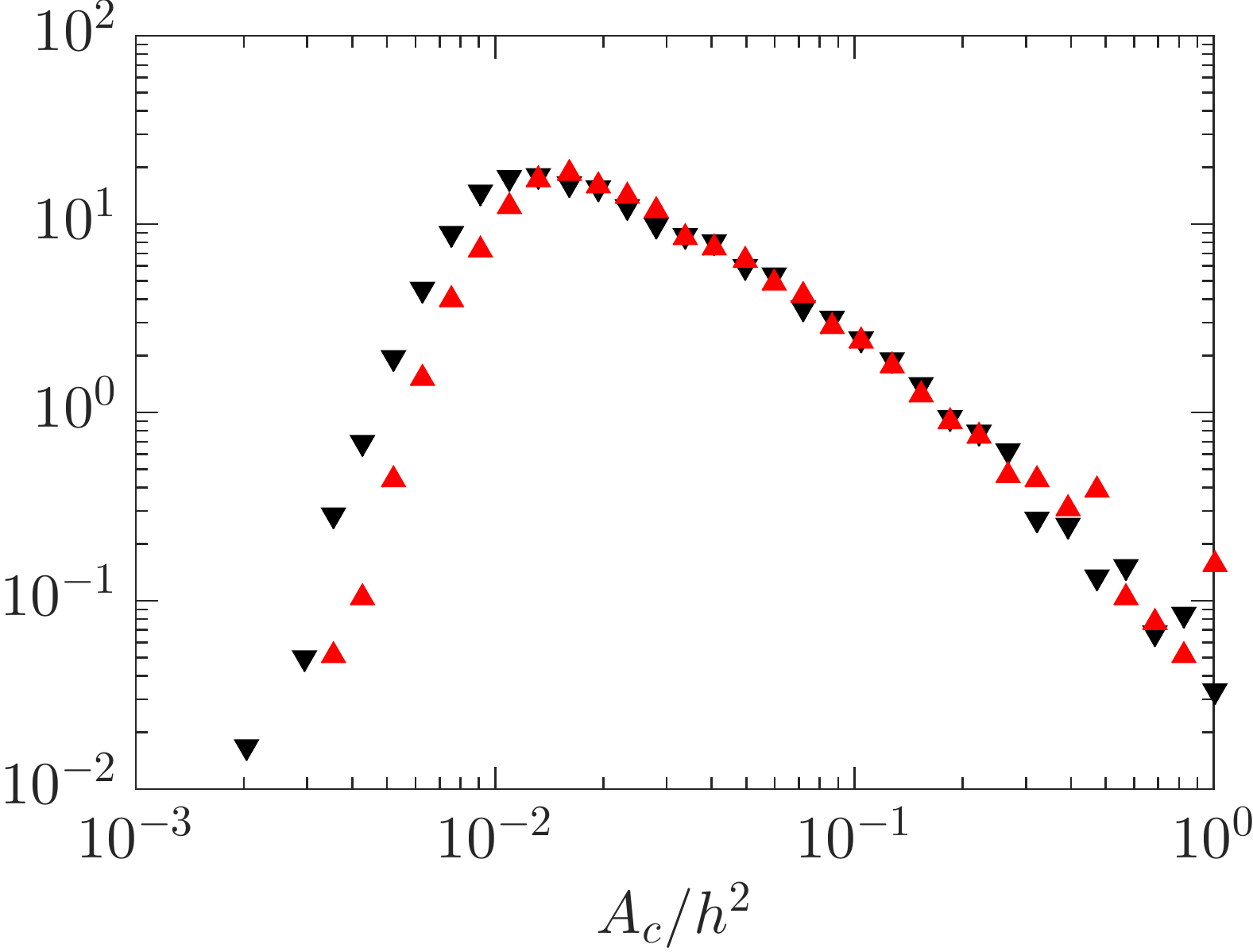}}
  \caption{}
 \end{subfigure}
  \caption{PDF of the cluster areas \textcolor{black}{for the LoSt-HiVF and HiSt-HiVF cases in the channel center plane, normalized by (a) the Kolmogorov scales corresponding to the flow velocity, and (b) the square of the channel half-height, $h^2$.}}
\label{fig:clust-area-pdf}
\end{figure}

Figure \ref{fig:clust-area-pdf} shows the probability distribution of cluster areas $P(A_C)$ along the center-plane for LoSt and HiSt, \textcolor{black}{the PDF in (a) normalized by the corresponding Kolmogorov length scale and (b) normalized by the square of the channel half-height, $h^2$.} Consistent with previous experimental studies, typical sizes are $\mathcal{O}$(10$\eta$) \citep{aliseda2002}, although such \textcolor{black}{estimates} may be affected by the number of particles in the system \citep{petersen2019}. HiSt particles tend to cluster over larger sets, consistent with their \textcolor{black}{limitation of only responding to the larger scales of turbulence.} Above a certain size the probability distributions approach a power-law decay, which is a \textcolor{black}{feature of fractality or} geometric self-similarity, \textcolor{black}{indicating a fractal-like formation process due to turbulence} \citep{baker2017}. This was clearer in the homogeneous turbulence studies of \citet{sumbekova2017} and \citet{petersen2019}, probably due to a combination of limited number of particles in the field of view and limited dynamic spatial range at the present Reynolds numbers. \textcolor{black}{Figure \ref{fig:clust-area-pdf}b shows that normalizing by the channel half-height produces a remarkable collapse of the LoSt and HiSt cases. This may suggest that, while particle clustering at the present $St$ is influenced by small-scale turbulence, the energetic scales of fluid motion also play a major role (as recently argued, e.g., by \citealt{petersen2019}). Moreover, this may indicate that the cluster size is significantly influenced by the channel geometry. Indeed, an \textit{a priori} estimate of the controlling effect of channel walls on clustering is often deduced from the ratio of the channel height to a characteristic cluster size, $L_c = \tau_p^2g$ (Capecelatro et al. 2014). Here $L_c = 2.8$ mm and $2h/L_c = 10.6$, and the value of $L_c$ is close to the peak of PDF($A_c$). We remark, however, that the definition of $L_c$ is usually adopted in much denser regimes than the present one, being independent of the air flow characteristics. Further studies with different particle properties and turbulence conditions shall discriminate whether $L_c$ is an appropriate scale for highly dilute systems.} 

Following \citet{baker2017}, we define “coherent clusters” as those objects large enough to display a scale-invariant topology, i.e. in the range of $P(A_C)$ that approximates the power-law decay. Smaller objects are considered as randomly occurring groups of particles, not necessarily brought together by the underlying turbulent flow. We conventionally set the cutoff at the respective maxima of $P(A_C)$ for both LoSt and HiSt, noting that the choice of twice larger cutoffs does not qualitatively change the observed trends. Besides the physical interpretation discussed in \citet{baker2017}, this step allows us to discard clusters formed by only a few particles (too small for a meaningful topological description).

\begin{figure}
\centering
\begin{subfigure}{.45\textwidth}
  \centerline{\includegraphics[scale=0.41]{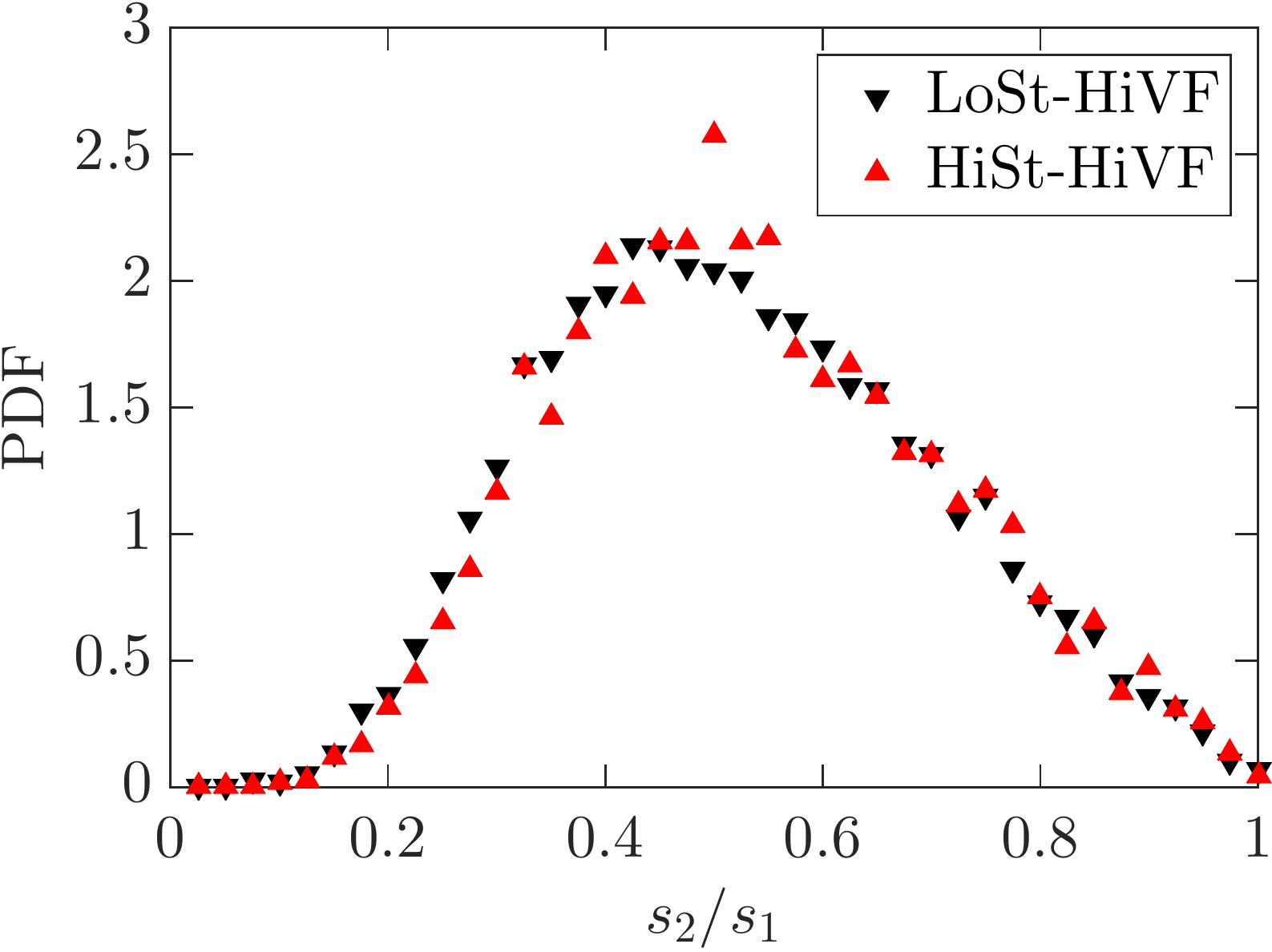}}
  \caption{}
 \end{subfigure}
 \begin{subfigure}{.54\textwidth}
   \centerline{\includegraphics[scale=0.41]{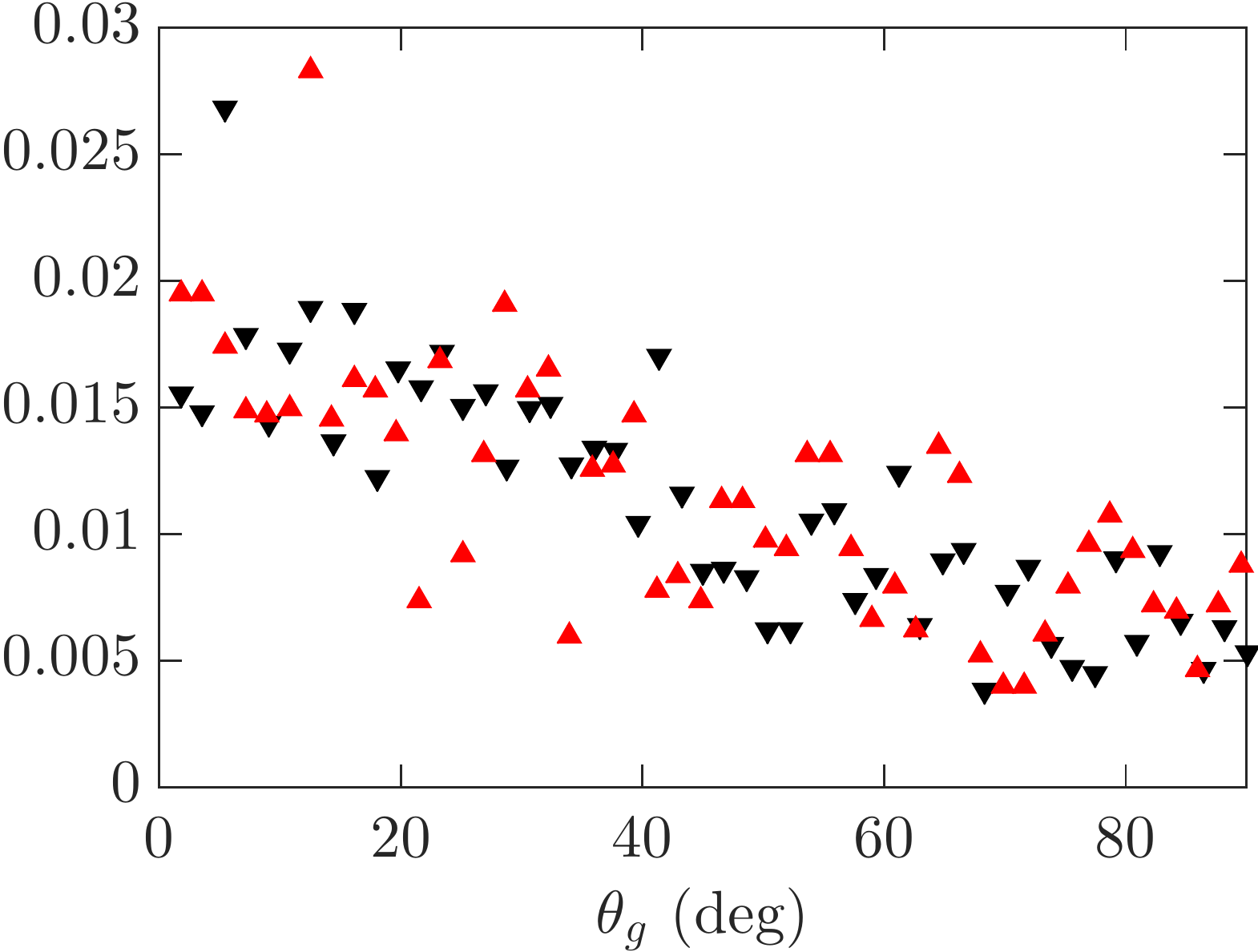}}
  \caption{}
 \end{subfigure}
  \caption{PDFs of (a) the SVD-based aspect ratio and (b) the angle between the primary axis and the vertical for the LoSt-HiVF and HiSt-HiVF cases.}
\label{fig:AR-orientation}
\end{figure}

We use the singular value decomposition (SVD) method introduced by \citet{baker2017} to probe the shape and spatial orientation of the coherent clusters. The SVD provides the principal axes and corresponding singular values for a particle set: the primary axis lies along the direction of greatest particle spread from the cluster centroid, the secondary axis being orthogonal to it. The corresponding singular values $s_1$ and $s_2$ measure the spread along the respective axes. In figure \ref{fig:AR-orientation}a the PDF of the aspect ratio $s_2/s_1$ is plotted for LoSt (differences with HiSt are marginal). The limit values 0 and 1 correspond to particles arranged in a straight line and in a perfect circle, respectively. The distribution is quantitatively similar to what reported by \citet{petersen2019} for clusters settling in homogeneous turbulence. The peak ratio between 0.4 and 0.55 reflects a tendency to form somewhat elongated objects. Figure \ref{fig:AR-orientation}b illustrates the probability distribution of the primary axis orientation, measured by the angle of the latter with the vertical ($\theta_g$). The peak at \textcolor{black}{$\theta_g = 0$} indicates a tendency of the (coherent) clusters to be aligned with gravity (and thus the direction of motion). This behavior was also reported in homogeneous turbulence studies (\citealt{baker2017};  \citealt{petersen2019}) and is consistent with the directional RDFs and ADFs presented above. 

\begin{figure}
\centering
\begin{subfigure}{.45\textwidth}
  \centerline{\includegraphics[scale=0.41]{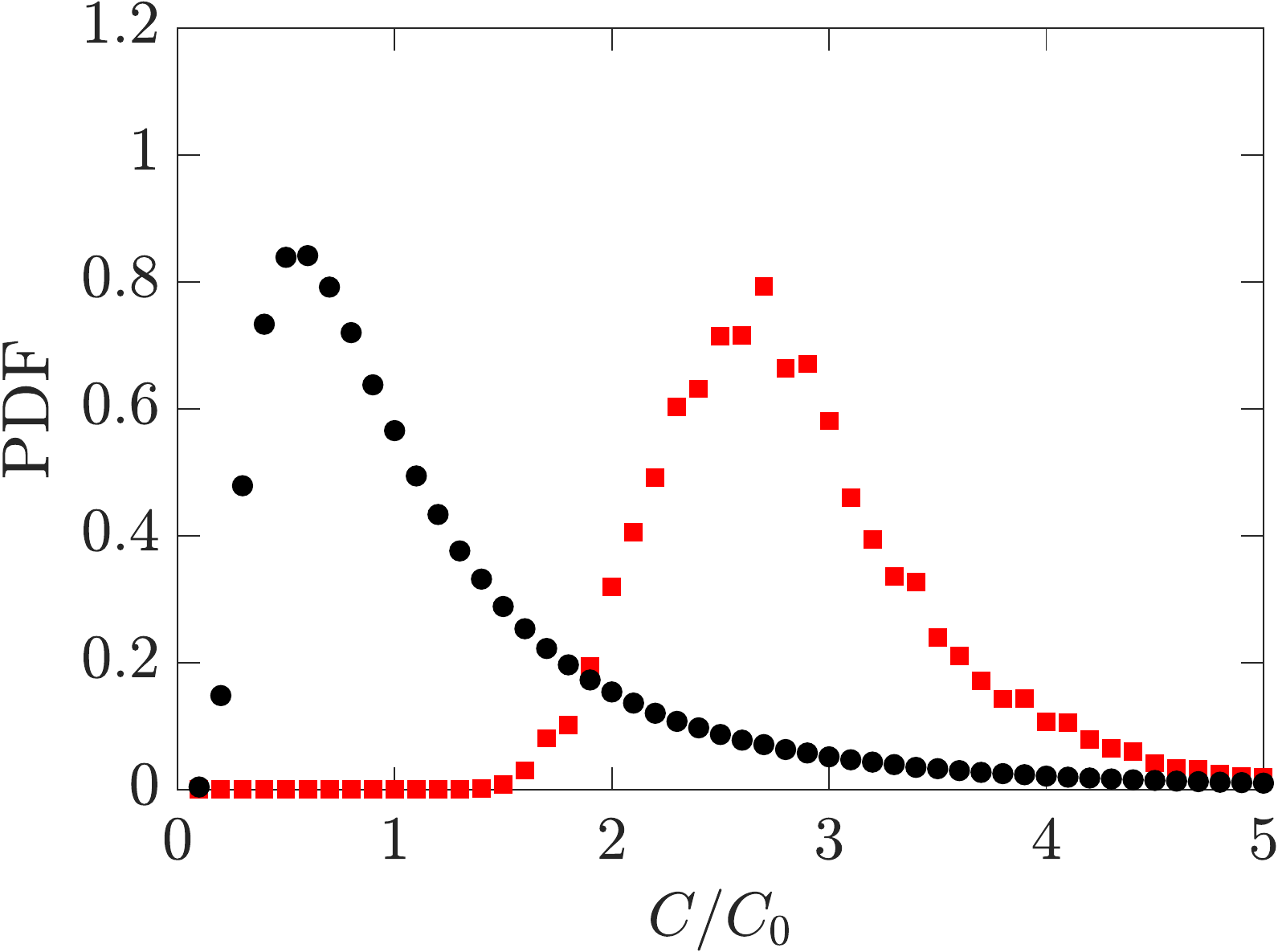}}
  \caption{}
 \end{subfigure}
 \begin{subfigure}{.54\textwidth}
   \centerline{\includegraphics[scale=0.41]{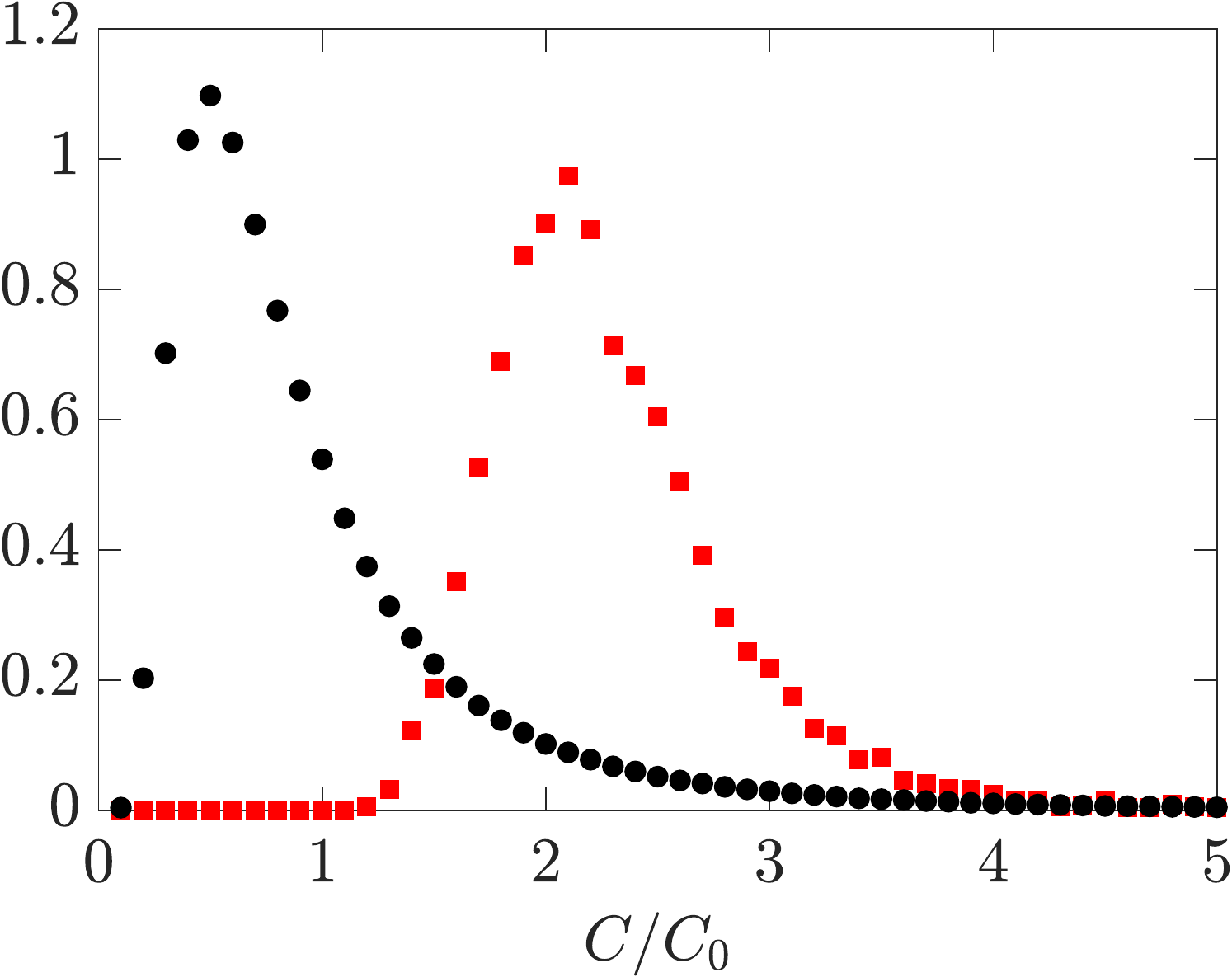}}
  \caption{}
 \end{subfigure}
  \caption{PDFs of normalized in-cluster (shown in red squares) and global concentrations (in black circles) for (a) LoSt-HiVF and (b) HiSt-HiVF.}
\label{fig:clust-conc-pdf}
\end{figure}

The local particle concentration within clusters can be significantly higher compared to the global value $C_0$ \citep{baker2017}.  Figure \ref{fig:clust-conc-pdf} shows PDFs of $C/C_0$ associated \textcolor{black}{with} the particles within coherent clusters, compared to the unconditional distribution: the \textcolor{black}{peaks in the PDF} for the clustered particles are about four times higher in HiSt and almost five times higher in LoSt. Besides reaffirming that the latter case displays more intense clustering, these plots indicate how the in-cluster concentration can be substantial, such that two-way-coupling (and possibly four-way-coupling) effects may be at play. 

\begin{figure}
\centering
\begin{subfigure}{.45\textwidth}
  \centerline{\includegraphics[scale=0.5]{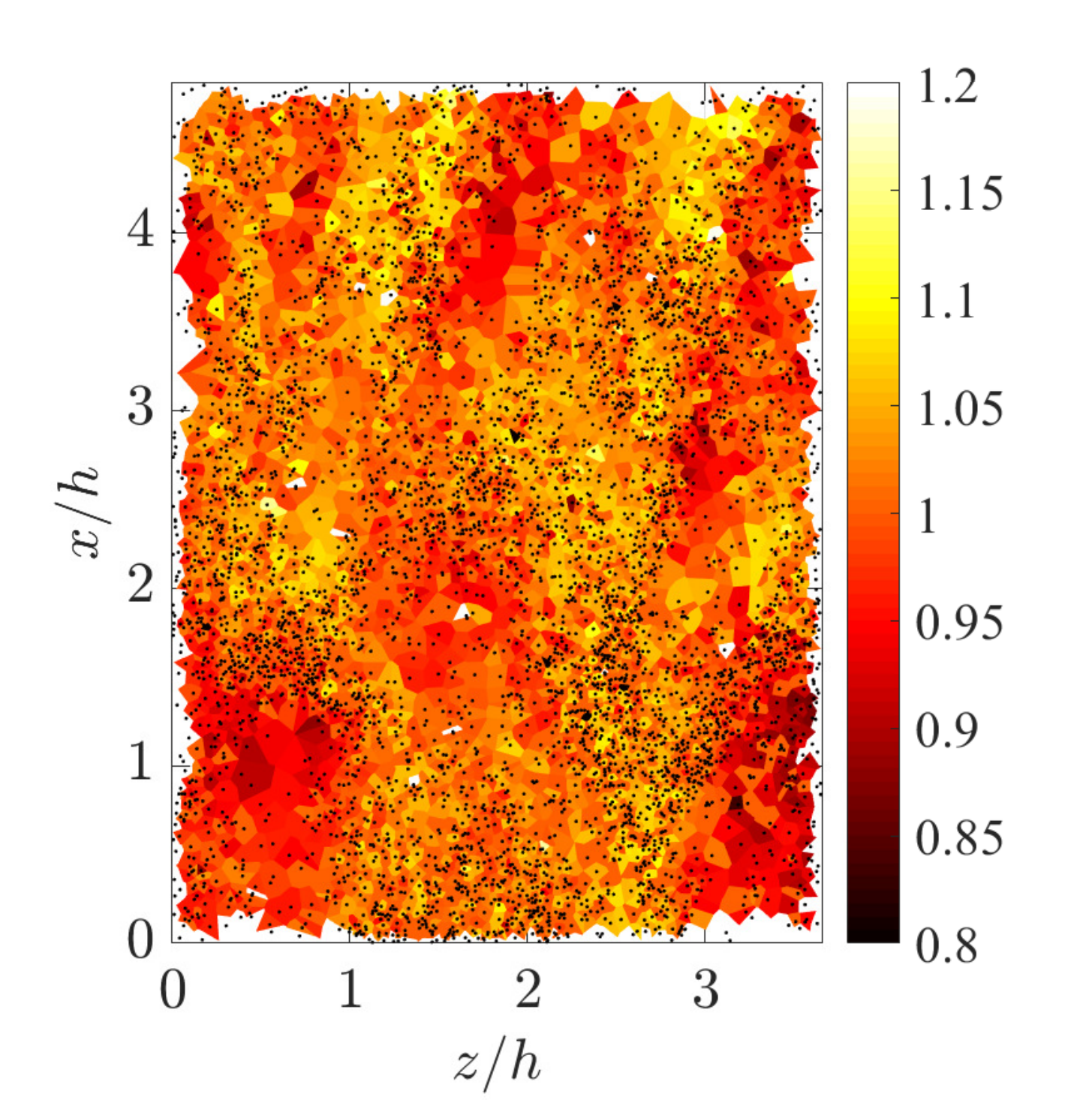}}
  \caption{}
 \end{subfigure}
 \begin{subfigure}{.54\textwidth}
   \centerline{\includegraphics[scale=0.4]{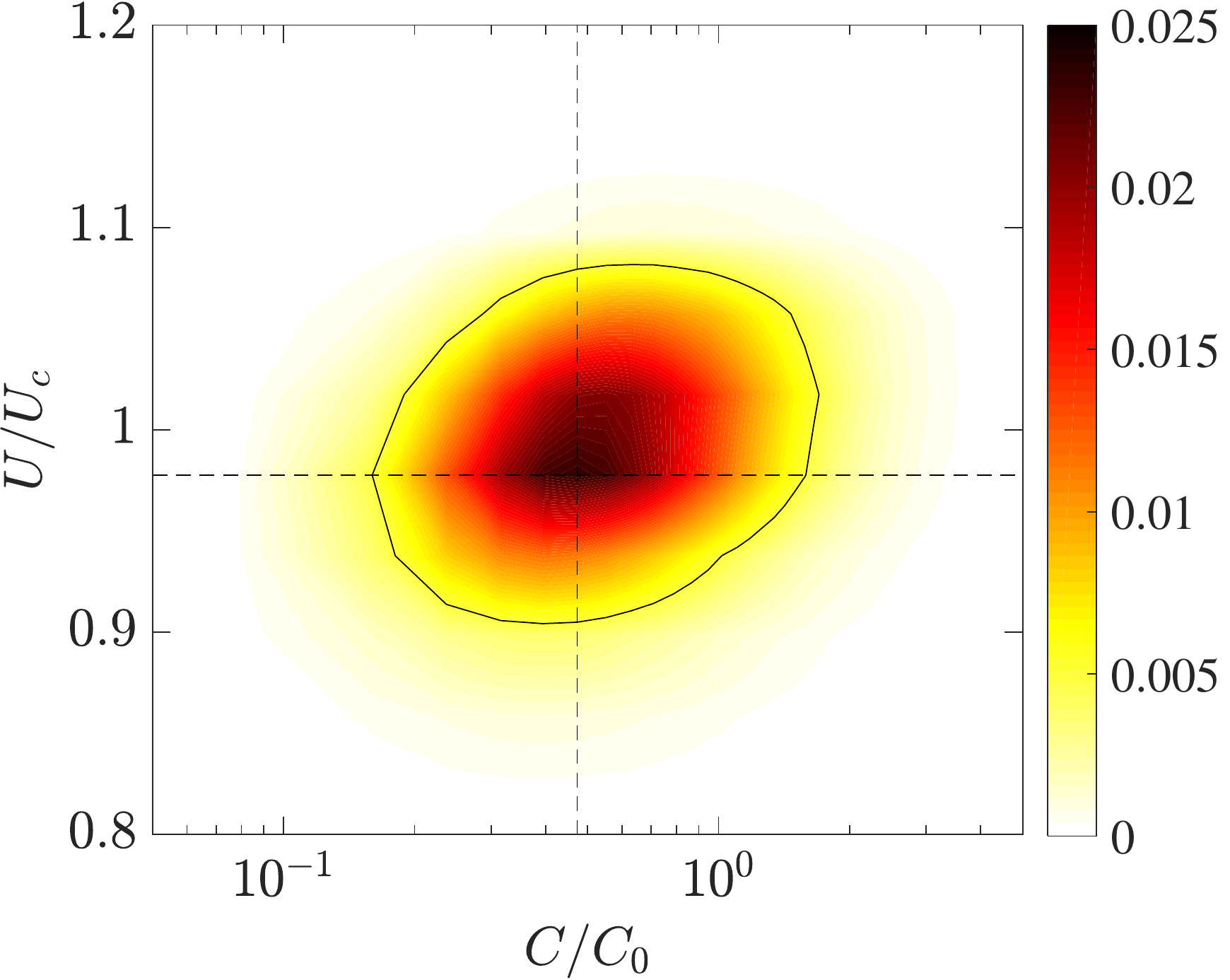}}
  \caption{}
 \end{subfigure}
  \caption{(a) The same instantaneous realization as in figure \ref{fig:voronoi-ctr}, with the Voronoi cell color-coded by the local particle streamwise velocity. (b) Joint PDF of streamwise velocity and concentration at the center-plane for the LoSt-HiVF case. The dashed lines indicate the velocity and concentration values averaged along the center-plane. The black contour line indicates the PDF level at 0.005.}
\label{fig:jpdf-ctr}
\end{figure}

\begin{figure}
\centering
\begin{subfigure}{.45\textwidth}
  \centerline{\includegraphics[scale=0.41]{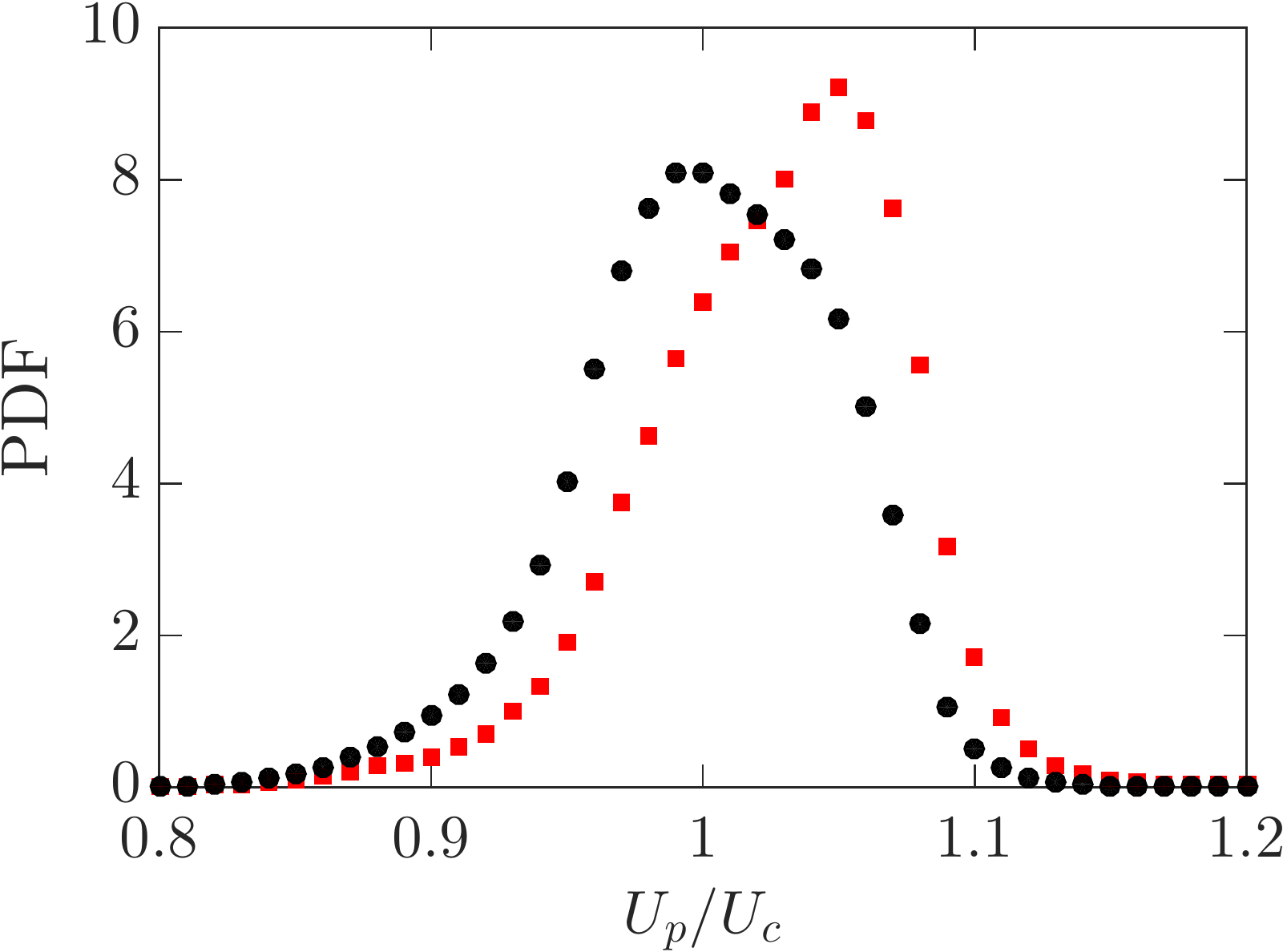}}
  \caption{}
 \end{subfigure}
 \begin{subfigure}{.54\textwidth}
   \centerline{\includegraphics[scale=0.41]{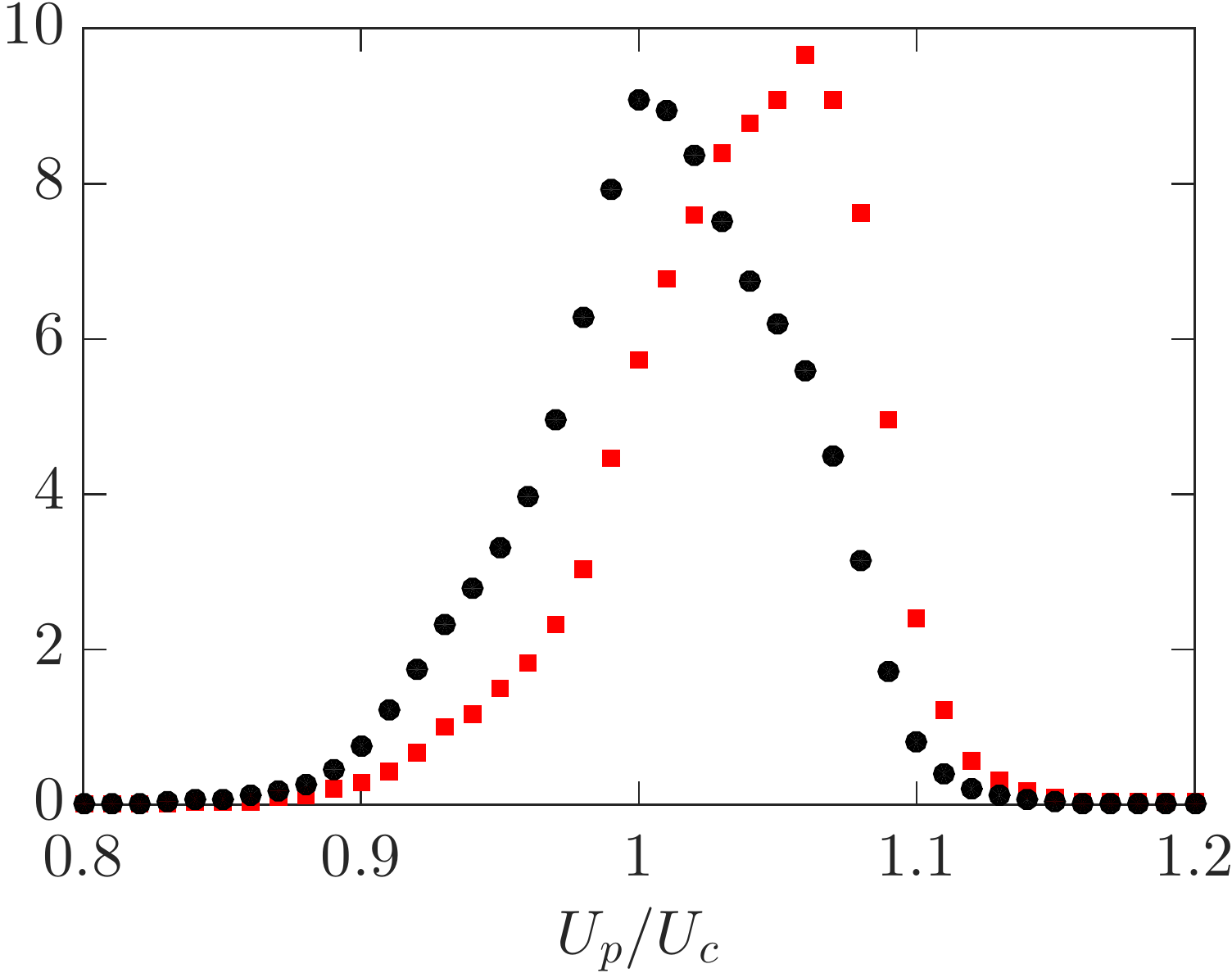}}
  \caption{}
 \end{subfigure}
  \caption{PDFs of normalized in-cluster (shown in red squares) and global velocities (in black circles) for (a) LoSt-HiVF and (b) HiSt-HiVF in the channel center plane.}
\label{fig:clust-vel-pdf}
\end{figure}

The traveling velocity of the clustered particles is also important to the transport process. Figure \ref{fig:jpdf-ctr}a depicts the same instantaneous realization as in figure \ref{fig:voronoi-ctr}, with the Voronoi cells now color-coded by the streamwise velocity of the respective particles. The more concentrated regions appear associated with higher velocities, as confirmed by the joint PDF for LoSt-HiVF (figure \ref{fig:jpdf-ctr}b): the local particle concentration and streamwise velocity are positively correlated. For a more quantitative account, shown in figure \ref{fig:clust-vel-pdf} are the PDF of the streamwise velocity for particles belonging to coherent clusters along the center-plane, as well as for all particles in the field of view. The clustered particles travel downward measurably faster than the generic particles. The explanations may be two-fold: on one hand, particles may be favoring the downwash side of turbulent eddies, according to the picture of preferential sweeping originally proposed by \citet{maxey1987} and later demonstrated by the simulations of \citet{wang1993} and recently by the experiments of \citet{petersen2019}; on the other hand, the highly concentrated clusters may be exerting a collective drag force on the fluid, in turn enhancing their vertical velocity as shown in the numerical simulations of \citet{bosse2006} and \citet{frankel2016}. 

\begin{figure}
\centering
\begin{subfigure}{.45\textwidth}
  \centerline{\includegraphics[scale=0.35]{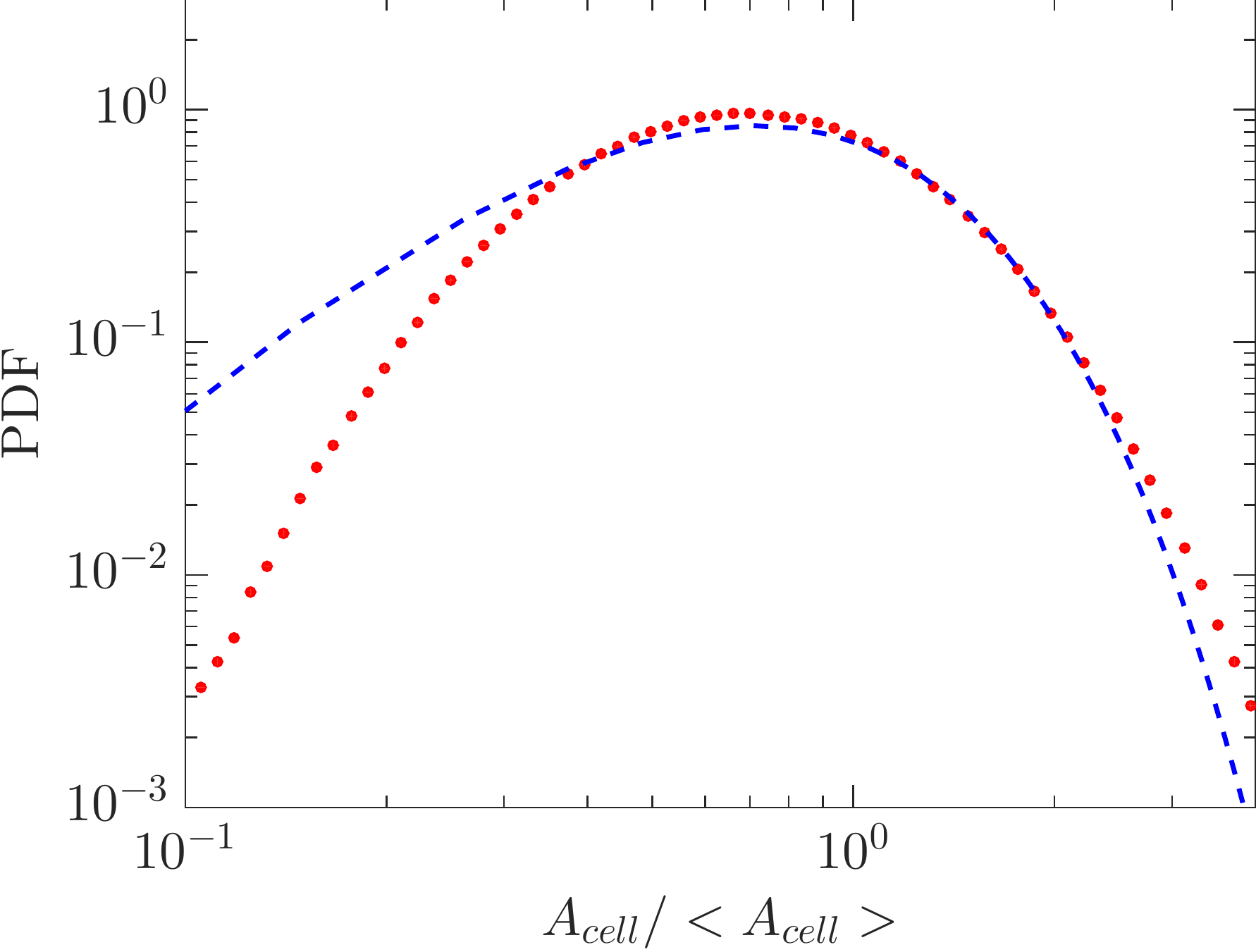}}
  \caption{}
 \end{subfigure}
 \begin{subfigure}{.54\textwidth}
   \centerline{\includegraphics[scale=0.41]{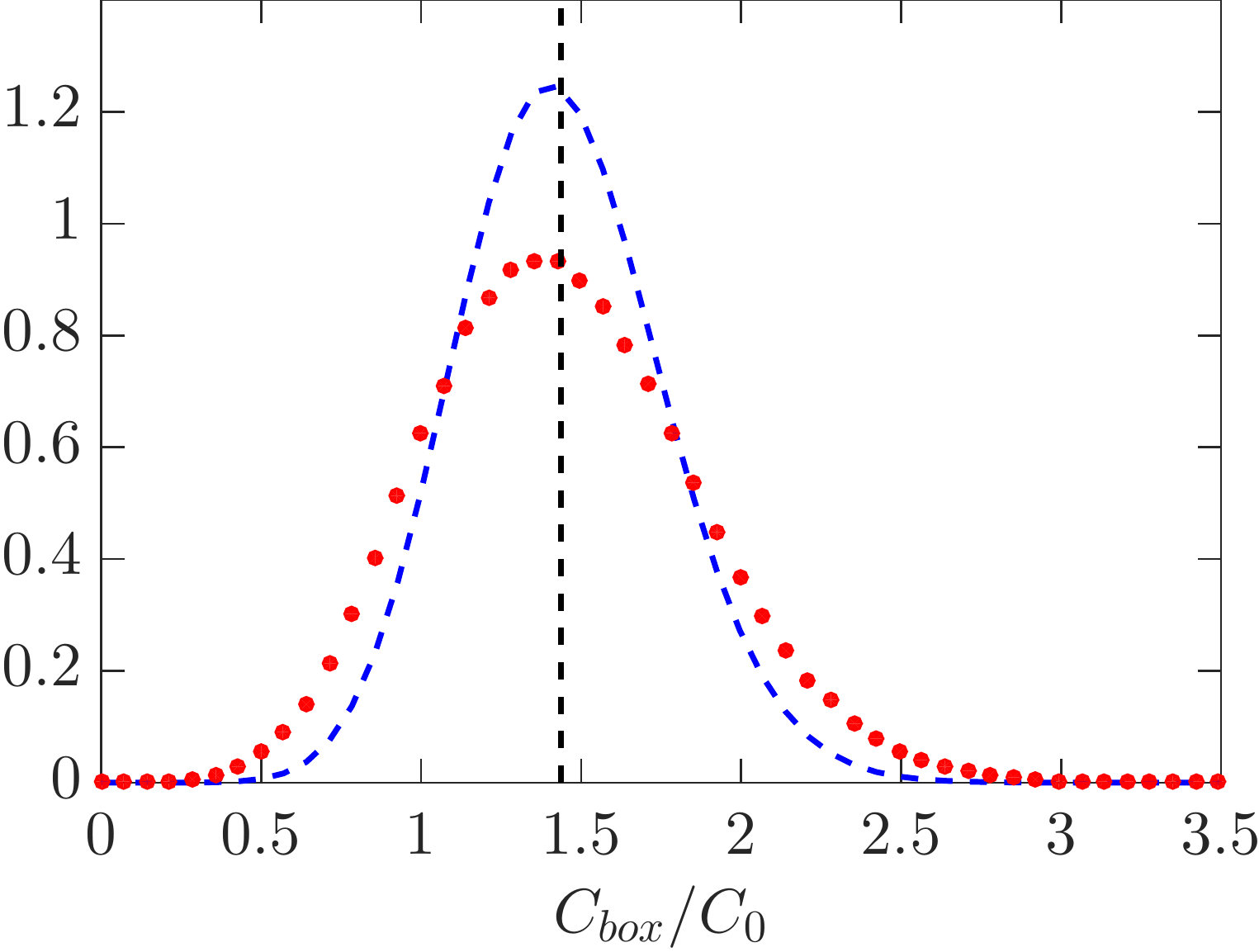}}
  \caption{}
 \end{subfigure}
  \caption{(a) PDF of the Voronoi cell areas $A_{cell}$ along the near-wall plane (red cicles), compared with a $\Gamma$ distribution (blue dashed line). (b) PDF of particle concentration in square boxes of size 0.18$h$ used to tessellate the \textcolor{black}{near-wall plane} (red circles) compared to a Poisson distribution (blue dashed line). }
\label{fig:box-nearwall}
\end{figure}

In principle, the Voronoi method may also be used to identify the highly concentrated structures near the wall. However, figure \ref{fig:box-nearwall}a shows how the PDF of the Voronoi cell areas measured along the near-wall plane is in fact narrower than for randomly distributed particles. This may be an artifact due to the relatively high concentration near the wall: particles very close to each other might be identified as one, reducing the probability of detecting small cell areas. Alternatively, the actual topology of the particle field, expected to be organized in streaks, could result in a “crystallized” pattern with a relatively regular arrangement of cells (and thus a narrow PDF of their area). Either way, the Voronoi tessellation method (in its standard form) does not appear as a suitable tool to study clustering in the present near-wall particle fields. We therefore resort to the box-counting approach: we tessellate the domain with square boxes of size 60 wall units or $\sim 0.18h$, and in figure \ref{fig:box-nearwall}b we plot the PDF of the concentration in each box $C_{box}$, comparing it with the Poisson distribution expected for randomly located particles. The relatively broad distribution indicates that the particles are indeed clustered over scales of the order of the box size. The choice of the latter is informed by the width of the particle streaks as estimated from the RDF analysis in the previous subsection, although it is verified that varying it by a factor two yields similar conclusions. The box index $BI = (\sigma\,–\,\sigma_{Poisson})/\mu$ (where $\mu$ and $\sigma$ indicate mean and standard deviation of the distribution, respectively, and $\sigma_{Poisson}$ is the standard deviation of the Poisson distribution) is a comparative measure of clustering. Near the wall we find $BI = 0.05$, while $BI = 0.2$ at the center-plane. The latter (consistent with the results of \citealt{fessler1994}) confirms the indication from the RDF analysis that clustering in the channel core is significantly more intense.

\begin{figure}
\centering
\begin{subfigure}{.45\textwidth}
  \centerline{\includegraphics[scale=0.45]{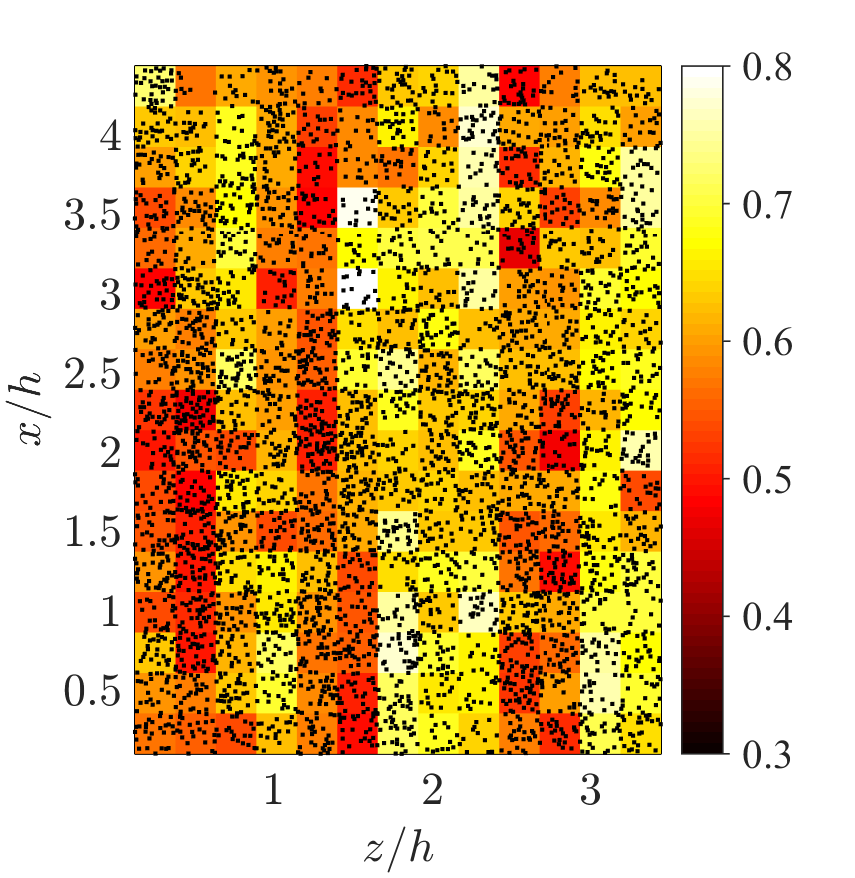}}
  \caption{}
 \end{subfigure}
 \begin{subfigure}{.54\textwidth}
   \centerline{\includegraphics[scale=0.4]{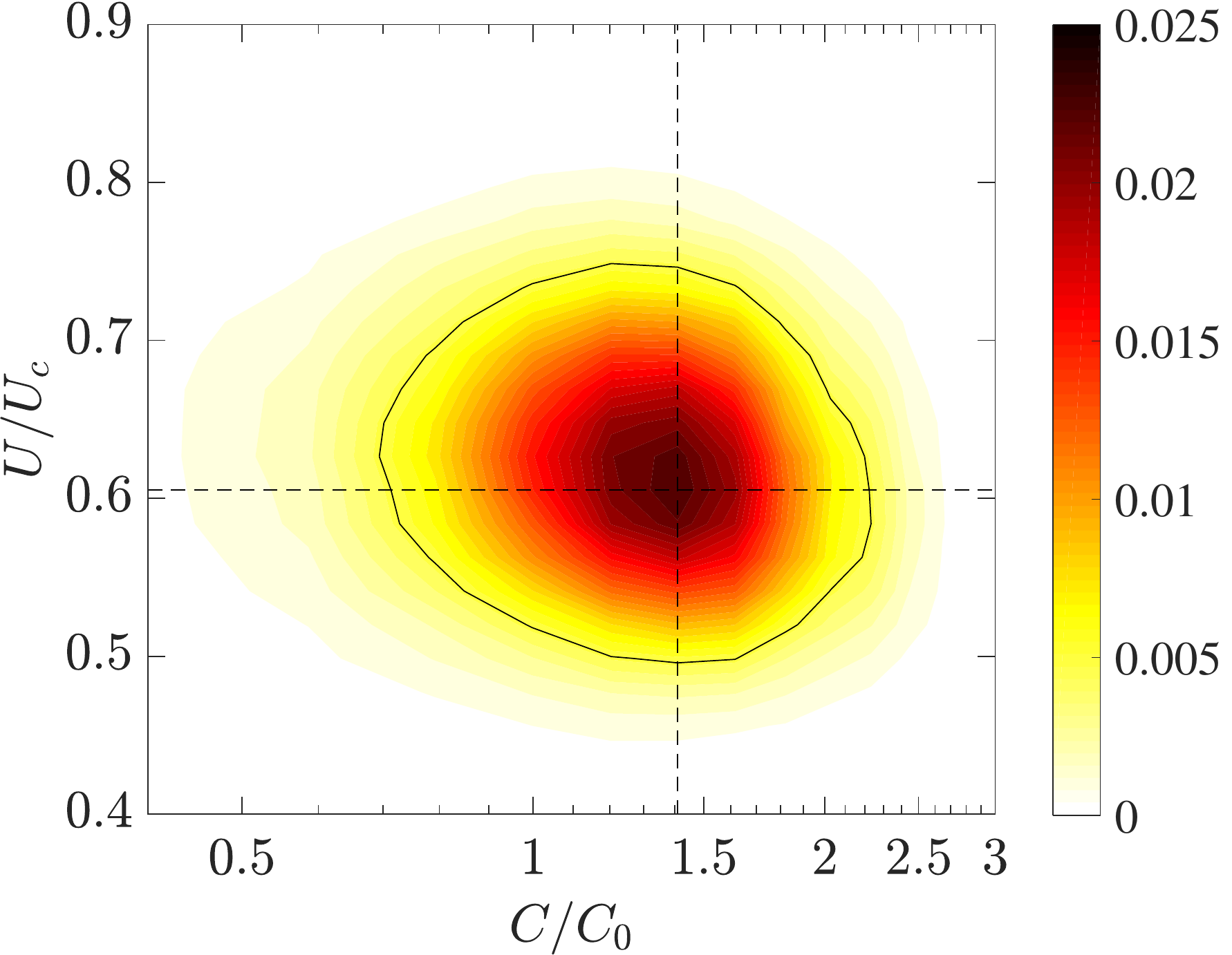}}
  \caption{}
 \end{subfigure}
  \caption{(a) Sample realization at the near-wall plane, tessellated by 0.18$h\,\times\,$0.18$h$ boxes and color-coded by the mean local streamwise particle velocity within the boxes. (b) Joint PDF of streamwise velocity and box-based concentration at the near-wall plane for the LoSt-HiVF case. The dashed lines indicate the velocity and concentration averaged along the near-wall plane. The black contour line indicates the PDF level at 0.005.}
\label{fig:jpdf-nearwall}
\end{figure}

Figure \ref{fig:jpdf-nearwall}a shows an instantaneous realization of the near-wall particle field, color-coded with the mean particle velocity in each square box of the tessellation. The visual impression of elongated slow-velocity streaks is in line with the RDF results. In figure \ref{fig:jpdf-nearwall}b we plot a joint PDF of local particle concentration and streamwise velocity, based on values averaged in each box. The apparent negative correlation contrasts with the center-plane trend in figure \ref{fig:jpdf-ctr}b, indicating a tendency of the highly concentrated particles to travel slower than the average. Combined with the RDFs, ADFs and two-point correlations reported above, this confirms the picture of particles accumulating in slow-moving streaks. 

\section{Conclusions}
We have reported on a series of experiments conducted on a vertical turbulent channel flow at $Re_{\tau} = 235$ and 335, in which particle-laden air flows downward. Several aspects of the configuration are chosen to provide a canonical case: smooth walls, streamwise development, relatively large aspect ratio of the cross-section, and small size-selected particles with $St_{\eta}$ = $\mathcal{O}$(10) and $St^+$ = $\mathcal{O}$(100). Care is taken to keep the experimental parameters under strict control, including the smoothness of the walls and the particle mass loading. The latter is varied across what is usually considered the boundary between one-way and two-way coupling. Using laser imaging we have investigated in detail the particle spatial distribution and velocity, gaining new insight expected to be useful to reach a predictive understanding of particle-laden wall bounded flows. Here we summarize and discuss the main findings.

At volume fractions $\mathcal{O}$($10^{-6}$), the particles show a distinct, although relatively mild, tendency to segregate near the wall, with the concentration displaying a second maximum at the channel centerline. The results are consistent with turbophoresis acting down the gradients of turbulence intensity, and in particular of $V_{rms}$ as proposed by \citet{young1997}. At volume fractions $\mathcal{O}$($10^{-5}$), the near-wall peak is much more pronounced and the centerline maximum is absent, indicating more vigorous turbophoretic drift towards the wall. The increase of near-wall segregation with mass loading is opposite to what reported in previous two-way-coupled point-particle simulations. 

The mean velocity profiles show particles traveling faster than the unladen fluid in the immediate vicinity of the wall, resulting in an effective slip velocity. Away from the wall, the more dilute case has particles following a profile similar to the unladen air velocity. In the denser case, the particles are measurably slower up to $y/h \sim$ 0.4 (0.6) for the lower (higher) Stokes number. Both streamwise and wall-normal velocity fluctuations of the particles exceed those of the unladen fluid near the wall; in the denser case the effect is much more significant and extended to larger wall distances. The wall-normal fluctuations do not vanish close to the wall, and lead to estimates of the deposition velocity in line with the expectations in the dilute case, but several times larger in the dense case. The particle Reynolds shear stress follows a similar behavior as the normal stresses: it equals the unladen fluid stress in the channel core, but exceeds it when approaching the wall to a degree that depends on the loading. A quadrant analysis reveals that the prevalence of Q4 events (the equivalent of “sweeps” for the fluid motion) is enhanced in the near-wall region compared to the unladen fluid, suggesting that fluid sweeps are key in the particle segregation process. The effects above are similarly displayed by all cases investigated here, but are more evident for the higher St.

For the denser cases, the spatial distributions of the particle positions and velocities are analyzed over wall-parallel planes using both two-point statistics and tessellation techniques, providing further details on the particle organization and dynamics. In the channel core, the particles show a strong propensity to cluster, forming somewhat elongated objects preferentially aligned in the vertical streamwise direction. Clustering is more intense for the cases closer to the condition $St_{\eta} \sim$ 1, although at higher $St$ the clusters tend to be larger \citep{petersen2019}. Groups of particles above a certain size range (“coherent clusters”, \citealt{baker2017}) reach concentrations several times higher than the global mean, and tend to travel/fall faster than the non-clustered particles. This suggests that, although the flow in the channel core is a poor approximation of homogeneous isotropic turbulence, the classic phenomena of preferential concentration (\citealt{squires1991}; or other more recent mechanisms that explain clustering, see \citealt{goto2008}; \citealt{bragg2014}) and preferential sweeping \citep{wang1993} are at play. In the near-wall region the particles are observed to form elongated streaks, several channel height in length and spaced by $\mathcal{O}$(100) wall units (although the limited range of Reynolds numbers does not allow us to determine a conclusive scaling). Those streaks tend to move slower than the generic particles, supporting the view that they are coupled to fluid low-speed streaks typical of wall turbulence. The particle velocity contains a significant component of random uncorrelated motion. In agreement with the mesoscopic Eulerian formalism introduced by \citet{fevrier2005}, this Brownian-like motion is more prominent for higher $St$ and in the near-wall region, where the particle response time is much larger than the turbulent time scales. In general, we note that the differences between the particle behavior near and far from the wall, already remarkable in the present regimes, are expected to be magnified at higher $Re_{\tau}$.

Taken together, these results are consistent with a scenario in which the increase in volume fraction from $\mathcal{O}$($10^{-6}$) to $\mathcal{O}$($10^{-5}$) triggers two-way and (locally) four-way coupling effects. In particular, the particle backreaction may alter the turbulence structure in ways that enhance the turbophoretic drift towards the wall (e.g., by enhancing the peak of turbulence intensity and/or displacing it away from the wall). In turn, the higher near-wall concentration may promote inter-particle and wall-particle collisions. These would \textcolor{black}{damp} the particle kinetic energy, causing them to travel more slowly but enhancing their velocity fluctuations, as observed. Moreover, the inelastic collisions may prevent the particles from escaping the near-wall region \citep{hrenya1997}. In the absence of simultaneous fluid measurements, this and other possible scenarios remain speculative. Future studies shall fill this gap, for example using phase separation techniques (\citealt{kiger2000}; \citealt{khalitov2002}; \citealt{capone2015}; \citealt{petersen2019}) to accurately measure fluid statistics in the near-wall region. This, however is expected to be a challenging task: small inertial particles accumulating near the wall pose a major problem to imaging surrounding tracers. \textcolor{black}{This can be partly alleviated by augmenting the spatial resolution, which is becoming possible thanks to the steady increase of the sensor size of available cameras \citep{discetti2018}}. 

Despite the relevance for practical applications and the understanding of particle-laden turbulence, previous experiments focused on the present regime have been scarce. The lack is unfortunate, especially given the need to validate point-particle models and the exorbitant cost of particle-resolved simulations. Contrasting our observations with previous numerical studies suggests that, while point-particle simulations capture many key features of the particle transport (e.g., \citealt{soldati2009}), the underlying hypotheses may be missing or misestimating some important aspects, especially concerning the two-way and four-way coupling. This is exemplified by the increased near-wall concentration measured for increasing mass loading, which is opposite to the trend found in past simulations. The limitations of point-particle methods in capturing the two-way coupling are well known (\citealt{eaton2009}; \citealt{balachandar2010}) and have been assessed in detail by recent studies on homogeneous turbulence (\citealt{mehrabadi2018}; \citealt{petersen2019}). Wall-bounded turbulence may pose even harder problems due to the spatially varying resolution. However, as mentioned in the Introduction, recent approaches are showing promising improvements. For example, \citet{capecelatro2015} used a volume-filtering method to simulate the configuration of \citet{benson2005} and found satisfactory agreement. That case contained \textcolor{black}{heavier} particles than the present one, with a Stokes number likely too high to produce strong near-wall segregation and clustering. It would be interesting to evaluate this and other modeling strategies against the present data, also considering that the dilute and dense cases we investigated should allow isolating one-way and two-way coupling issues. Importantly, given the significant role played by the dense regions in the momentum coupling, future comparisons with simulations should ideally include information on the spatial correlation of the particle field and the clustering properties. The moderate Reynolds number in our study may allow for a comparison with future particle-resolved simulations, which are becoming feasible even for relatively small particles \citep{schneiders2017}.

Finally, we remark how the effect of inter-particle and wall-particle collisions could also contribute to experimental-numerical discrepancies. As mentioned in §2.1, the restitution coefficient may vary significantly depending on the collision velocity (\citealt{joseph2001}; \citealt{gondret2002}). Following \citet{gondret2002}, we can define a wall-collision Stokes number as $St_p = (1/9)Re_p\rho_p/\rho_f$ (where we used the fact that $V_t = \tau_pg$ is of the order of the particle $V_{rms}$ close to the wall, the latter giving a scale for the wall-normal collision velocity). One finds $St_p$ = $\mathcal{O}$(100), which is in the range for which $e$ is a strong function of $St_p$ \citep{gondret2002}. Given the possibly large variance of the collision velocity and the role of inelastic collisions to enhance near-wall accumulation \citep{hrenya1997}, the common assumption of a constant restitution coefficient might be inadequate.

\begin{acknowledgments}
\section*{Acknowledgements}
This project is funded partly by the U.S. Army Research Office (Division of Fluid Dynamics, grants W911NF-17-1-0366 and W911NF-18-1-0354) and partly by the Environment and Natural Resources Trust Fund of Minnesota.
\end{acknowledgments}

\begin{appendix}
\section*{Appendix A. Electrostatic dissipative acrylic}

\begin{figure}
\begin{subfigure}{.44\textwidth}
  \centerline{\includegraphics[scale=0.4]{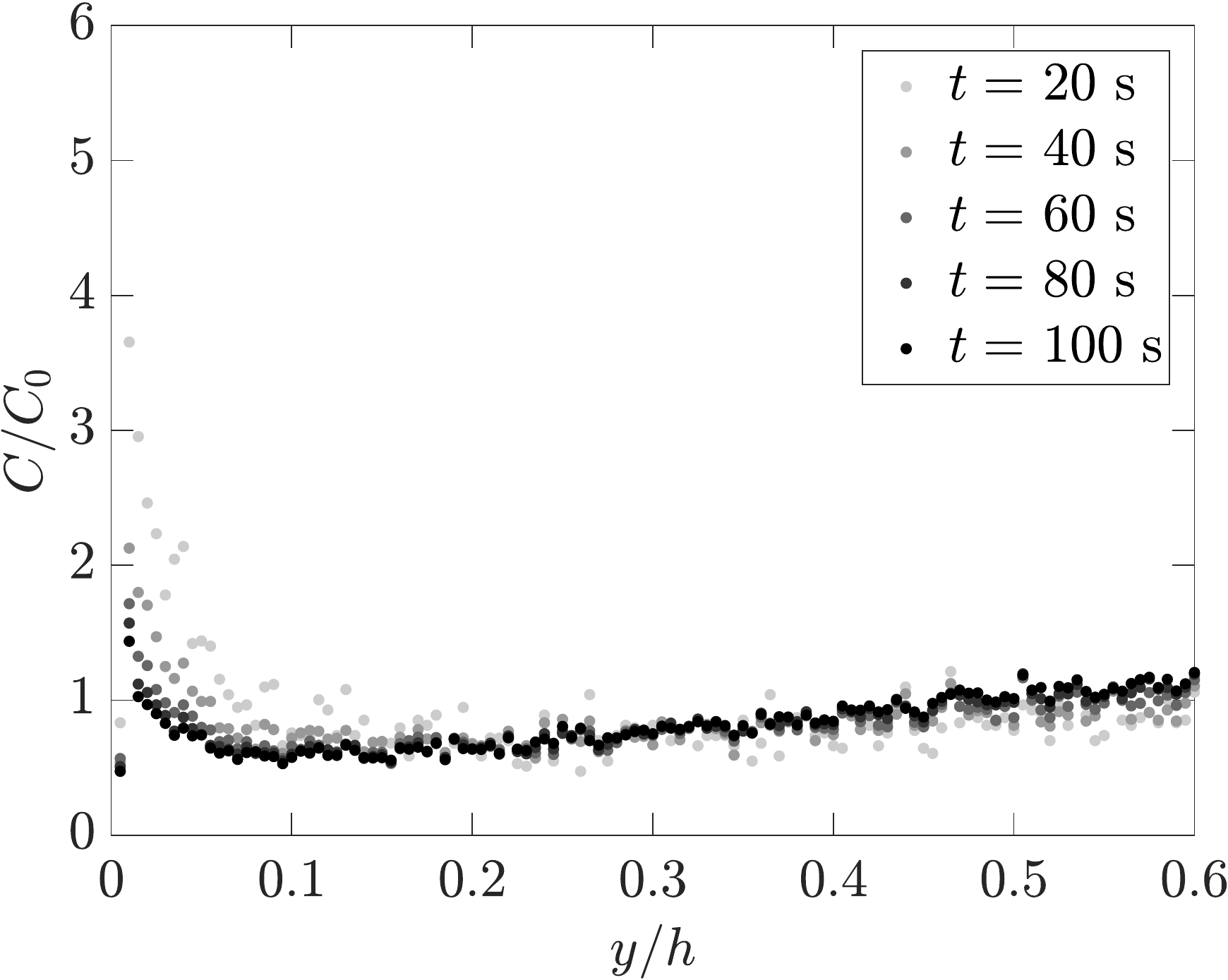}}
  \caption{}
\end{subfigure}
\begin{subfigure}{.6\textwidth}
  \centerline{\includegraphics[scale=0.4]{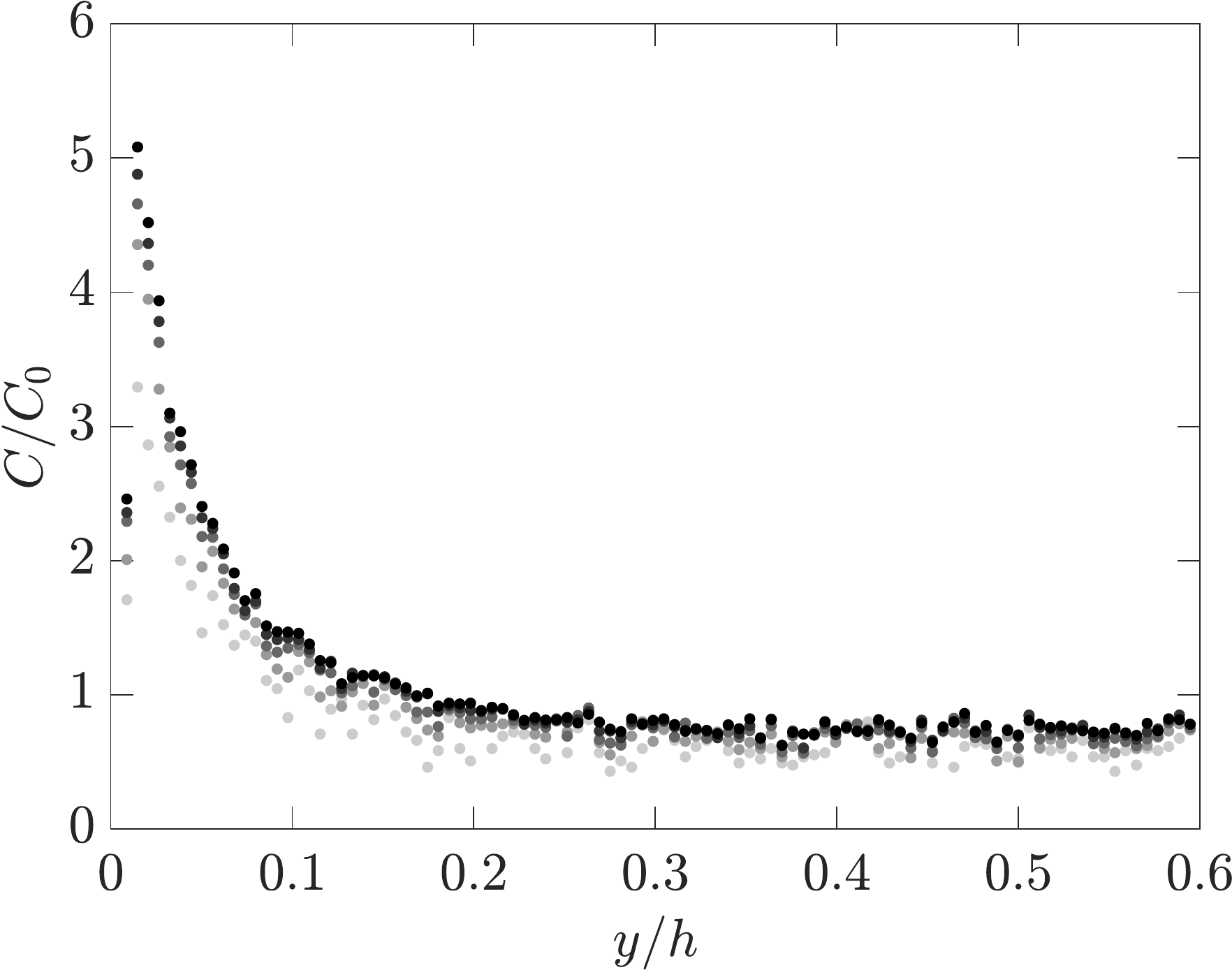}}
  \caption{}
\end{subfigure}
  \caption{Wall-normal profiles of mean concentration for recordings of different durations, using (a) standard acrylic and (b) electrostatic dissipative acrylic. $C_0$ = 880 particles/cm\textsuperscript{3}, corresponding to $\phi_v = 5\times10^{-5} $ for both cases. Refer to table \ref{tab:params} for definitions on $y,h$.}
\label{fig:roughness}
\end{figure}

Electrostatic dissipative acrylic (SciCron Technologies) is used to build the optical test section, and the channel is provided with static discharge wires grounded to structural supports. This prevents the particles from accumulating upon impaction and building up unwanted roughness, an effect that has impacted past experiments \citep{benson2005}. The importance of such precautions is illustrated in figure \ref{fig:roughness}, where cumulative wall-normal profiles of particle concentrations are plotted in the higher volume fraction case. The details of how those measurements are obtained are given in §2.2. When using standard acrylic walls, the concentration profiles start with a strong near-wall peak but drift in time, with particles migrating away from the wall. This is likely the consequence of wall deposition which leads to significant roughness (as verified \textcolor{black}{by} inspecting the wall surface at the end of the experiments), in turn altering the collision dynamics and ultimately enhancing particle dispersion towards the channel core. Using electrostatic dissipative acrylic virtually eliminates particle deposition (as also verified by direct wall inspection) and warrants consistent concentration profiles during each run and between different runs.

\section*{Appendix B. Coefficient of restitution}

\begin{figure}
  \centerline{\includegraphics[width=0.65\textwidth]{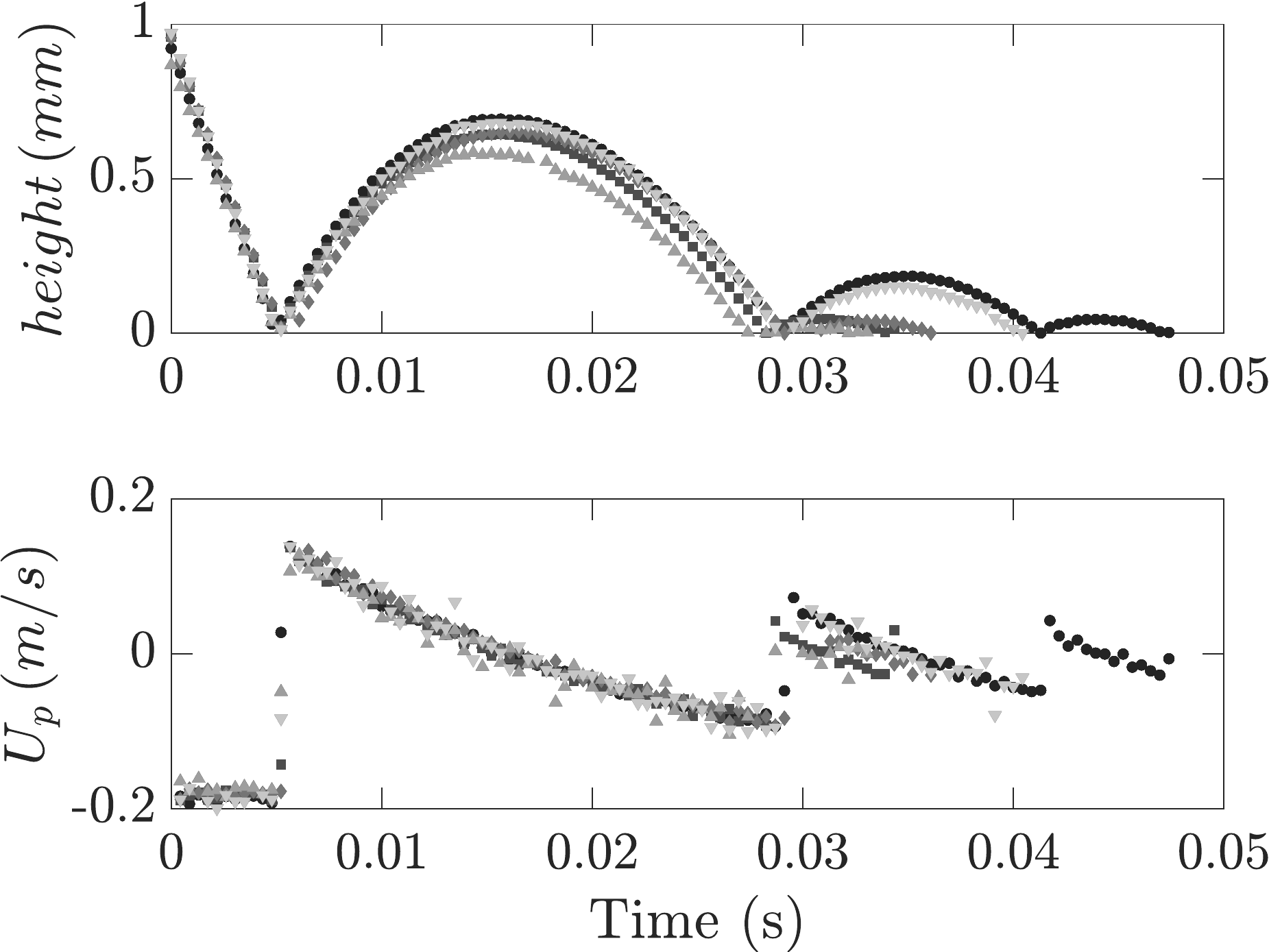}}
  \caption{Particle height and vertical velocity data used for determining the restitution coefficient \textcolor{black}{for particle-wall collisions} over five trials. Positive velocity indicates upward motion.Tracks for each trial are shown in markers of different shapes and colors.}
\label{fig:coeftest}
\end{figure}

The coefficient of restitution, $e$ \textcolor{black}{for particle-wall collisions} is measured by dropping individual glass beads from a height of 350 mm above a horizontal plate made of the same acrylic used for the test section. The particles reach steady-state terminal velocity before bouncing on the plate. This is achieved independently from the method of release due to the short free-fall stopping distance (of order $\tau_p^2g \sim 3$ mm). Particles are imaged at 2300 fps with a high-speed CMOS camera (VEO 640) paired with a 200 mm lens at f/4, and tracked using the same method used for the particle-laden flow measurements. The coefficient of restitution $e$ is defined as the ratio of the wall-normal particle velocity just after and just before the collision, and is calculated averaging over five trials as shown in figure \ref{fig:coeftest}. 

\end{appendix}
\bibliographystyle{jfm}
\bibliography{keeonn}

\begin{thebibliography}{122}
\expandafter\ifx\csname natexlab\endcsname\relax\def\natexlab#1{#1}\fi
\def\au#1{#1} \def\ed#1{#1} \def\yr#1{#1}\def\at#1{#1}\def\jt#1{\textit{#1}}
  \def\bt#1{#1}\def\bvol#1{\textbf{#1}} \def\vol#1{#1} \def\pg#1{#1}
  \def\publ#1{#1}\def\arxiv#1{#1}\def\org#1{#1}\def\st#1{\textit{#1}}

\bibitem[Aliseda {\em et~al.\/}(2002)Aliseda, Cartellier, Hainaux \&
  Lasheras]{aliseda2002}
{\sc \au{Aliseda, Alberto}, \au{Cartellier, Alain}, \au{Hainaux, F} \&
  \au{Lasheras, Juan~C}} \yr{2002}  \at{Effect of preferential concentration on
  the settling velocity of heavy particles in homogeneous isotropic
  turbulence}.  \jt{Journal of Fluid Mechanics}  \bvol{468},  \pg{77--105}.

\bibitem[Baek \& Lee(1996)]{baek1996}
{\sc \au{Baek, SJ} \& \au{Lee, SJ}} \yr{1996}  \at{A new two-frame particle
  tracking algorithm using match probability}.  \jt{Experiments in Fluids}
  \bvol{22}~(1),  \pg{23--32}.

\bibitem[Baker {\em et~al.\/}(2017)Baker, Frankel, Mani \& Coletti]{baker2017}
{\sc \au{Baker, Lucia}, \au{Frankel, Ari}, \au{Mani, Ali} \& \au{Coletti,
  Filippo}} \yr{2017}  \at{Coherent clusters of inertial particles in
  homogeneous turbulence}.  \jt{Journal of Fluid Mechanics}  \bvol{833},
  \pg{364--398}.

\bibitem[Balachandar \& Eaton(2010)]{balachandar2010}
{\sc \au{Balachandar, S.} \& \au{Eaton, J.~K.}} \yr{2010}  \at{Turbulent
  dispersed multiphase flow}.  \jt{Annu. Rev. Fluid Mech.}  \bvol{42},
  \pg{111--133}.

\bibitem[Balachandar {\em et~al.\/}(2019)Balachandar, Liu \&
  Lakhote]{balachandar2019}
{\sc \au{Balachandar, S}, \au{Liu, Kai} \& \au{Lakhote, Mandar}} \yr{2019}
  \at{Self-induced velocity correction for improved drag estimation in
  euler--lagrange point-particle simulations}.  \jt{Journal of Computational
  Physics}  \bvol{376},  \pg{160--185}.

\bibitem[Bendat \& Piersol(2011)]{bendat2011}
{\sc \au{Bendat, Julius~S} \& \au{Piersol, Allan~G}} \yr{2011} {\em Random
  data: analysis and measurement procedures\/}, ,  \vol{vol. 729}.  \publ{John
  Wiley \& Sons}.

\bibitem[Benson {\em et~al.\/}(2005)Benson, Tanaka \& Eaton]{benson2005}
{\sc \au{Benson, Michael}, \au{Tanaka, Tomohiko} \& \au{Eaton, John~K}}
  \yr{2005}  \at{Effects of wall roughness on particle velocities in a
  turbulent channel flow}.  \jt{Journal of fluids engineering}  \bvol{127}~(2),
   \pg{250--256}.

\bibitem[Bernardini(2014)]{bernardini2014}
{\sc \au{Bernardini, Matteo}} \yr{2014}  \at{Reynolds number scaling of
  inertial particle statistics in turbulent channel flows}.  \jt{Journal of
  Fluid Mechanics}  \bvol{758}.

\bibitem[Bewley {\em et~al.\/}(2013)Bewley, Saw \& Bodenschatz]{bewley2013}
{\sc \au{Bewley, Gregory~P}, \au{Saw, Ewe-Wei} \& \au{Bodenschatz, Eberhard}}
  \yr{2013}  \at{Observation of the sling effect}.  \jt{New Journal of Physics}
   \bvol{15}~(8),  \pg{083051}.

\bibitem[Bosse {\em et~al.\/}(2006)Bosse, Kleiser \& Meiburg]{bosse2006}
{\sc \au{Bosse, Thorsten}, \au{Kleiser, Leonhard} \& \au{Meiburg, Eckart}}
  \yr{2006}  \at{Small particles in homogeneous turbulence: Settling velocity
  enhancement by two-way coupling}.  \jt{Physics of Fluids}  \bvol{18}~(2),
  \pg{027102}.

\bibitem[Bragg \& Collins(2014)]{bragg2014}
{\sc \au{Bragg, Andrew~D} \& \au{Collins, Lance~R}} \yr{2014}  \at{New insights
  from comparing statistical theories for inertial particles in turbulence: I.
  spatial distribution of particles}.  \jt{New Journal of Physics}
  \bvol{16}~(5),  \pg{055013}.

\bibitem[Capecelatro \& Desjardins(2013)]{capecelatro2013}
{\sc \au{Capecelatro, Jesse} \& \au{Desjardins, Olivier}} \yr{2013}  \at{An
  euler--lagrange strategy for simulating particle-laden flows}.  \jt{Journal
  of Computational Physics}  \bvol{238},  \pg{1--31}.

\bibitem[Capecelatro \& Desjardins(2015)]{capecelatro2015}
{\sc \au{Capecelatro, Jesse} \& \au{Desjardins, Olivier}} \yr{2015}  \at{Mass
  loading effects on turbulence modulation by particle clustering in dilute and
  moderately dilute channel flows}.  \jt{Journal of Fluids Engineering}
  \bvol{137}~(11),  \pg{111102}.

\bibitem[Capecelatro {\em et~al.\/}(2016)Capecelatro, Desjardins \&
  Fox]{capecelatro2016}
{\sc \au{Capecelatro, Jesse}, \au{Desjardins, Olivier} \& \au{Fox, Rodney~O}}
  \yr{2016}  \at{Strongly coupled fluid-particle flows in vertical channels. i.
  reynolds-averaged two-phase turbulence statistics}.  \jt{Physics of Fluids}
  \bvol{28}~(3),  \pg{033306}.

\bibitem[Capecelatro {\em et~al.\/}(2018)Capecelatro, Desjardins \&
  Fox]{capecelatro2018}
{\sc \au{Capecelatro, Jesse}, \au{Desjardins, Olivier} \& \au{Fox, Rodney~O}}
  \yr{2018}  \at{On the transition between turbulence regimes in particle-laden
  channel flows}.  \jt{Journal of Fluid Mechanics}  \bvol{845},  \pg{499--519}.

\bibitem[Capecelatro {\em et~al.\/}(2014)Capecelatro, Pepiot \&
  Desjardins]{capecelatro2014}
{\sc \au{Capecelatro, Jesse}, \au{Pepiot, Perrine} \& \au{Desjardins, Olivier}}
  \yr{2014}  \at{Numerical characterization and modeling of particle clustering
  in wall-bounded vertical risers}.  \jt{Chemical Engineering Journal}
  \bvol{245},  \pg{295--310}.

\bibitem[Capone {\em et~al.\/}(2015)Capone, Romano \& Soldati]{capone2015}
{\sc \au{Capone, Alessandro}, \au{Romano, Giovanni~Paolo} \& \au{Soldati,
  Alfredo}} \yr{2015}  \at{Experimental investigation on interactions among
  fluid and rod-like particles in a turbulent pipe jet by means of particle
  image velocimetry}.  \jt{Experiments in Fluids}  \bvol{56}~(1),  \pg{1}.

\bibitem[Caporaloni {\em et~al.\/}(1975)Caporaloni, Tampieri, Trombetti \&
  Vittori]{caporaloni1975}
{\sc \au{Caporaloni, M}, \au{Tampieri, F}, \au{Trombetti, F} \& \au{Vittori,
  O}} \yr{1975}  \at{Transfer of particles in nonisotropic air turbulence}.
  \jt{Journal of the atmospheric sciences}  \bvol{32}~(3),  \pg{565--568}.

\bibitem[Caraman {\em et~al.\/}(2003)Caraman, Bor{\'e}e \&
  Simonin]{caraman2003}
{\sc \au{Caraman, N}, \au{Bor{\'e}e, J} \& \au{Simonin, Olivier}} \yr{2003}
  \at{Effect of collisions on the dispersed phase fluctuation in a dilute tube
  flow: Experimental and theoretical analysis}.  \jt{Physics of Fluids}
  \bvol{15}~(12),  \pg{3602--3612}.

\bibitem[Clauser(1956)]{clauser1956}
{\sc \au{Clauser, Francis~H}} \yr{1956}  \at{The turbulent boundary layer}.
  \jt{Advances in applied mechanics}  \bvol{4},  \pg{1--51}.

\bibitem[Clift {\em et~al.\/}(2005)Clift, Grace \& Weber]{clift2005}
{\sc \au{Clift, Roland}, \au{Grace, John~R} \& \au{Weber, Martin~E}} \yr{2005}
  {\em Bubbles, drops, and particles\/}.  \publ{Courier Corporation}.

\bibitem[Discetti \& Coletti(2018)]{discetti2018}
{\sc \au{Discetti, Stefano} \& \au{Coletti, Filippo}} \yr{2018}  \at{Volumetric
  velocimetry for fluid flows}.  \jt{Measurement Science and Technology}
  \bvol{29}~(4),  \pg{042001}.

\bibitem[Dritselis \& Vlachos(2011)]{dritselis2011}
{\sc \au{Dritselis, Chris~D} \& \au{Vlachos, Nicholas~S}} \yr{2011}
  \at{Numerical investigation of momentum exchange between particles and
  coherent structures in low re turbulent channel flow}.  \jt{Physics of
  Fluids}  \bvol{23}~(2),  \pg{025103}.

\bibitem[Eaton(2009)]{eaton2009}
{\sc \au{Eaton, John~K}} \yr{2009}  \at{Two-way coupled turbulence simulations
  of gas-particle flows using point-particle tracking}.  \jt{International
  Journal of Multiphase Flow}  \bvol{35}~(9),  \pg{792--800}.

\bibitem[Eaton \& Fessler(1994)]{eaton1994}
{\sc \au{Eaton, John~K} \& \au{Fessler, JR}} \yr{1994}  \at{Preferential
  concentration of particles by turbulence}.  \jt{International Journal of
  Multiphase Flow}  \bvol{20},  \pg{169--209}.

\bibitem[Elghobashi(1994)]{elghobashi1994}
{\sc \au{Elghobashi, S.}} \yr{1994}  \at{On predicting particle-laden turbulent
  flows}.  \jt{Appl. Sci. Res.}  \bvol{52}~(4),  \pg{309--329}.

\bibitem[Ferenc \& N{\'e}da(2007)]{ferenc2007}
{\sc \au{Ferenc, J{\'a}rai-Szab{\'o}} \& \au{N{\'e}da, Zolt{\'a}n}} \yr{2007}
  \at{On the size distribution of poisson voronoi cells}.  \jt{Physica A:
  Statistical Mechanics and its Applications}  \bvol{385}~(2),  \pg{518--526}.

\bibitem[Fessler {\em et~al.\/}(1994)Fessler, Kulick \& Eaton]{fessler1994}
{\sc \au{Fessler, John~R}, \au{Kulick, Jonathan~D} \& \au{Eaton, John~K}}
  \yr{1994}  \at{Preferential concentration of heavy particles in a turbulent
  channel flow}.  \jt{Physics of Fluids}  \bvol{6}~(11),  \pg{3742--3749}.

\bibitem[Fevrier {\em et~al.\/}(2005)Fevrier, Simonin \& Squires]{fevrier2005}
{\sc \au{Fevrier, Pierre}, \au{Simonin, Olivier} \& \au{Squires, Kyle~D}}
  \yr{2005}  \at{Partitioning of particle velocities in gas--solid turbulent
  flows into a continuous field and a spatially uncorrelated random
  distribution: theoretical formalism and numerical study}.  \jt{Journal of
  Fluid Mechanics}  \bvol{533},  \pg{1--46}.

\bibitem[Fouxon {\em et~al.\/}(2018)Fouxon, Schmidt, Ditlevsen, van Reeuwijk \&
  Holzner]{fouxon2018}
{\sc \au{Fouxon, Itzhak}, \au{Schmidt, Lukas}, \au{Ditlevsen, Peter}, \au{van
  Reeuwijk, Maarten} \& \au{Holzner, Markus}} \yr{2018}  \at{Inhomogeneous
  growth of fluctuations of concentration of inertial particles in channel
  turbulence}.  \jt{Physical Review Fluids}  \bvol{3}~(6),  \pg{064301}.

\bibitem[Frankel {\em et~al.\/}(2016)Frankel, Pouransari, Coletti \&
  Mani]{frankel2016}
{\sc \au{Frankel, Ari}, \au{Pouransari, Hadi}, \au{Coletti, Filippo} \&
  \au{Mani, Ali}} \yr{2016}  \at{Settling of heated particles in homogeneous
  turbulence}.  \jt{Journal of Fluid Mechanics}  \bvol{792},  \pg{869--893}.

\bibitem[Garcia-Villalba {\em et~al.\/}(2012)Garcia-Villalba, Kidanemariam \&
  Uhlmann]{garcia2012}
{\sc \au{Garcia-Villalba, Manuel}, \au{Kidanemariam, Aman~G} \& \au{Uhlmann,
  Markus}} \yr{2012}  \at{Dns of vertical plane channel flow with finite-size
  particles: Voronoi analysis, acceleration statistics and particle-conditioned
  averaging}.  \jt{International Journal of Multiphase Flow}  \bvol{46},
  \pg{54--74}.

\bibitem[Gondret {\em et~al.\/}(2002)Gondret, Lance \& Petit]{gondret2002}
{\sc \au{Gondret, P}, \au{Lance, M} \& \au{Petit, L}} \yr{2002}  \at{Bouncing
  motion of spherical particles in fluids}.  \jt{Physics of fluids}
  \bvol{14}~(2),  \pg{643--652}.

\bibitem[Goto \& Vassilicos(2008)]{goto2008}
{\sc \au{Goto, Susumu} \& \au{Vassilicos, JC}} \yr{2008}  \at{Sweep-stick
  mechanism of heavy particle clustering in fluid turbulence}.  \jt{Physical
  review letters}  \bvol{100}~(5),  \pg{054503}.

\bibitem[Gualtieri {\em et~al.\/}(2009)Gualtieri, Picano \&
  Casciola]{gualtieri2009}
{\sc \au{Gualtieri, Picano}, \au{Picano, F} \& \au{Casciola, CM}} \yr{2009}
  \at{Anisotropic clustering of inertial particles in homogeneous shear flow}.
  \jt{Journal of Fluid Mechanics}  \bvol{629},  \pg{25--39}.

\bibitem[Gualtieri {\em et~al.\/}(2015)Gualtieri, Picano, Sardina \&
  Casciola]{gualtieri2015}
{\sc \au{Gualtieri, Paolo}, \au{Picano, F}, \au{Sardina, Gaetano} \&
  \au{Casciola, Carlo~Massimo}} \yr{2015}  \at{Exact regularized point particle
  method for multiphase flows in the two-way coupling regime}.  \jt{Journal of
  Fluid Mechanics}  \bvol{773},  \pg{520--561}.

\bibitem[Guha(2008)]{guha2008}
{\sc \au{Guha, Abhijit}} \yr{2008}  \at{Transport and deposition of particles
  in turbulent and laminar flow}.  \jt{Annu. Rev. Fluid Mech.}  \bvol{40},
  \pg{311--341}.

\bibitem[Gustavsson \& Mehlig(2016)]{gustavsson2016}
{\sc \au{Gustavsson, K} \& \au{Mehlig, B}} \yr{2016}  \at{Statistical models
  for spatial patterns of heavy particles in turbulence}.  \jt{Advances in
  Physics}  \bvol{65}~(1),  \pg{1--57}.

\bibitem[Hadinoto {\em et~al.\/}(2005)Hadinoto, Jones, Yurteri \&
  Curtis]{hadinoto2005}
{\sc \au{Hadinoto, K}, \au{Jones, EN}, \au{Yurteri, C} \& \au{Curtis, JS}}
  \yr{2005}  \at{Reynolds number dependence of gas-phase turbulence in
  gas--particle flows}.  \jt{International journal of multiphase flow}
  \bvol{31}~(4),  \pg{416--434}.

\bibitem[Hardalupas {\em et~al.\/}(1989)Hardalupas, Taylor \&
  Whitelaw]{hardalupas1989}
{\sc \au{Hardalupas, Y}, \au{Taylor, AMKP} \& \au{Whitelaw, James~Hunter}}
  \yr{1989}  \at{Velocity and particle-flux characteristics of turbulent
  particle-laden jets}.  \jt{Proc. R. Soc. Lond. A}  \bvol{426}~(1870),
  \pg{31--78}.

\bibitem[Hassan {\em et~al.\/}(1992)Hassan, Blanchat, Seeley~Jr \&
  Canaan]{hassan1992}
{\sc \au{Hassan, YA}, \au{Blanchat, TK}, \au{Seeley~Jr, CH} \& \au{Canaan, RE}}
  \yr{1992}  \at{Simultaneous velocity measurements of both components of a
  two-phase flow using particle image velocimetry}.  \jt{International Journal
  of Multiphase Flow}  \bvol{18}~(3),  \pg{371--395}.

\bibitem[Holtzer \& Collins(2002)]{holtzer2002}
{\sc \au{Holtzer, Gretchen~L} \& \au{Collins, Lance~R}} \yr{2002}
  \at{Relationship between the intrinsic radial distribution function for an
  isotropic field of particles and lower-dimensional measurements}.
  \jt{Journal of Fluid Mechanics}  \bvol{459},  \pg{93--102}.

\bibitem[Horwitz \& Mani(2016)]{horwitz2016}
{\sc \au{Horwitz, JAK} \& \au{Mani, Ali}} \yr{2016}  \at{Accurate calculation
  of stokes drag for point--particle tracking in two-way coupled flows}.
  \jt{Journal of Computational Physics}  \bvol{318},  \pg{85--109}.

\bibitem[Hrenya \& Sinclair(1997)]{hrenya1997}
{\sc \au{Hrenya, Christine~M} \& \au{Sinclair, Jennifer~L}} \yr{1997}
  \at{Effects of particle-phase turbulence in gas-solid flows}.  \jt{AIChE
  Journal}  \bvol{43}~(4),  \pg{853--869}.

\bibitem[Ireland \& Desjardins(2017)]{ireland2017}
{\sc \au{Ireland, Peter~J} \& \au{Desjardins, Olivier}} \yr{2017}
  \at{Improving particle drag predictions in euler--lagrange simulations with
  two-way coupling}.  \jt{Journal of Computational Physics}  \bvol{338},
  \pg{405--430}.

\bibitem[de~Jong {\em et~al.\/}(2010)de~Jong, Salazar, Woodward, Collins \&
  Meng]{dejong2010}
{\sc \au{de~Jong, J}, \au{Salazar, JPLC}, \au{Woodward, SH}, \au{Collins, LR}
  \& \au{Meng, H}} \yr{2010}  \at{Measurement of inertial particle clustering
  and relative velocity statistics in isotropic turbulence using holographic
  imaging}.  \jt{International Journal of Multiphase Flow}  \bvol{36}~(4),
  \pg{324--332}.

\bibitem[Joseph {\em et~al.\/}(2001)Joseph, Zenit, Hunt \&
  Rosenwinkel]{joseph2001}
{\sc \au{Joseph, GG}, \au{Zenit, R}, \au{Hunt, ML} \& \au{Rosenwinkel, AM}}
  \yr{2001}  \at{Particle--wall collisions in a viscous fluid}.  \jt{Journal of
  Fluid Mechanics}  \bvol{433},  \pg{329--346}.

\bibitem[Kaftori {\em et~al.\/}(1995{\natexlab{{\em a\/}}})Kaftori, Hetsroni \&
  Banerjee]{kaftori1995a}
{\sc \au{Kaftori, D}, \au{Hetsroni, G} \& \au{Banerjee, S}}
  \yr{1995{\natexlab{{\em a\/}}}}  \at{Particle behavior in the turbulent
  boundary layer. i. motion, deposition, and entrainment}.  \jt{Physics of
  Fluids}  \bvol{7}~(5),  \pg{1095--1106}.

\bibitem[Kaftori {\em et~al.\/}(1995{\natexlab{{\em b\/}}})Kaftori, Hetsroni \&
  Banerjee]{kaftori1995b}
{\sc \au{Kaftori, D}, \au{Hetsroni, G} \& \au{Banerjee, S}}
  \yr{1995{\natexlab{{\em b\/}}}}  \at{Particle behavior in the turbulent
  boundary layer. ii. velocity and distribution profiles}.  \jt{Physics of
  Fluids}  \bvol{7}~(5),  \pg{1107--1121}.

\bibitem[Khalitov \& Longmire(2002)]{khalitov2002}
{\sc \au{Khalitov, DA} \& \au{Longmire, EK}} \yr{2002}  \at{Simultaneous
  two-phase piv by two-parameter phase discrimination}.  \jt{Experiments in
  fluids}  \bvol{32}~(2),  \pg{252--268}.

\bibitem[Khalitov \& Longmire(2003)]{khalitov2003}
{\sc \au{Khalitov, Daniel~A} \& \au{Longmire, Ellen~K}} \yr{2003} Effect of
  particle size on velocity correlations in turbulent channel flow.  \bt{In
  {\em ASME/JSME 2003 4th Joint Fluids Summer Engineering Conference\/}},
  \pg{pp. 445--453}. American Society of Mechanical Engineers.

\bibitem[Kiger \& Pan(2000)]{kiger2000}
{\sc \au{Kiger, KT} \& \au{Pan, C}} \yr{2000}  \at{Piv technique for the
  simultaneous measurement of dilute two-phase flows}.  \jt{Journal of fluids
  engineering}  \bvol{122}~(4),  \pg{811--818}.

\bibitem[Kiger \& Pan(2002)]{kiger2002}
{\sc \au{Kiger, KT} \& \au{Pan, C}} \yr{2002}  \at{Suspension and turbulence
  modification effects of solid particulates on a horizontal turbulent channel
  flow}.  \jt{J. Turbulence}  \bvol{3}~(19),  \pg{1--17}.

\bibitem[Kim {\em et~al.\/}(1987)Kim, Moin \& Moser]{kim1987}
{\sc \au{Kim, John}, \au{Moin, Parviz} \& \au{Moser, Robert}} \yr{1987}
  \at{Turbulence statistics in fully developed channel flow at low reynolds
  number}.  \jt{Journal of fluid mechanics}  \bvol{177},  \pg{133--166}.

\bibitem[Kleinstreuer \& Zhang(2010)]{kleinstreuer2010}
{\sc \au{Kleinstreuer, C} \& \au{Zhang, Z}} \yr{2010}  \at{Airflow and particle
  transport in the human respiratory system}.  \jt{Annual review of fluid
  mechanics}  \bvol{42},  \pg{301--334}.

\bibitem[Knowles \& Kiger(2012)]{knowles2012}
{\sc \au{Knowles, Philip~L} \& \au{Kiger, Ken~T}} \yr{2012}  \at{Quantification
  of dispersed phase concentration using light sheet imaging methods}.
  \jt{Experiments in fluids}  \bvol{52}~(3),  \pg{697--708}.

\bibitem[Kuerten \& Vreman(2015)]{kuerten2015}
{\sc \au{Kuerten, Johannes~GM} \& \au{Vreman, AW}} \yr{2015}  \at{Effect of
  droplet interaction on droplet-laden turbulent channel flow}.  \jt{Physics of
  fluids}  \bvol{27}~(5),  \pg{053304}.

\bibitem[Kulick {\em et~al.\/}(1994)Kulick, Fessler \& Eaton]{kulick1994}
{\sc \au{Kulick, Jonathan~D}, \au{Fessler, John~R} \& \au{Eaton, John~K}}
  \yr{1994}  \at{Particle response and turbulence modification in fully
  developed channel flow}.  \jt{Journal of Fluid Mechanics}  \bvol{277},
  \pg{109--134}.

\bibitem[Kussin \& Sommerfeld(2002)]{kussin2002}
{\sc \au{Kussin, J} \& \au{Sommerfeld, M}} \yr{2002}  \at{Experimental studies
  on particle behaviour and turbulence modification in horizontal channel flow
  with different wall roughness}.  \jt{Experiments in Fluids}  \bvol{33}~(1),
  \pg{143--159}.

\bibitem[Li {\em et~al.\/}(2016)Li, Luo \& Fan]{li2016}
{\sc \au{Li, Dong}, \au{Luo, Kun} \& \au{Fan, Jianren}} \yr{2016}
  \at{Modulation of turbulence by dispersed solid particles in a spatially
  developing flat-plate boundary layer}.  \jt{Journal of Fluid Mechanics}
  \bvol{802},  \pg{359--394}.

\bibitem[Li {\em et~al.\/}(2012)Li, Wang, Liu, Chen \& Zheng]{li2012}
{\sc \au{Li, Jing}, \au{Wang, Hanfeng}, \au{Liu, Zhaohui}, \au{Chen, Sheng} \&
  \au{Zheng, Chuguang}} \yr{2012}  \at{An experimental study on turbulence
  modification in the near-wall boundary layer of a dilute gas-particle channel
  flow}.  \jt{Experiments in fluids}  \bvol{53}~(5),  \pg{1385--1403}.

\bibitem[Li {\em et~al.\/}(2001)Li, McLaughlin, Kontomaris \& Portela]{li2001}
{\sc \au{Li, Yiming}, \au{McLaughlin, John~B}, \au{Kontomaris, K} \&
  \au{Portela, L}} \yr{2001}  \at{Numerical simulation of particle-laden
  turbulent channel flow}.  \jt{Physics of Fluids}  \bvol{13}~(10),
  \pg{2957--2967}.

\bibitem[Lin {\em et~al.\/}(2017)Lin, Shao, Yu \& Wang]{lin2017}
{\sc \au{Lin, Zhao-wu}, \au{Shao, Xue-ming}, \au{Yu, Zhao-sheng} \& \au{Wang,
  Lian-ping}} \yr{2017}  \at{Effects of finite-size heavy particles on the
  turbulent flows in a square duct}.  \jt{Journal of Hydrodynamics}
  \bvol{29}~(2),  \pg{272--282}.

\bibitem[Liu \& Agarwal(1974)]{liu1974}
{\sc \au{Liu, Benjamin~YH} \& \au{Agarwal, Jugal~K}} \yr{1974}
  \at{Experimental observation of aerosol deposition in turbulent flow}.
  \jt{Journal of Aerosol Science}  \bvol{5}~(2),  \pg{145--155}.

\bibitem[Marchioli \& Soldati(2002)]{marchioli2002}
{\sc \au{Marchioli, Cristian} \& \au{Soldati, Alfredo}} \yr{2002}
  \at{Mechanisms for particle transfer and segregation in a turbulent boundary
  layer}.  \jt{Journal of fluid Mechanics}  \bvol{468},  \pg{283--315}.

\bibitem[Marchioli {\em et~al.\/}(2008)Marchioli, Soldati, Kuerten, Arcen,
  Taniere, Goldensoph, Squires, Cargnelutti \& Portela]{marchioli2008}
{\sc \au{Marchioli, Ch}, \au{Soldati, A}, \au{Kuerten, JGM}, \au{Arcen, B},
  \au{Taniere, A}, \au{Goldensoph, G}, \au{Squires, KD}, \au{Cargnelutti, MF}
  \& \au{Portela, LM}} \yr{2008}  \at{Statistics of particle dispersion in
  direct numerical simulations of wall-bounded turbulence: results of an
  international collaborative benchmark test}.  \jt{International Journal of
  Multiphase Flow}  \bvol{34}~(9),  \pg{879--893}.

\bibitem[Masi {\em et~al.\/}(2014)Masi, Simonin, Riber, Sierra \&
  Gicquel]{masi2014}
{\sc \au{Masi, Enrica}, \au{Simonin, Olivier}, \au{Riber, Eleonore},
  \au{Sierra, P} \& \au{Gicquel, Laurent~YM}} \yr{2014}  \at{Development of an
  algebraic-closure-based moment method for unsteady eulerian simulations of
  particle-laden turbulent flows in very dilute regime}.  \jt{International
  Journal of Multiphase Flow}  \bvol{58},  \pg{257--278}.

\bibitem[Maxey(1987)]{maxey1987}
{\sc \au{Maxey, MR}} \yr{1987}  \at{The gravitational settling of aerosol
  particles in homogeneous turbulence and random flow fields}.  \jt{Journal of
  Fluid Mechanics}  \bvol{174},  \pg{441--465}.

\bibitem[McLaughlin(1989)]{mclaughlin1989}
{\sc \au{McLaughlin, John~B}} \yr{1989}  \at{Aerosol particle deposition in
  numerically simulated channel flow}.  \jt{Physics of Fluids A: Fluid
  Dynamics}  \bvol{1}~(7),  \pg{1211--1224}.

\bibitem[Mehrabadi {\em et~al.\/}(2018)Mehrabadi, Horwitz, Subramaniam \&
  Mani]{mehrabadi2018}
{\sc \au{Mehrabadi, M}, \au{Horwitz, JAK}, \au{Subramaniam, S} \& \au{Mani, A}}
  \yr{2018}  \at{A direct comparison of particle-resolved and point-particle
  methods in decaying turbulence}.  \jt{Journal of Fluid Mechanics}
  \bvol{850},  \pg{336--369}.

\bibitem[Meneguz \& Reeks(2011)]{meneguz2011}
{\sc \au{Meneguz, Elena} \& \au{Reeks, Michael~W}} \yr{2011}  \at{Statistical
  properties of particle segregation in homogeneous isotropic turbulence}.
  \jt{Journal of Fluid Mechanics}  \bvol{686},  \pg{338--351}.

\bibitem[Monchaux {\em et~al.\/}(2010)Monchaux, Bourgoin \&
  Cartellier]{monchaux2010}
{\sc \au{Monchaux, Romain}, \au{Bourgoin, Micka{\"e}l} \& \au{Cartellier,
  Alain}} \yr{2010}  \at{Preferential concentration of heavy particles: a
  vorono{\"\i} analysis}.  \jt{Physics of Fluids}  \bvol{22}~(10),
  \pg{103304}.

\bibitem[Monchaux {\em et~al.\/}(2012)Monchaux, Bourgoin \&
  Cartellier]{monchaux2012}
{\sc \au{Monchaux, Romain}, \au{Bourgoin, Mickael} \& \au{Cartellier, Alain}}
  \yr{2012}  \at{Analyzing preferential concentration and clustering of
  inertial particles in turbulence}.  \jt{International Journal of Multiphase
  Flow}  \bvol{40},  \pg{1--18}.

\bibitem[Moser {\em et~al.\/}(1999)Moser, Kim \& Mansour]{moser1999}
{\sc \au{Moser, Robert~D}, \au{Kim, John} \& \au{Mansour, Nagi~N}} \yr{1999}
  \at{Direct numerical simulation of turbulent channel flow up to re $\tau$=
  590}.  \jt{Physics of fluids}  \bvol{11}~(4),  \pg{943--945}.

\bibitem[Nasr {\em et~al.\/}(2009)Nasr, Ahmadi \& McLaughlin]{nasr2009}
{\sc \au{Nasr, Hojjat}, \au{Ahmadi, Goodarz} \& \au{McLaughlin, John~B}}
  \yr{2009}  \at{A dns study of effects of particle--particle collisions and
  two-way coupling on particle deposition and phasic fluctuations}.
  \jt{Journal of Fluid Mechanics}  \bvol{640},  \pg{507--536}.

\bibitem[Nicolai {\em et~al.\/}(2013)Nicolai, Jacob \& Piva]{nicolai2013}
{\sc \au{Nicolai, Claudia}, \au{Jacob, B} \& \au{Piva, Renzo}} \yr{2013}
  \at{On the spatial distribution of small heavy particles in homogeneous shear
  turbulence}.  \jt{Physics of Fluids}  \bvol{25}~(8),  \pg{083301}.

\bibitem[Nilsen {\em et~al.\/}(2013)Nilsen, Andersson \& Zhao]{nilsen2013}
{\sc \au{Nilsen, Christopher}, \au{Andersson, Helge~I} \& \au{Zhao, Lihao}}
  \yr{2013}  \at{A vorono{\"\i} analysis of preferential concentration in a
  vertical channel flow}.  \jt{Physics of Fluids}  \bvol{25}~(11),
  \pg{115108}.

\bibitem[Nino \& Garcia(1996)]{nino1996}
{\sc \au{Nino, Y} \& \au{Garcia, MH}} \yr{1996}  \at{Experiments on
  particle—turbulence interactions in the near--wall region of an open
  channel flow: implications for sediment transport}.  \jt{Journal of Fluid
  Mechanics}  \bvol{326},  \pg{285--319}.

\bibitem[Ohmi \& Li(2000)]{ohmi2000}
{\sc \au{Ohmi, Kazuo} \& \au{Li, Hang-Yu}} \yr{2000}  \at{Particle-tracking
  velocimetry with new algorithms}.  \jt{Measurement Science and Technology}
  \bvol{11}~(6),  \pg{603}.

\bibitem[Oliveira {\em et~al.\/}(2017)Oliveira, van~der Geld \&
  Kuerten]{oliveira2017}
{\sc \au{Oliveira, JL~Goes}, \au{van~der Geld, CWM} \& \au{Kuerten,
  Johannes~GM}} \yr{2017}  \at{Concentration and velocity statistics of
  inertial particles in upward and downward pipe flow}.  \jt{Journal of fluid
  mechanics}  \bvol{822},  \pg{640--663}.

\bibitem[Pan \& Banerjee(1996)]{pan1996}
{\sc \au{Pan, Y.} \& \au{Banerjee, S.}} \yr{1996}  \at{Numerical simulation of
  particle interactions with wall turbulence}.  \jt{Phys. Fluids}
  \bvol{8}~(10),  \pg{2733--2755}.

\bibitem[Paris(2001)]{paris2001}
{\sc \au{Paris, Anthony~Dana}} \yr{2001}  \at{Turbulence attenuation in a
  particle-laden channel flow} .

\bibitem[Petersen {\em et~al.\/}(2019)Petersen, Baker \& Coletti]{petersen2019}
{\sc \au{Petersen, Alec~J}, \au{Baker, Lucia} \& \au{Coletti, Filippo}}
  \yr{2019}  \at{Experimental study of inertial particles clustering and
  settling in homogeneous turbulence}.  \jt{Journal of Fluid Mechanics}
  \bvol{864},  \pg{925--970}.

\bibitem[Picano {\em et~al.\/}(2015)Picano, Breugem \& Brandt]{picano2015}
{\sc \au{Picano, Francesco}, \au{Breugem, Wim-Paul} \& \au{Brandt, Luca}}
  \yr{2015}  \at{Turbulent channel flow of dense suspensions of neutrally
  buoyant spheres}.  \jt{Journal of Fluid Mechanics}  \bvol{764},
  \pg{463--487}.

\bibitem[Pope(2000)]{pope2000}
{\sc \au{Pope, S.~B.}} \yr{2000} {\em Turbulent Flows\/}.  \publ{Cambridge, UK:
  Cambridge Univ. Press}.

\bibitem[Rabencov {\em et~al.\/}(2014)Rabencov, Arca \& van Hout]{rabencov2014}
{\sc \au{Rabencov, B}, \au{Arca, J} \& \au{van Hout, R}} \yr{2014}
  \at{Measurement of polystyrene beads suspended in a turbulent square channel
  flow: Spatial distributions of velocity and number density}.
  \jt{International Journal of Multiphase Flow}  \bvol{62},  \pg{110--122}.

\bibitem[Reeks(1983)]{reeks1983}
{\sc \au{Reeks, MW}} \yr{1983}  \at{The transport of discrete particles in
  inhomogeneous turbulence}.  \jt{Journal of aerosol science}  \bvol{14}~(6),
  \pg{729--739}.

\bibitem[Reeks(2014)]{reeks2014}
{\sc \au{Reeks, Michael~W}} \yr{2014} Transport, mixing and agglomeration of
  particles in turbulent flows.  \bt{In {\em Journal of Physics: Conference
  Series\/}}, ,  \vol{vol. 530},  \pg{p. 012003}. IOP Publishing.

\bibitem[Richter \& Sullivan(2013)]{richter2013}
{\sc \au{Richter, David~H} \& \au{Sullivan, Peter~P}} \yr{2013}  \at{Momentum
  transfer in a turbulent, particle-laden couette flow}.  \jt{Physics of
  Fluids}  \bvol{25}~(5),  \pg{053304}.

\bibitem[Richter \& Sullivan(2014)]{richter2014}
{\sc \au{Richter, David~H} \& \au{Sullivan, Peter~P}} \yr{2014}
  \at{Modification of near-wall coherent structures by inertial particles}.
  \jt{Physics of Fluids}  \bvol{26}~(10),  \pg{103304}.

\bibitem[Righetti \& Romano(2004)]{righetti2004}
{\sc \au{Righetti, M} \& \au{Romano, Giovanni~Paolo}} \yr{2004}
  \at{Particle--fluid interactions in a plane near-wall turbulent flow}.
  \jt{Journal of Fluid Mechanics}  \bvol{505},  \pg{93--121}.

\bibitem[Robinson(1991)]{robinson1991}
{\sc \au{Robinson, Stephen~K}} \yr{1991}  \at{Coherent motions in the turbulent
  boundary layer}.  \jt{Annual Review of Fluid Mechanics}  \bvol{23}~(1),
  \pg{601--639}.

\bibitem[Rouson \& Eaton(2001)]{rouson2001}
{\sc \au{Rouson, Damian~WI} \& \au{Eaton, John~K}} \yr{2001}  \at{On the
  preferential concentration of solid particles in turbulent channel flow}.
  \jt{Journal of Fluid Mechanics}  \bvol{428},  \pg{149--169}.

\bibitem[Sahu {\em et~al.\/}(2014)Sahu, Hardalupas \& Taylor]{sahu2014}
{\sc \au{Sahu, S}, \au{Hardalupas, Y} \& \au{Taylor, AMKP}} \yr{2014}
  \at{Droplet--turbulence interaction in a confined polydispersed spray: effect
  of droplet size and flow length scales on spatial droplet--gas velocity
  correlations}.  \jt{Journal of Fluid Mechanics}  \bvol{741},  \pg{98--138}.

\bibitem[Sahu {\em et~al.\/}(2016)Sahu, Hardalupas \& Taylor]{sahu2016}
{\sc \au{Sahu, S}, \au{Hardalupas, Y} \& \au{Taylor, AMKP}} \yr{2016}
  \at{Droplet--turbulence interaction in a confined polydispersed spray: effect
  of turbulence on droplet dispersion}.  \jt{Journal of Fluid Mechanics}
  \bvol{794},  \pg{267--309}.

\bibitem[Salazar {\em et~al.\/}(2008)Salazar, De~Jong, Cao, Woodward, Meng \&
  Collins]{salazar2008}
{\sc \au{Salazar, Juan~PLC}, \au{De~Jong, Jeremy}, \au{Cao, Lujie},
  \au{Woodward, Scott~H}, \au{Meng, Hui} \& \au{Collins, Lance~R}} \yr{2008}
  \at{Experimental and numerical investigation of inertial particle clustering
  in isotropic turbulence}.  \jt{Journal of Fluid Mechanics}  \bvol{600},
  \pg{245--256}.

\bibitem[Sardina {\em et~al.\/}(2012{\natexlab{{\em a\/}}})Sardina, Schlatter,
  Brandt, Picano \& Casciola]{sardina2012a}
{\sc \au{Sardina, G}, \au{Schlatter, Philipp}, \au{Brandt, Luca}, \au{Picano,
  F} \& \au{Casciola, Carlo~Massimo}} \yr{2012{\natexlab{{\em a\/}}}}  \at{Wall
  accumulation and spatial localization in particle-laden wall flows}.
  \jt{Journal of Fluid Mechanics}  \bvol{699},  \pg{50--78}.

\bibitem[Sardina {\em et~al.\/}(2012{\natexlab{{\em b\/}}})Sardina, Schlatter,
  Picano, Casciola, Brandt \& Henningson]{sardina2012b}
{\sc \au{Sardina, Gaetano}, \au{Schlatter, Philipp}, \au{Picano, Francesco},
  \au{Casciola, CM}, \au{Brandt, Luca} \& \au{Henningson, Dan~Stafan}}
  \yr{2012{\natexlab{{\em b\/}}}}  \at{Self-similar transport of inertial
  particles in a turbulent boundary layer}.  \jt{Journal of Fluid Mechanics}
  \bvol{706},  \pg{584--596}.

\bibitem[Schneiders {\em et~al.\/}(2017)Schneiders, Meinke \&
  Schr{\"o}der]{schneiders2017}
{\sc \au{Schneiders, Lennart}, \au{Meinke, Matthias} \& \au{Schr{\"o}der,
  Wolfgang}} \yr{2017}  \at{Direct particle--fluid simulation of
  kolmogorov-length-scale size particles in decaying isotropic turbulence}.
  \jt{Journal of Fluid Mechanics}  \bvol{819},  \pg{188--227}.

\bibitem[Shokri {\em et~al.\/}(2017)Shokri, Ghaemi, Nobes \&
  Sanders]{shokri2017}
{\sc \au{Shokri, R}, \au{Ghaemi, S}, \au{Nobes, DS} \& \au{Sanders, RS}}
  \yr{2017}  \at{Investigation of particle-laden turbulent pipe flow at
  high-reynolds-number using particle image/tracking velocimetry (piv/ptv)}.
  \jt{International Journal of Multiphase Flow}  \bvol{89},  \pg{136--149}.

\bibitem[Soldati \& Marchioli(2009)]{soldati2009}
{\sc \au{Soldati, Alfredo} \& \au{Marchioli, Cristian}} \yr{2009}  \at{Physics
  and modelling of turbulent particle deposition and entrainment: Review of a
  systematic study}.  \jt{International Journal of Multiphase Flow}
  \bvol{35}~(9),  \pg{827--839}.

\bibitem[Squires \& Eaton(1991)]{squires1991}
{\sc \au{Squires, Kyle~D} \& \au{Eaton, John~K}} \yr{1991}  \at{Preferential
  concentration of particles by turbulence}.  \jt{Physics of Fluids A: Fluid
  Dynamics}  \bvol{3}~(5),  \pg{1169--1178}.

\bibitem[Sumbekova {\em et~al.\/}(2017)Sumbekova, Cartellier, Aliseda \&
  Bourgoin]{sumbekova2017}
{\sc \au{Sumbekova, Sholpan}, \au{Cartellier, Alain}, \au{Aliseda, Alberto} \&
  \au{Bourgoin, Mickael}} \yr{2017}  \at{Preferential concentration of inertial
  sub-kolmogorov particles: The roles of mass loading of particles, stokes
  numbers, and reynolds numbers}.  \jt{Physical Review Fluids}  \bvol{2}~(2),
  \pg{024302}.

\bibitem[Sundaram \& Collins(1997)]{sundaram1997}
{\sc \au{Sundaram, Shivshankar} \& \au{Collins, Lance~R}} \yr{1997}
  \at{Collision statistics in an isotropic particle-laden turbulent suspension.
  part 1. direct numerical simulations}.  \jt{Journal of Fluid Mechanics}
  \bvol{335},  \pg{75--109}.

\bibitem[Sundaram \& Collins(1999)]{sundaram1999}
{\sc \au{Sundaram, Shivshankar} \& \au{Collins, Lance~R}} \yr{1999}  \at{A
  numerical study of the modulation of isotropic turbulence by suspended
  particles}.  \jt{Journal of Fluid Mechanics}  \bvol{379},  \pg{105--143}.

\bibitem[Taniere {\em et~al.\/}(1997)Taniere, Oesterle \& Monnier]{taniere1997}
{\sc \au{Taniere, A}, \au{Oesterle, B} \& \au{Monnier, JC}} \yr{1997}  \at{On
  the behaviour of solid particles in a horizontal boundary layer with
  turbulence and saltation effects}.  \jt{Experiments in Fluids}
  \bvol{23}~(6),  \pg{463--471}.

\bibitem[Vance {\em et~al.\/}(2006)Vance, Squires \& Simonin]{vance2006}
{\sc \au{Vance, Marion~W}, \au{Squires, Kyle~D} \& \au{Simonin, Olivier}}
  \yr{2006}  \at{Properties of the particle velocity field in gas-solid
  turbulent channel flow}.  \jt{Physics of Fluids}  \bvol{18}~(6),
  \pg{063302}.

\bibitem[Varaksin {\em et~al.\/}(2000)Varaksin, Polezhaev \&
  Polyakov]{varaksin2000}
{\sc \au{Varaksin, A~Yu}, \au{Polezhaev, Yu~V} \& \au{Polyakov, Anatoly~F}}
  \yr{2000}  \at{Effect of particle concentration on fluctuating velocity of
  the disperse phase for turbulent pipe flow}.  \jt{International journal of
  heat and fluid flow}  \bvol{21}~(5),  \pg{562--567}.

\bibitem[Vreman(2007)]{vreman2007}
{\sc \au{Vreman, AW}} \yr{2007}  \at{Turbulence characteristics of
  particle-laden pipe flow}.  \jt{Journal of fluid mechanics}  \bvol{584},
  \pg{235--279}.

\bibitem[Vreman(2015)]{vreman2015}
{\sc \au{Vreman, AW}} \yr{2015}  \at{Turbulence attenuation in particle-laden
  flow in smooth and rough channels}.  \jt{Journal of Fluid Mechanics}
  \bvol{773},  \pg{103--136}.

\bibitem[Wang {\em et~al.\/}(2017)Wang, Abbas \& Climent]{wang2017}
{\sc \au{Wang, Guiquan}, \au{Abbas, Micheline} \& \au{Climent, {\'E}ric}}
  \yr{2017}  \at{Modulation of large-scale structures by neutrally buoyant and
  inertial finite-size particles in turbulent couette flow}.  \jt{Physical
  Review Fluids}  \bvol{2}~(8),  \pg{084302}.

\bibitem[Wang \& Richter(2018)]{wang2018}
{\sc \au{Wang, Guiquan} \& \au{Richter, David}} \yr{2018}  \at{Modulation of
  the turbulence regeneration cycle by inertial particles in planar couette
  flow}.  \jt{arXiv preprint arXiv:1807.02107} .

\bibitem[Wang \& Maxey(1993)]{wang1993}
{\sc \au{Wang, Lian-Ping} \& \au{Maxey, Martin~R}} \yr{1993}  \at{Settling
  velocity and concentration distribution of heavy particles in homogeneous
  isotropic turbulence}.  \jt{Journal of fluid mechanics}  \bvol{256},
  \pg{27--68}.

\bibitem[Wei {\em et~al.\/}(2005)Wei, Schmidt \& McMurtry]{wei2005}
{\sc \au{Wei, Tie}, \au{Schmidt, Rodney} \& \au{McMurtry, Patrick}} \yr{2005}
  \at{Comment on the clauser chart method for determining the friction
  velocity}.  \jt{Experiments in fluids}  \bvol{38}~(5),  \pg{695--699}.

\bibitem[Wilkinson \& Mehlig(2005)]{wilkinson2005}
{\sc \au{Wilkinson, M} \& \au{Mehlig, Bernhard}} \yr{2005}  \at{Caustics in
  turbulent aerosols}.  \jt{EPL (Europhysics Letters)}  \bvol{71}~(2),
  \pg{186}.

\bibitem[Wood {\em et~al.\/}(2005)Wood, Hwang \& Eaton]{wood2005}
{\sc \au{Wood, AM}, \au{Hwang, W} \& \au{Eaton, JK}} \yr{2005}
  \at{Preferential concentration of particles in homogeneous and isotropic
  turbulence}.  \jt{International journal of multiphase flow}
  \bvol{31}~(10-11),  \pg{1220--1230}.

\bibitem[Wu {\em et~al.\/}(2006)Wu, Wang, Liu, Li, Zhang \& Zheng]{wu2006}
{\sc \au{Wu, Yi}, \au{Wang, Hangfeng}, \au{Liu, Zhaohui}, \au{Li, Jing},
  \au{Zhang, Liqi} \& \au{Zheng, Chuguang}} \yr{2006}  \at{Experimental
  investigation on turbulence modification in a horizontal channel flow at
  relatively low mass loading}.  \jt{Acta Mechanica Sinica}  \bvol{22}~(2),
  \pg{99--108}.

\bibitem[Yang \& Shy(2005)]{yang2005}
{\sc \au{Yang, TS} \& \au{Shy, SS}} \yr{2005}  \at{Two-way interaction between
  solid particles and homogeneous air turbulence: particle settling rate and
  turbulence modification measurements}.  \jt{Journal of fluid mechanics}
  \bvol{526},  \pg{171--216}.

\bibitem[Young \& Leeming(1997)]{young1997}
{\sc \au{Young, John} \& \au{Leeming, Angus}} \yr{1997}  \at{A theory of
  particle deposition in turbulent pipe flow}.  \jt{Journal of Fluid Mechanics}
   \bvol{340},  \pg{129--159}.

\bibitem[Zamansky {\em et~al.\/}(2016)Zamansky, Coletti, Massot \&
  Mani]{zamansky2016}
{\sc \au{Zamansky, R{\'e}mi}, \au{Coletti, Filippo}, \au{Massot, Marc} \&
  \au{Mani, Ali}} \yr{2016}  \at{Turbulent thermal convection driven by heated
  inertial particles}.  \jt{Journal of Fluid Mechanics}  \bvol{809},
  \pg{390--437}.

\bibitem[Zhang \& Ahmadi(2000)]{zhang2000}
{\sc \au{Zhang, Haifeng} \& \au{Ahmadi, Goodarz}} \yr{2000}  \at{Aerosol
  particle transport and deposition in vertical and horizontal turbulent duct
  flows}.  \jt{Journal of Fluid Mechanics}  \bvol{406},  \pg{55--80}.

\bibitem[Zhao {\em et~al.\/}(2010)Zhao, Andersson \& Gillissen]{zhao2010}
{\sc \au{Zhao, LH}, \au{Andersson, Helge~I} \& \au{Gillissen, JJJ}} \yr{2010}
  \at{Turbulence modulation and drag reduction by spherical particles}.
  \jt{Physics of Fluids}  \bvol{22}~(8),  \pg{081702}.

\end{thebibliography}

\end{document}